\documentclass[12pt]{article}

\usepackage[fleqn]{amsmath}
\usepackage{amsfonts}
\usepackage{amssymb}
\usepackage{fullpage}
\usepackage{epsfig}

\newcommand{\scr}{\cal}

\DeclareMathOperator{\Artanh}{artanh}
\DeclareMathOperator{\supp}{supp}

\newcommand{\dd}{\mathrm{d}}
\newcommand{\dds}{\frac{\dd\phantom{s}}{\dd \tau}}

\newcommand{\ddt}{\frac{\partial\phantom{t}}{\partial t}}

\DeclareMathOperator{\tr}{Tr}

\newcommand{\rotM}{{\scr M}_{\text b}}

\newcommand{\refeq}[1]{(\ref{#1})}

\newcommand{\mz}{m_{\text{b}}}
\newcommand{\Iz}{I_{\text{b}}}

\newcommand{\LoF}{{\scr F}_{\textsc{l}}}
\newcommand{\boF}{{\scr F}_{\text{body}}}
\newcommand{\FeWaF}{{\scr F}_{\textsc{fw}}}

\newcommand{\fe}{f_{\text{e}}}
\newcommand{\fm}{f_{\text{m}}}

\newcommand{\vect}[1] {\boldsymbol{{ #1}} }

\newcommand{\qv}[1]{{\textbf{\textsl{#1}}}}

\newcommand{\tenseur}[1]{{\textbf{\textsf{#1}}}}

\newcommand{\tens}{\otimes}     

\newcommand{\apriori}{\emph{a priori}}

\newcommand{\Rset}{\mathbb{R}}
\newcommand{\Xiq}{{\Xi_{\zQ}}}

\newcommand{\AQ}{\qv{A}}                

\newcommand{\FQ}{\tenseur{F}}           
\newcommand{\FQred}{{\FQ^\prime}}	
\newcommand{\gQ}{\tenseur{g}}           
\newcommand{\LQ}{\tenseur{L}}           
\newcommand{\IQb}{\tenseur{I}_{\text{b}}}    
\newcommand{\NQ}{\tenseur{N}}           
\newcommand{\SQ}{\tenseur{S}}           
\newcommand{\SQb}{{\tenseur{S}_{\text{b}}}} 
\newcommand{\SQe}{{\tenseur{S}_{\text{f}}}} 
\newcommand{\tQT}{\tenseur{t}}          

\newcommand{\aQ}{\qv{a}}                
\newcommand{\eQ}{\qv{e}}                
\newcommand{\fQ}{\qv{f}}        	
\newcommand{\JQ}{\qv{J}}                
\newcommand{\pQ}{\qv{p}}                
\newcommand{\pQb}{{\qv{p}}_{\text{b}}}  
\newcommand{\PQ}{\qv{P}}                
\newcommand{\sQb}{{\qv{s}}_{\text{b}}}  
\newcommand{\sQe}{{\qv{s}}_{\text{f}}}  
\newcommand{\tQ}{\qv{t}}        	
\newcommand{\uQ}{\qv{u}}                
\newcommand{\vQ}{\qv{v}}                
\newcommand{\xQ}{\qv{x}}                
\newcommand{\yQ}{\qv{y}}                
\newcommand{\zQ}{\qv{q}}                

\newcommand{\aV}{\vect{a}}              
\newcommand{\jV}{{\vect{j}}}		
\newcommand{\pV}{{\vect{p}}}            
\newcommand{\pVb}{{\vect{p}}_{\text{b}}} 
\newcommand{\qV}{{\vect{q}}}            
\newcommand{\sV}{{\vect{s}}}            
\newcommand{\sVb}{{\vect{s}}_{\text{b}}} 
\newcommand{\sVe}{{\vect{s}}_{\text{f}}}
\newcommand{\tVE}{\vect{t}^E}           
\newcommand{\vV}{{\vect{v}}}            
\newcommand{\xV}{\vect{x}}              
\newcommand{\yV}{\vect{y}}              

\newcommand{\muV}{\vect{\mu}}           
\newcommand{\sigmaVB}{\vect{\sigma}^B}  

\newcommand{\AV}{\pmb{{\cal A}}}
\newcommand{\BV}{\pmb{{\cal B}}}
\newcommand{\EV}{\pmb{{\cal E}}}
\newcommand{\GV}{\pmb{{\cal G}}}
\newcommand{\LV}{\pmb{{\cal L}}}
\newcommand{\PV}{\pmb{{\cal P}}}

\newcommand{\nab}{\vect{\nabla}}
\newcommand{\nabQ}{\vect{\nabla}_{\text{g}}}
\newcommand{\lapQ}{{\Delta}_{\text{g}}}
\newcommand{\wop}{\vect{\square}}

\newcommand{\EulerQ}{\mbox{$\vect{\Omega}_{\text{\textsc{e}}}$}}

\newcommand{\eulerQ}{\mbox{$\qv{w}_{\text{\textsc{e}}}$}}
\newcommand{\eulerV}{\mbox{$\vect\omega_{\text{\textsc{e}}}$}}
\newcommand{\euler}{\mbox{$\omega_{\text{\textsc{e}}}$}}
\newcommand{\ThomasQ}{\mbox{$\vect{\Omega}_{\text{\textsc{fw}}}{}$}}

\newcommand{\thomasV}{\mbox{$\vect\omega_{\text{\textsc{t}}}{}$}}

\newcommand{\OmegaQ}{\vect{\Omega}}

\newcommand{\SigmaQ}{\mbox{$\vect{\Sigma}$}}
\newcommand{\thetaQ}{\vect{\theta}}
\newcommand{\varthetaQ}{\vect{\vartheta}}
\newcommand{\Mfp}{\tenseur{M}_{\text{fp}}} 	
\newcommand{\Mmink}{\tenseur{M}} 		
\newcommand{\Mnodv}{\tenseur{M}_{\text{\textsc{n}}}} 
\newcommand{\Mbare}{\tenseur{M}_{\text{b}}} 	
\newcommand{\RC}{R_{\textsc{c}}}        

\newcommand{\qdot}[1]{{\stackrel{\,\circ}{{#1}}}}
\newcommand{\qddot}[1]{{\stackrel{\circ\circ}{{#1}}}}
\newcommand{\qdddot}[1]{{\stackrel{\circ\circ\circ}{{#1}}}}

\newcommand{\abs}[1]{\left| #1 \right|}
\newcommand{\norm}[1]{\left\| #1 \right\| }

\newcommand{\defeg}{\stackrel{\textrm{\tiny def}}{=}}

\renewcommand{\leq}{\leqslant}
\renewcommand{\geq}{\geqslant}

\newcommand{\Ab}{{\scr A}_{\text{bare}}}  
\newcommand{\Af}{{\scr A}_{\text{field}}} 
\newcommand{\Abf}{{\scr A}_{\text{coupl}}} 

\newcommand{\Lm}{ {\cal L}_{\text{bare}}}  
\newcommand{\Lf}{ {\cal L}_{\text{field}}} 
\newcommand{\Lmf}{{\cal L}_{\text{coupl}}} 

\newcommand{\Lb}{{ L}_{\text{b}}}       
\newcommand{\Lc}{{ L}_{\text{c}}}       

\newcommand{\sssp}{{\scriptscriptstyle\parallel}}

\reversemarginpar

\numberwithin{equation}{section}

\begin{document}


\title{\uppercase{Mass and spin renormalization \\
                  in Lorentz electrodynamics}}

\author{\textbf{Walter APPEL}$^*$ and \textbf{Michael K.-H. KIESSLING}\\
                Department of Mathematics, Rutgers University\\
                110 Frelinghuysen Rd., PISCATAWAY, NJ 08854, USA\\ 
        $^*$\emph{On leave from}: Laboratoire de Physique\\
                Unit\'e de recherche 1325 associ\'ee au CNRS\\
                \'Ecole normale sup\'erieure de Lyon\\
                46 all\'ee d'Italie, 69\,364 LYON Cedex 07 France\\ \\
\textbf{Revised version of October 04, 2000.}\\
\textrm{Typeset with \LaTeX\ on:}}
\maketitle
\thispagestyle{empty}

\begin{abstract}
{\noindent 
        Lorentz' objection to the Uhlenbeck-Goudsmit proposal of  
a spin magnetic moment of the electron, namely that the electron's 
equatorial rotation speed would exceed the speed of light by a 
factor $\approx 10$, mutated into an objection against 
Lorentz electrodynamics (LED) itself when the spin magnetic 
moment became established. 
        However, Lorentz' renormalization calculation, based on 
the  early $20^{\rm th}$ century's notion of 
a purely electromagnetic electron, does not qualify as proper from
a modern perspective.
        This paper shows that  renormalization treated properly 
does lead to a  mathematically consistent and physically viable LED.
	A new, relativistically covariant massive LED is presented in 
which the bare particle has a finite positive bare rest mass and 
moment of inertia.
        The particle's electromagnetic self-interaction renormalizes 
its mass and spin. 
        Most crucially, the  renormalized particle is a 
\emph{soliton}: after any scattering process
its rest mass and spin magnitude are dynamically restored to their 
pre-scattering values.
        This guarantees that ``an electron remains an electron,''
poetically speaking.
        A renormalization flow study of the limit 
of vanishing bare rest mass is conducted for this model.
        This limit yields a  purely electromagnetic classical field theory 
with ultra-violet cutoff at about the electron's Compton wavelength!
        The renormalized limit model matches the empirical electron data 
as orderly as one can hope for at the level of Lorentz theory. 
        In particular, no superluminal equatorial speeds occur.
}
\end{abstract}

\vfill
\hrule 
\smallskip
 {\copyright} (2000) The authors. 
 Reproduction for non-commercial purposes of this article, 
 in its entirety and by any means, is permitted.

\newpage

                \section{Introduction}

        The roots of the mass renormalization program of
quantum electrodynamics (QED) trace back to the late 
$19^{\rm th}$ and early $20^{\rm th}$ centuries' notion 
that the electron's inertia has a purely electromagnetic origin.
        This notion was central to the pre-quantum attempts 
of the time to construct a consistent electromagnetic theory 
of matter and radiation that would unify electrodynamics and 
mechanics, now known as classical electron theory. 
	When  Lorentz proposed that Abraham's~\cite{abraham, abrahamBOOK} 
purely electromagnetic, strictly rigid spherical electron (see Appendix A.2) 
had to be replaced by an electron that undergoes the Lorentz-Fitzgerald 
contraction, the laws of special relativity began to take shape in the 
papers of 
Lorentz~\cite{lorentzACAD}, and  Poincar\'e~\cite{poincare};
see~\cite{millerBOOK} for an excellent account of the historical developments. 
        Ironically, special relativity and renormalization 
later contributed to the demise of relativistic 
Lorentz electrodynamics (LED) more than quantum mechanics.
	Of course, quantum mechanics removed LED from the 
list of contenders for a truly fundamental electromagnetic 
theory of matter and radiation, but it did not jeopardize  
the viability of LED as a physical theory in the classical limit. 
        After all, Bohr's correspondence principle was undisputed 
in the quest for the fundamental quantum theory, and 
as such one would not have been, nor would one be unhappy with 
a consistent classical microscopic relativistic theory of 
electromagnetic matter and radiation that could serve as the 
empirically correct classical limit of a consistent quantum theory.
        Special relativity on the other hand had emancipated itself 
in the works of Einstein~\cite{einsteinA, einsteinB} 
and Minkowski~\cite{minkowski}
as a theory of spacetime structure independent of LED or any other
particular matter model~\cite{millerBOOK}.
	Thus, when Lorentz objected to the 
Uhlenbeck--Goudsmit~\cite{uhlenbeckgoudsmit} 
proposal of an electron spin magnetic moment by pointing out that LED 
would forbid the required magnitude of about one Bohr magneton because 
the purely electromagnetic electron would have to reach an impossible 
equatorial rotation speed of about 10 times the speed of light 
(see \cite{tomonagaBOOK}, p. 35), he had furnished an argument that
--- eventually ---  weighed heavily against the physical sensibility 
of LED itself, and which has persisted ever since.

        Judged from our modern perspective on mass
renormalization~\cite{glimmjaffeBOOK, fernandezfroehlichsokalBOOK},
the purely electromagnetic calculations by 
Lorentz~\cite{lorentzACAD, lorentzBOOKb}
and his contemporaries do, however, not qualify as proper.
        A proper renormalization analysis would investigate the `flow' 
toward vanishing bare rest mass for a consistent relativistic LED 
\emph{with} positive bare rest mass (`massive' LED) in which the bare 
particle parameters of the model can be chosen such that the particle charge, 
the particle spin magnetic moment, and the renormalized particle mass match 
the empirical electron data without involving superluminal gyration speeds 
of the model electron.
	If the limit of vanishing bare rest mass exists, this limit would
define a renormalized LED which in a sense completes Lorentz' program of a 
purely electromagnetic electron theory, and whose physical viability 
is no longer objectable to on basis of the superluminal arguments that
grounded the original Lorentz theory.

	In this paper we will show that such a properly renormalized approach 
to the Lorentz program of electrodynamics does indeed lead to a satisfactory
Lorentz electrodynamics. 
	For simplicity we consider only a single Lorentz electron coupled 
to the Maxwell--Lorentz fields in the whole spacetime.
	The extension to a many particle theory does not pose any 
difficulties.

	We first present a novel, manifestly Lorentz-covariant massive LED
which displays several crucial features expected of a realistic, 
consistent classical electrodynamics. In particular:
 
\begin{itemize}
        \item
the dynamical equations constitute a Cauchy problem for
the evolution of the physical state of massive LED;

        \item
scattering does not change the nature of the renormalized particle, which 
moves as a \emph{soliton};  

        \item
the parameters of the bare particle can be chosen such that charge, 
magnetic moment, and renormalized mass of the `free' particle can be 
matched to the physical electron data without involving superluminal
gyration speeds.
\end{itemize}

	We then study the renormalization flow to vanishing bare rest
mass of our empirically matched massive LED. 
	The `renormalized LED' which emerges in the limit 
is characterized by the following additional features:
\begin{itemize}

        \item
the equatorial gyration speed reaches the speed of light $c$
precisely in the limit of vanishing bare rest mass so that 
the bare {gyrational} mass converges to a nonvanishing contribution. 
It is fitting to call this a `photonic' mass;

        \item
 the renormalized spin magnitude is a \emph{derived} property in our model.
It turns out to be $3\hbar/2$ rather than $\hbar/2$,
with corrections of order $\alpha$ (Sommerfeld's fine structure constant),
and we have $g \approx 2/3$ rather than
$g \approx g_{\text{\tiny QM}}\defeg 2$ or
$g \approx g_{\text{\tiny Cl}}\defeg 1$.

        \item
 our renormalized LED is, strictly speaking, 
a classical field theory with ultraviolet cutoff corresponding
to the  physical electron's Compton length! 
\end{itemize}

	It is not unexpected that in our classical theory the 
relation between spin and magnetic moment holds with a  
Land\'e factor $g \not\approx 2$.
        On the other hand, since we operate here with a minimalist model, 
it is not \apriori\ clear whether a slightly more intelligently
constructed classical model would not account also for the correct 
$g \approx 2$.
	In any event, the issue of the $g$ factor should not be 
confused with the superluminal speed problem of Lorentz.

	Among the many physically correct features of the model, the
truly rewarding one is its soliton dynamics. 
	Without it the rest mass, magnetic moment and spin of the 
renormalized particle would typically be altered by each scattering 
process, and the theory would then simply fail to account for the classical
motion of electrons and their electromagnetic fields. 

	Another gratifying feature is the size of the renormalized, purely 
electromagnetic model electron, whose radius comes out to $1.5$ physical 
electron Compton lengths, about $200$ times larger than the somewhat 
misnamed ``classical electron radius'' $e^2/mc^2$.
	Our renormalized LED is thus intrinsically 
equipped with the ultraviolet cutoff at which heuristic
discussions~\cite{landaulifshitz} have traditionally set the spatial limit 
of applicability of classical electrodynamics as a physical theory.

	Remarkable is furthermore that the Cauchy problem for the state 
evolution bears a certain resemblance to the problem of motion of a 
black hole in general relativity. 
	Unexpectedly, massive LED thus  provides a playground 
for this much harder problem of general relativity.

        Our model owes much to the monumental work of Nodvik~\cite{nodvik}. 
	However, although in Nodvik's Lorentz-covariant massive LED
one can match the particle parameters to the empirical electron data 
for charge, mass, and magnetic moment without encountering superluminal 
gyration speeds, one can do this only for sufficiently large bare rest mass.
	More specifically, the renormalization flow toward smaller bare rest 
mass terminates at about $1/2$ of the empirical electron mass, i.e. 
before reaching the purely electromagnetic limit of vanishing bare rest mass, 
and its spin is then merely about $10^{-3}\hbar$. 
	In addition to this deficiency, Nodvik's massive LED suffers from a
serious dynamical defect which it shares with the purely electromagnetic 
models of Abraham and Lorentz; namely, the initial value problem is singular.
	Interestingly, both deficiencies of Nodvik's massive 
LED are caused by his 
choice of bare rest mass distribution, which concentrates all the bare 
inertia at the center of the particle so that its bare moment of inertia 
vanishes.
	By endowing the particle with a bare rest mass distribution 
that gives it a strictly positive  bare moment of inertia our
massive LED overcomes both deficiencies and, in this sense, improves 
decisively on Nodvik's massive LED. 

	The presentation of our results is organized as follows.
        To begin with, in section 2 we stipulate the relativistic 
notation used in this paper. 
        In section 3 we discuss the relativistic kinematical
prerequisites of the model.
	Massive LED is defined by a Lorentz invariant action
functional that `minimally couples' the dynamics of the bare 
particle and the electromagnetic fields.
	We first formulate the relativistic mechanics of a gyrating 
bare particle for given Minkowski force and torque of
unspecified origin (section 4) and then, in section 5, present 
Nodvik's~\cite{nodvik} covariant Maxwell--Lorentz equations which 
determine the electromagnetic fields also in our model, given the 
kinematical history of the charged particle.  
	Our action functional for massive LED is introduced in section 6,
where we also discuss its symmetries and associated conservation laws.
	The dynamical equations of our massive LED are presented in
section 7.
	In section 8 we address the Cauchy problem for the state evolution. 
	The scattering problem is discussed in section 9, where we 
will see that the renormalized particle behaves like a soliton.
        In section 10 the renormalization group flow toward vanishing 
bare rest mass is studied, in detail for the stationary particle, 
and in an outline for the dynamical particle.
	We conclude the main body of our paper in section 11  
with a summary and an outlook on follow-up work and open problems.
	In particular, we also speculate about possible implications of 
our results in the quest for a consistent QED.

	In order not to interrupt the flow of the presentation, 
some technical material adapted from~\cite{nodvik}, e.g. the 
derivations of the Euler-Lagrange equations, has been relegated 
to Appendix A.1. 
	To facilitate the comparison with Nodvik's model~\cite{nodvik}, 
Appendix A.2 contains a brief discussion of the point mass limit in 
which our massive LED formally reduces to the massive LED of 
Nodvik~\cite{nodvik}. 
	In particular, we explain that this limit is singular 
and severely overdetermines the initial value problem. 
	Incidentally, a similar assessment holds for
Abraham's~\cite{abraham, abrahamBOOK} purely electromagnetic 
versions of semi-relativistic LED with and without spin.
	Since the ill-posedness of the initial value problem for the
purely electromagnetic Abraham model is itself a new observation,
Appendix A.3 gives a concise summary of the semi-relativistic 
massive LED and its purely electromagnetic, singular limit.
	There we start with the massive model with spin and show 
next that the more familiar semi-relativistic massive `LED without spin'  
actually is a theory with spin in `hidden' form, obtained from the massive
model with spin in a non-singular limit.
	We then address the singular purely electromagnetic Abraham limit.
	A fair list of references has been selected from the vast 
pertinent literature.
    \section{Minkowski space: vectors, tensors, and all that}
        \label{sec:PREP}

        Various equivalent conventions for the Minkowski space
formalism exist in the literature.  
        By and large, we here follow the conventions of the
book by Misner, Thorne and Wheeler~\cite{misnerthornewheeler}.  
        However, since the model is essentially a flat space model, 
we will not introduce forms.  
        Instead, throughout this work we use four-vectors and tensors 
constructed from them.

        Thus, abstract Minkowski space ${\mathbb M}^{4}$ is a flat 
four-dimensional pseudo-Riemannian manifold with Lorentzian metric 
of signature $+2$.  
        The invariance group acting on Minkowski space is the 
inhomogeneous \emph{Poincar\'e group}; its homogeneous subgroup is
called the \emph{Lorentz group}; for good discussions,
see~\cite{streaterwightman, barutBOOKa}.
        We can choose any point $P\in{\mathbb M}^{4}$
and identify ${\mathbb M}^{4}$ with the tangent 
space ${\mathbb T}_P({\mathbb M}^{4})$ at $P$.  
        A tetrad
$\LoF = \{\eQ_0,\eQ_1,\eQ_2,\eQ_3\}$ of fixed unit four-vectors 
in ${\mathbb T}_P({\mathbb M}^{4})$ that satisfy the elementary 
inner product rules
\begin{equation}
        \eQ_\mu\cdot \eQ_\nu 
        = 
        \left\{
        \begin{array}{rl}
                -1 & \text{for $\mu =   \nu = 0 $}\\
                 1 & \text{for $\mu =   \nu > 0 $}\\
                 0 & \text{for $\mu\neq \nu     $  }
        \end{array}
        \right.
        \, 
        \label{eq:ipr} 
\end{equation}
is a basis for ${\mathbb T}_P({\mathbb M}^{4})$, called a 
\emph{Lorentz frame}.
        With respect to a Lorentz frame, any real-valued vector 
$\vQ \in {\mathbb T}_P({\mathbb M}^{4})$ has the representation
\begin{equation}
        \vQ
        =       
        \sum_{\mu=0}^3 v^\mu \eQ_\mu
        \label{eq:qvte} 
\end{equation}
with $v^\mu\in {\mathbb R}$ given by $v^0 = - \vQ \cdot \qv{e}_0$ and
by $v^\mu = \vQ\cdot \qv{e}_\mu$ for $\mu= 1,2,3$.
        Therefore, ${\mathbb T}_P({\mathbb M}^{4})$ can be identified with 
${\mathbb R}^{1,3}$, the set of ordered real 
$4$-tuples $\vQ =(v^0,v^1,v^2,v^3)$, which is a global
coordinate chart for ${\mathbb M}^{4}$.

        To emphasize \emph{spacetime} among the various physical
realizations of abstract Minkowski space, spacetime points
will be denoted by $E$ and called events, and their four-vector 
representation w.r.t. $\LoF$ will be denoted as $\qv{x}$ or $\qv{y}$ etc. 
        The coordinates of $\qv{x}$ in a Lorentz frame $\LoF$  are sometimes 
grouped as
\begin{equation}
        \qv{x}
        =       
        (ct, \vect{x}) \, ,
\end{equation}
where $\vect{x}= (x^1,x^2,x^3)$ is called a `point in space,' 
and $t=x^0/c$ an `instant of time,' with respect to 
the given Lorentz frame. Here, $c$ is the speed of light \emph{in vacuo}. 
        Event four-vectors $\xQ$ are classified into spacelike, 
lightlike, and timelike according as $\xQ\cdot\xQ >0$, $\xQ\cdot\xQ =0$, 
or $\xQ\cdot\xQ <0$, respectively, where the inner product of any two event 
four-vectors $\xQ$ and $\yQ$ is evaluated using 
\refeq{eq:qvte}, distributivity and the elementary 
inner product rules \refeq{eq:ipr}. 
        A spacelike four-vector is connected through an orbit of the
Lorentz group with a four-vector of the form $(0,\xV)$,
a timelike one with a four-vector of the form $(ct,\vect{0})$. 
        The lightlike vectors form the `light cone,' a double
cone in ${\mathbb R}^{1,3}$ which separates the spacelike from the timelike
four-vectors.
	This classification is then carried over to four-vectors $\qv{v}$
in abstract Minkowski space.

        For convenience,  $\qv{v}\cdot\qv{v}$ will sometimes be 
abbreviated thus, \begin{equation}
        \qv{v}\cdot\qv{v} 
        =        
        \norm{\qv{v}}^2  
        \label{eq:qvnorm} 
\end{equation}
where $\norm{\qv{v}}$ is the Minkowski `norm' of $\qv{v}$, defined as the
principal value of $(\qv{v}\cdot\qv{v})^{1/2}$.  
        Notice that $\norm{\qv{v}}^2$ is negative for timelike vectors.

        The tensor product between any two four-vectors $\qv{a}$ and
$\qv{b}$ is a tensor of rank two, denoted by $\qv{a}\otimes\qv{b}$, 
and defined by its inner-product action on four-vectors, thus
\begin{align}
        (\qv{a}\otimes\qv{b})\cdot \qv{c} 
        \defeg  
        \qv{a} (\qv{b}\cdot \qv{c}) \\ 
        \qv{c} \cdot (\qv{a}\otimes\qv{b})
        \defeg  
        (\qv{a}\cdot \qv{c}) \qv{b}
\end{align}
        (Notice that our convention here differs from the one used, for
instance, in \cite{gitteletal}.)
	Given a frame $\{\eQ_\mu\}$, any tensor of rank two,
$\tenseur{T}$, can be uniquely written as
\begin{equation}
        \tenseur{T} 
        = 
        \sum_{0\leq\mu,\nu\leq 3} T^{\mu\nu}\eQ_\mu \otimes \eQ_\nu 
\, .\label{eq:genT} 
\end{equation}
	A rank-two tensor $\tenseur{T}$ for which
$T^{\mu\nu}= \pm T^{\nu\mu}$ in \refeq{eq:genT} is called
\emph{symmetric} ($+$ sign), respectively \emph{anti-symmetric} ($-$ sign).
	For symmetric ($+$) and anti-symmetric ($-$) tensors we have
\begin{equation}
	\tenseur{T}_\pm\cdot\vQ 
=
	\pm \vQ\cdot\tenseur{T}_\pm
\, , 
\end{equation}
but for a general tensor
\begin{equation}
	\tenseur{T}_{\text{general}}\cdot\vQ 
\neq 
	\pm \vQ\cdot\tenseur{T}_{\text{general}}
\, .
\end{equation}

	A special example for an anti-symmetric tensor of rank two is 
the exterior product between two four-vectors, denoted by a wedge-up 
product $\qv{a}\wedge\qv{b}$, and defined by
\begin{equation}
        \qv{a}\wedge\qv{b}
 \defeg  
        \qv{a}\otimes \qv{b} -  \qv{b}\otimes \qv{a} 
\,.
\end{equation}
	Examples for symmetric tensors of rank two are, 
(i) the symmetrized tensor product between two four-vectors, denoted by a 
wedge-down product $\qv{a}\vee\qv{b}$, and defined by
\begin{equation}
        \qv{a}\vee\qv{b}
        \defeg  
        \qv{a}\otimes \qv{b} +  \qv{b}\otimes \qv{a} 
\,,
\end{equation}
and (ii) the \emph{metric tensor} 
\begin{equation}
        \gQ
  = 
        \sum_{0\leq\mu,\nu\leq 3} g^{\mu\nu}\eQ_\mu \otimes \eQ_\nu 
\label{eq:gT} 
\end{equation}
with $g^{\mu\nu} = \eQ_\mu\cdot \eQ_\nu$ given in \refeq{eq:ipr}.
	Notice that $\gQ$ has the same
components $g^{\mu\nu}$ in all Lorentz frames because the 
$\eQ_\mu\cdot \eQ_\nu$ are Lorentz scalars. 
	Notice in particular also that $\gQ$ acts as identity on 
four-vectors, i.e. $\gQ\cdot\qv{v} = \qv{v}$.

        The \emph{(anti-)commutator} of any two tensors of rank two, 
$\tenseur{A}$ and $\tenseur{B}$ is defined as usual, 
\begin{equation}
        \left[\tenseur{A},\tenseur{B}\right]_\pm\defeg
        \tenseur{A}\cdot\tenseur{B} \pm \tenseur{B}\cdot\tenseur{A}
\, .
\end{equation}
	The action of tensors on a vector is associative, i.e.
\begin{equation}
	(\tenseur{A}\cdot\tenseur{B})\cdot\vQ 
= 
	\tenseur{A}\cdot(\tenseur{B}\cdot\vQ)
\end{equation}
so that
\begin{equation}
        \left[\tenseur{A},\tenseur{B}\right]_\pm\cdot\vQ 
=
        \tenseur{A}\cdot(\tenseur{B}\cdot\vQ) 
	\pm \tenseur{B}\cdot(\tenseur{A}\cdot\vQ)
\, .
\end{equation}

	\emph{Warning:}
	In general, 
\begin{equation}
	\tenseur{A}\cdot(\vQ\cdot\tenseur{B})
\neq 
	(\tenseur{A}\cdot\vQ)\cdot\tenseur{B}
\, .
\end{equation}

        The four-trace, or contraction, of a tensor of rank two is given by 
\begin{equation}
        \tr{\tenseur{T}}
        =
        \sum_{0\leq\mu \leq 3} g^{\mu\mu}T^{\mu\mu}.
\label{eq:trace} 
\end{equation}
        In particular, $\tr{\tenseur{g}}=4$, while 
 $\tr{\tenseur{A}} =0$ for any anti-symmetric tensor ${\tenseur{A}}$.

        For a differentiable function $f(\qv{x})$ we denote its 
four-gradient with respect to the metric $\tenseur{g}$ 
by $\nabQ f$.
	In  local Lorentz frame coordinates it is given by
\begin{equation}
        \nabQ f(\xQ)
        =       
        \left(-\partial_{_{x^0}}f,\nab f\right).
\end{equation}
where $\nab$ is the usual three-gradient. 
	In analogy with the conventional curl, we also define the 
four-curl of a differentiable four-vector function as the 
anti-symmetric four tensor function
\begin{equation}
        \nabQ \wedge \qv{A}(\xQ)\
        =       
        \sum_{0\leq\mu,\nu,\lambda,\eta\leq 3} 
\varepsilon^{\mu\nu\lambda\eta}
\eQ_\mu \otimes \eQ_\nu (\eQ_\lambda\cdot\nabQ)( \eQ_\eta \cdot\qv{A})
\end{equation}
where the $\varepsilon^{\mu\nu\lambda\eta}$ are the  entries of the 
conventional rank-four Levi-Civita tensor.
        The four-Laplacian with respect to $\tenseur{g}$
is  just the (negative of the) d'Alembertian, or
wave operator, given by
\begin{equation}
        \lapQ
\defeg
        \nabQ \cdot \nabQ =  -\wop
\end{equation}

                \section{Particle kinematics}
                \label{sec:kinematics}

        A spherically extended, `rigidly' gyrating particle 
in spacetime is kinematically described by a pair of curves 
in Minkowski space. 
        We shall assume that both curves are of class $C^2$.

        One curve is the so-called \emph{world-line} of the particle,
a timelike future-oriented curve in spacetime itself.
        The concept of the world-line is familiar from the
relativistic mechanics of point particles. 
        For an extended `rigid' particle an event on its world-line 
marks the location of the particle's center in spacetime. 
        The world-line has associated with it an invariant 
`arc-length' element  $\dd \tau$. 
        In units such that the speed of light $c=1$, which henceforth
we shall use, this invariant arc-length element becomes the 
\emph{proper-time} element. 
        Choosing any particular event $E_0$ on the
world-line as reference world point and assigning to it the proper-time
$\tau(E_0) =0$, and stipulating the `future' direction as the one 
along which $\dd\tau$ increases, then by integrating $\dd\tau$ along the world
line automatically determines a unique invariant proper-time 
$\tau(E)\in \Rset$  for any other event $E$ on the world-line.
        Since the world-line is timelike future-oriented, its proper-time 
assignment is one-to-one onto and order-preserving, whence 
serves as a natural parameterization for the world-line.
        In any particular Lorentz frame $\LoF$, the particle world-line is 
then given by a mapping $\tau\mapsto \xQ = \zQ(\tau)\in {\mathbb R}^{1,3}$.
        The proper-time element accordingly is given by 
$\dd \tau\ = \sqrt{- \dd \xQ\cdot\dd \xQ}$, 
where $\dd\xQ$ is taken along the world-line. 

        With this parameterization of the world-line, 
the four-velocity of the particle at the event
$\zQ(\tau)$ is obtained as 
$\dd \xQ/\dd \tau =  {} \qdot{\zQ}(\tau){}\defeg{}  \uQ(\tau)$.
        The map  $\tau\mapsto \uQ(\tau)$, called `four-hodograph,' 
traces out a complicated curve in \emph{four-velocity space}, located
on the future unit hyperboloid which is the genus-zero three-dimensional
sub-manifold defined by $\uQ\cdot\uQ =-1$ and $u^0 >0$. 
        When $\tau\mapsto\qv{u}$ is constant, the four-hodograph 
degenerates into a Minkowski point.
        We will sometimes group the four-velocity into its time and
space components w.r.t. the chosen Lorentz frame, as
\begin{equation}
        \uQ  =  \left(\gamma,\gamma \vV\right),
\end{equation}
where $\vV$ is the usual three-velocity in the space part of the chosen
Lorentz frame, and where 
\begin{equation}
        \gamma = \dfrac{1}{\sqrt{1-\abs{\vV}^2}}.
\end{equation}
        Finally, by taking the second derivative w.r.t. $\tau$ we
obtain the four-acceleration of the particle at the
event represented by $\zQ(\tau)$, thus
$ \dd^2 \xQ/\dd \tau^2 = \qdot{\uQ}(\tau){}\defeg{} \aQ(\tau)$.
        The map $\tau\mapsto \aQ(\tau)$ is a section in the
tangent bundle to the future unit hyperboloid of four-velocity space.

        The world-line does not contain any kinematical
information about the gyration of the extended rigid particle. 
        That information is contained in what may be called
the \emph{gyrograph} of the particle, a spacelike-oriented curve 
in {four angular velocity space}, or \emph{gyrospace} for short.
        The introduction of gyrospace and the gyrograph map
requires the consideration of two non-inertial reference frames:
\begin{itemize}
        \item
the body frame~$\boF$;
        \item
the Fermi--Walker frame~$\FeWaF$.
\end{itemize}


        The body frame $\boF$ is a co-moving and co-rotating frame, 
fixed in the particle. 
        It is given by a tetrad of orthonormal
basis vectors $\{\tilde{\eQ}_\mu\}_{\mu=0, \ldots,3}$, 
with $\tilde{\eQ}_0 =\uQ(\tau)$, 
that satisfy the equation \cite{misnerthornewheeler}
\begin{equation}
        \dds{\tilde{\eQ}}_\mu = - \OmegaQ\cdot\tilde{\eQ}_\mu\, ,
\end{equation}
where $\OmegaQ$ is the anti-symmetric tensor of the 
instantaneous rate of four-gyration of the particle's body frame 
$\boF$ as seen in the Lorentz frame $\LoF$.
        The components of four-vectors with respect to $\boF$ will be
marked by primes: 
$\xQ= \sum_{0\leq\mu\leq 3}\tilde{x}{}^\mu \tilde{\eQ}_\mu$.


        The Fermi--Walker frame $\FeWaF$ is  the natural co-moving tetrad 
of orthonormal basis vectors $\{\bar{\eQ}_\mu\}_{\mu=0, \ldots,3}$, with
$\bar{\eQ}_0= \uQ(\tau)$, satisfying the law of 
\emph{Fermi--Walker transport}~\cite{misnerthornewheeler},
\begin{equation}
        \dds{\bar{\eQ}}_\mu \,
        = 
        - \ThomasQ\cdot\bar{\eQ}_\mu\, ,
\end{equation}
where 
\begin{equation}
        \ThomasQ 
\defeg
	\, \qdot{\uQ}\, \wedge\,\uQ 
\end{equation}
is the anti-symmetric Fermi--Walker tensor.
        The components of vectors with respect to $\FeWaF$ will be marked by
overbars, i.e.  $\qv{x}=\sum_{0\leq\mu\leq 3}\bar{x}^\mu \bar{\eQ}_\mu$.

        Fermi--Walker transport is a kinematical effect which 
does not discriminate between objects of different physical
characteristics like inertia or charge. 
        The most prominent effect associated with Fermi--Walker transport
is the celebrated \emph{Thomas precession}~\cite{thomas}
of the space axes of $\FeWaF$ relative to the Lorentz frame $\LoF$,
described by the angular velocity three-vector~\cite{tomonagaBOOK,
thomas, bargmannmicheltelegdi},
\begin{equation}
        \thomasV
= 
        (\gamma - 1) \frac{1}{\abs{\vV}^{2}}\aV\times  \vV \, .
\label{eq:tpv} 
\end{equation} 
        Here $\aV$ is the usual three-acceleration of the origin of $\FeWaF$
as seen from $\LoF$, and $\vV$ its three-velocity.

        If the four-gyration of the body frame w.r.t. the 
Lorentz frame differs from the Thomas precession, one ascribes
this difference to  the particle's \emph{intrinsic four-gyration}, 
which is captured by the anti-symmetric tensor 
\begin{equation}
        \OmegaQ - \ThomasQ \defeg \EulerQ 
\, .
\end{equation}
        The subscript `$_E$' here is meant as reference to Euler's 
monumental analysis of Newtonian gyration of an inert rigid body.
        Beside being anti-symmetric, the tensor $\EulerQ$ satisfies
\begin{equation}
        \EulerQ\cdot \qv{u} =\qv{0}
\, ,
\end{equation}
so that $\EulerQ$ has only three independent components. 
        It follows that $\EulerQ$ is dual to a spacelike four-vector 
$\eulerQ$ which satisfies 
\begin{equation}
        \EulerQ\cdot \eulerQ =\qv{0}
\end{equation}
and $\eulerQ\cdot\qv{u} = \qv{0}$.
        In $\FeWaF$, $\eulerQ$ is given in the form $\eulerQ =(0,\eulerV)$, 
where ${\eulerV}$ is the usual angular velocity three-vector which is
directed along the instantaneous axis of body gyration in the spacelike 
three-slice of the frame~$\FeWaF$.

        At each instant of proper-time $\tau$, the 
instantaneous state of intrinsic four-gyration of the particle 
is captured by the four-vector $\eulerQ(\tau)$.
        As $\tau \in {\mathbb R}$ varies along the particle's world-line,
the map $\tau\mapsto \eulerQ(\tau)\in {\mathbb R}^{1,3}$ traces out 
the gyrograph of the particle in gyrospace, which 
is the Minkowski space consisting of all possible four-vectors 
with the meaning of an $\eulerQ$. 
        The gyrograph is spacelike, since for 
$\eulerV\neq \vect{0}$, we have 
\begin{equation}
        - \frac12 \tr ( \EulerQ\cdot\EulerQ) 
= 
        \|\eulerQ\|^2 
=
        |{\eulerV}|^2 
>
        0\, .
\end{equation}

                \section{Bare particle mechanics}
                \label{sec:bareparticle}

        Formally stripping the particle from its electromagnetic
features leaves one with the \emph{bare particle}. 
        Our bare particle is  characterized by a strictly
positive inertia against changes of its translational  
and  gyrational momenta, respectively.

                \subsection{Bare inertia}
                \label{sec:bareinertia}

        In our model, all forms of bare inertia are computed with 
a compactly supported, spherical `density' 
$\fm(|\ .\ |): {\mathbb R}^3 \to {\mathbb R}^+$ of inert mass
of unspecified non-electromagnetic origin.
        In general, $\fm: {\mathbb R}^+\to {\mathbb R}^+$ is 
a measure on $[0,R]$.
	We demand that $\fm$ generates a strictly positive
moment of inertia (see below).
	For instance, two admissible mass densities are:
\begin{itemize}
        \item
 \emph{(volume inertia)}\ 
$\fm(\norm{\xQ}) = \mz\displaystyle{\frac{3}{4\pi R^3}} \chi_{R}(\norm{\xQ})$,
        \item
 \emph{(surface inertia)}\
$\fm(\norm{\xQ}) = \mz\displaystyle{\frac{1}{4\pi R^2}}\,\delta(\norm{\xQ}-R)$ 
.\end{itemize}
        Here, $\delta(\;.\;)$ is the Dirac distribution, and 
$\chi_{_{R}}(\; .\;)$ the characteristic function of the interval~$[0,R]$
with $R>0$.
	A fraction of the mass in $\fm$ is allowed to be a point mass, 
but not all of it. 
	Notice that the complete point mass density 
$\fm(|\xV|) = \mz (1/4\pi |\xV|^2) \delta(|\xV|)$ violates our condition 
that $\fm$ must have a strictly positive moment of inertia.
	The singular limiting case of a point mass density is addressed 
in Appendix A.2.

        The most basic inertia is the \emph{bare rest mass}, given by
\begin{equation}
        \mz =  \int_{{\mathbb R}^3} {\fm}(|\xV|)\,\dd^3x \, ,
        \label{eq:mNULL}
\end{equation}
where $r=|\xV|$. 
        Since in general the bare particle is in a state of gyration, 
its instantaneous effective inert mass in the Fermi--Walker frame
is  the sum of its bare rest mass and the relativistic kinetic energy 
associated with its Euler gyration, which we call the
\emph{bare gyrational mass}, a Lorentz scalar given by
\begin{align}
	\rotM(\norm{\eulerQ})
& \defeg
	\int_{{\mathbb R}^{1,3}} 
		\frac{1}{\sqrt{\displaystyle 1-\norm{\EulerQ\cdot\qv{x}}^2}}
	\fm\big(\norm{\qv{x}}\big)\,\delta\big(\qv{u}\cdot\qv{x}\big)\,
	\dd^4 \qv{x}
\label{eq:gyroMASS}
\\
& =
	\int_0^R \frac{1}{|\eulerV|r} \Artanh (|\eulerV| r)\fm(r)4\pi r^2\dd r
\, . \label{eq:gyroMASSeval}
\end{align}
        The `integration over a simultaneity space slice'
in \refeq{eq:gyroMASS} follows the insight of Fermi~\cite{fermi}, 
who used this setup in his solution to the infamous $4/3$ problem 
(see~\cite{rohrlichAJPold, panofskyphillipsBOOK, feynmanLECTURES, 
jacksonBOOK, rohrlichBOOK, schwinger, yaghjianBOOK, barutpavsicB} 
for discussions of the $4/3$ problem).
        In the limit of small inertial angular velocity, $|\eulerV|\ll 1$,
a MacLaurin expansion of \refeq{eq:gyroMASSeval} in powers of $|\eulerV|$ 
gives the expected result
\begin{equation}
	\rotM(|\eulerV|) 
= 
	\mz  + \frac{1}{2} \Iz |\eulerV|^2 + O(|\eulerV|^4)
\, ,  \label{eq:MacLauM}
\end{equation}
where
\begin{equation}
  \Iz 
  =
  \frac23\int_{{\mathbb R}^3} |\xV|^2 {\fm}(\abs{\xV})\,\dd^3x
\end{equation}
is the bare particle's non-relativistic principal moment of inertia.  
	By the spherical symmetry of the particle in any of its rest 
frames the principal inertias are of course degenerate.  
	The corresponding non-relativistic moment-of-inertia 
three-tensor of the bare particle generalizes to the tensor of 
the \emph{Minkowski inertia (about $\zQ$)}, given by
\begin{equation}
  \IQb(\norm{\eulerQ})
  =
  \int_{{\mathbb R}^{1,3}}
  \frac{\|\qv{x}\|^2\gQ - \qv{x}\otimes\qv{x}}
  {\sqrt{\displaystyle 1-\norm{\EulerQ(\tau)\cdot\qv{x}}^2}}
  \fm\big(\norm{\qv{x}}\big)\,\delta\big(\qv{u}(\tau)\cdot\qv{x}\big)\,
  \dd^4 \qv{x}.
  \label{eq:baremomentofinertia}
\end{equation}
which captures the inertia associated with changes of the state of 
intrinsic four-gyration of the bare particle. 

                \subsection{Bare dynamics}
                \label{sec:baredynamics}

        The various inertias acquire their meaning from their
role as `coefficients of resistance' against changes of the 
bare particle's various Minkowski momenta.
        The bare particle's four-vector of \emph{Minkowski momentum}
is given by
\begin{equation}
        \pQb
=
        \rotM\uQ
\, ,
\end{equation}
and the four-vector of its \emph{Minkowski spin} by
\begin{equation}
        \sQb
=
        \IQb\cdot\eulerQ
\, . \label{eq:bareSPINvect}
\end{equation}
	For a given Minkowski force $\fQ(\tau)$ and
Minkowski torque $\tQ(\tau)$, here of unspecified origin,
but satisfying $\tQ\cdot\uQ =\qv{0}$ and $\fQ\cdot\uQ =\qv{0}$,
the evolution of these momenta is governed by the coupled dynamical equations
\begin{equation}
        \dds{\pQb} \,  =  \fQ
\end{equation}
and 
\begin{equation}
        \dds{\sQb} +  \ThomasQ\cdot{\sQb}
        = 
        \tQ\, .
\end{equation}
	Due to the occurrence of $\zQ(\tau)$ in the gyrational
mass, these dynamical equations are of second order for $\zQ$ 
and of first order for $\eulerQ$, so they need to be supplemented
with initial data for $\zQ$, $\uQ$, and $\eulerQ$. 
	A discussion of these equations is interestingly in itself,
but will not be pursued here. 


        \section{Nodvik's Maxwell--Lorentz equations}
        \label{sec:MLeq}

	If the charged particle's world-line $\zQ(\tau)$ and 
gyrograph $\EulerQ(\tau)$ are given, the anti-symmetric 
electromagnetic tensor field of rank two $\FQ$ can be computed 
by solving the linear, relativistic Maxwell--Lorentz equations
\begin{equation}
        \nabQ\, \cdot\, {^\star\FQ} 
=
	\qv{0}
\, ,\label{eq:homMLeq}
\end{equation}
\begin{equation}
        \nabQ\cdot \FQ 
=
        4\pi \JQ\, ,
\label{eq:inhMLeq}
\end{equation}
where ${^\star\FQ}$ is the (left) Hodge dual of $\FQ$, and where the covariant 
charge-current density four-vector field $\JQ$ is of the general 
form `charge density times four-velocity.' 
	The specific form of the charge-current density 
of course depends on the particle model under consideration. 
	Our massive LED inherits the charge-current density from 
Nodvik's~\cite{nodvik} model of a gyrating Lorentz electron, 
which generalizes the non-relativistic expression for a gyrating, 
strictly rigid Abraham electron (Appendix A.3) to the manifestly 
covariant expression
\begin{equation}
        \qv{J}(\xQ)
=
        \int_{-\infty}^{+\infty}
		\qv{U}(\qv{x};\tau)
        	\;{\fe}\big(\norm{\qv{x}-\zQ(\tau)}\big)
	        \,\delta\Big(\qv{u}(\tau)\cdot\big(\qv{x}-\zQ(\tau)\big)\Big)
	\,\dd \tau
\, .\label{eq:Jnodvik}
\end{equation}
	Here,
\begin{equation}
        \qv{U}(\qv{x};\tau) 
=
	\qv{u}(\tau) -\OmegaQ(\tau)\cdot \big(\qv{x}-\zQ(\tau)\big)
\, ,\label{eq:Ueq}
\end{equation}
is  the four-velocity of a particle element located at 
$\xQ$ in $\LoF$, with $\OmegaQ = \ThomasQ + \EulerQ$, and
$\fe(|\ .\ |): {\mathbb R}^3 \to {\mathbb R}^-$ is
the spherically symmetric electrical charge `density' (in general 
a finite measure) of the particle at rest. 
	To avoid infinite electromagnetic energy densities and 
forces, we request that $\fe$ does not feature any singular Dirac 
point charge distribution anywhere on its support; more general
Dirac distributions are however allowed.
	For instance, the following two possibilities are admissible:
\begin{itemize}
        \item
 \emph{(volume charge)}\
$\fe(\norm{\xQ}) 
	= -e \displaystyle{\frac{3}{4\pi R^3}} \chi_{_{R}}(\norm{\xQ})$,
        \item
 \emph{(surface charge)}\
$\fe(\norm{\xQ}) = -e\displaystyle{\frac{1}{4\pi R^2}}\,\delta(\norm{\xQ}-R)$. 
\end{itemize}
	Here, $e>0$ is the elementary charge as appropriate for a model
of an electron.

	Problems with infinite self energies and self forces aside, 
however, it should be noticed that in the formal limit 
$\fe(|\xV|)\to -e(4\pi|\xV|^2)^{-1}\delta(|\xV|)$,
together with the assumption that $\norm{\eulerQ}< C$, 
the Nodvik charge-current density \refeq{eq:Jnodvik} reduces to the familiar 
expression for the spinless point charge~\cite{jacksonBOOK, weinbergBOOKa},
\begin{equation}
	\JQ_{\text{point charge}}(\xQ) 
=
	-e\int_{-\infty}^{+\infty}
		\uQ(\tau)\delta\big(\xQ-\zQ(\tau)\big)
	\dd\tau
\, .
\end{equation}

        \section{Minimal coupling of particle and fields}

	In order to obtain the dynamical equations of massive LED, 
we have to couple the relativistic Maxwell--Lorentz equations for 
the electromagnetic fields and the relativistic equations for the 
particle's world-line and gyrograph in a consistent manner. 
	Since both the dynamics of the bare particle and the
dynamics of the Maxwell--Lorentz field derive separately from
some \emph{principle of `least' action}, we can follow
Nodvik~\cite{nodvik} and apply the standard procedure of 
minimal coupling to the action functionals and then postulate 
the principle of least action for this new, joint action.
	As a general reference for relativistic action principles
we recommend~\cite{christodoulou}.

                \subsection{The action of massive LED}

\begin{figure}
  \centerline{\epsfig{file=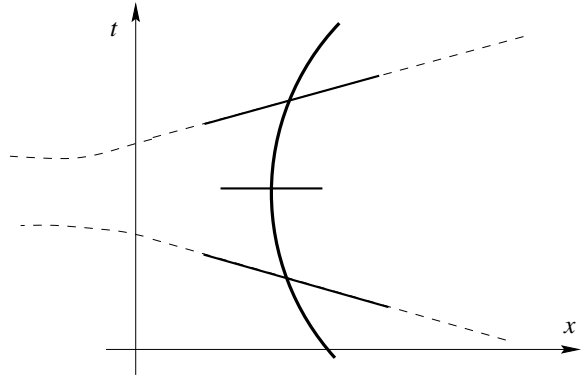,height=5cm}} 
  \caption{An example of an admissible four-domain $\Xi$,
  which contains an ordered history of the particle.
  The extra-bold line is the world-line of the particle. 
  The solid lines indicate the support of $\fe(\norm{\xQ-\zQ(\tau)})$
  restricted to the simultaneity space slices 
  $\uQ(\tau)\cdot\big(\xQ-\zQ(\tau)\big)=0$ at $\tau_1, \tau_2$ and
  an intermediate value of $\tau$. The dashed lines are the 
  space slices that bound $\Xi$.
  \label{fig:slice1}}
\end{figure}

\begin{figure}
  \centerline{\epsfig{file=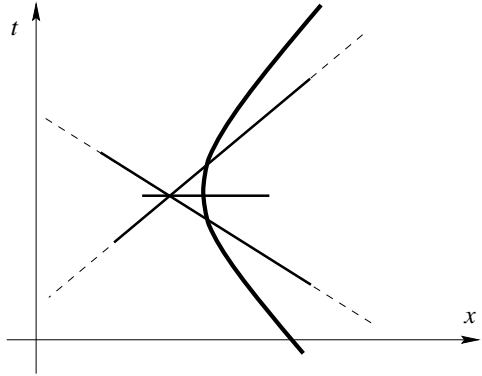,height=5cm}}
  \caption{An example of a non-admissible four-domain $\Xi$. 
   No ordered particle history is possible because the supports 
   of the functions $\xQ\mapsto \fe(\norm{\xQ-\zQ(\tau)})$
   restricted to different space slices of simultaneity
   $\uQ(\tau)\cdot \big(\xQ-\zQ(\tau)\big)=0$ must intersect. 
    \label{fig:slice2}}
\end{figure}

        The action is defined through an integral over any
four-domain $\Xi$ which is sandwiched between two disjoint 
spacelike three-subspaces that contain the intersections of the supports 
of $\fe(\norm{\qv{x}-\zQ(\tau)})$ (and of $\fm(\norm{\qv{x}-\zQ(\tau)})$)
with the spacelike slices $\qv{u}(\tau)\cdot\big(\qv{x}-\zQ(\tau)\big)=0$ 
of events which are simultaneous to the events at $\zQ(\tau)$
for $\tau =\tau_1$ and $\tau=\tau_2$, respectively.
	Such a four-domain is called `admissible.'
	It contains \emph{ordered particle histories}, 
defined by particle world-lines which connect the supports of
$\fe(\norm{\qv{x}-\zQ(\tau)})$ in the slices 
$\qv{u}(\tau)\cdot\big(\qv{x}-\zQ(\tau)\big)=0$ at $\tau =\tau_1$ 
and $\tau=\tau_2$ in such a way that successive space 
slices of simultaneity do not intersect in the support of 
$\fe(\norm{\qv{x}-\zQ(\tau)})$; see Figure~\ref{fig:slice1}.    
	Not every world-line corresponds to an ordered history.
	The space slices of simultaneity for two different values
of~$\tau$ may intersect in the support $\fe(\norm{\xQ-\zQ(\tau)})$
if the particle acceleration is too strong; 
see Figure~\ref{fig:slice2}.

	To formulate our minimally coupled action functional, we 
also need to introduce the four-vector of the electromagnetic potential. 
        We recall that \refeq{eq:homMLeq} implies that $\FQ$ is the 
exterior derivative of an electromagnetic potential four-vector $\AQ(\xQ)$,
i.e. $\FQ (\xQ) = \nabQ\wedge\, \AQ  (\xQ)$. 
        We will work within the class of $\AQ$ that satisfy the so-called 
Lorentz gauge condition $\nabQ\cdot\AQ = 0$, which still allows 
the gauge transformation $\AQ\to\AQ+\nabQ\psi$ for any $\psi$ 
satisfying the wave equation $\wop\psi =0$. 

	Given an admissible $\Xi$, the action functional ${\scr A}$ 
is given as integral 
\begin{equation}
        {\scr A} =  \int_{\Xi} { \cal L}(\xQ)\,\dd^4\xQ\, ,
\end{equation}
where ${\cal L}(\xQ)$ is the Lagrangian four-density 
for the minimally coupled field-particle system,
\begin{equation}
        {\cal L} =  \Lf + \Lmf + \Lm \, .
\end{equation}
        The first term is Schwarzschild's Lagrangian density for the free
electromagnetic field of Maxwell, i.e.
\begin{equation}
        \Lf (\xQ)
        \defeg          
        \, - \frac{1}{16\pi} \tr
        \big(
        \FQ(\xQ)\cdot\FQ(\xQ)
        \big)\, ,
        \label{eq:Sf}
\end{equation}
and the second term is the standard Lagrangian density for
the  minimal electromagnetic  coupling of particle and field 
degrees of freedom, i.e.
\begin{equation}
        \Lmf  (\xQ)
        \defeg   
        \JQ (\xQ)\cdot\qv{A}(\xQ)\, ,
        \label{eq:Smf}
\end{equation}
where  $\JQ$ is the Nodvik charge-current density given in 
\refeq{eq:Jnodvik}. 
        The last term is the Lagrangian density for our bare particle,
\begin{equation}
        \Lm (\xQ)
        \defeg  
        \,- \int_{\tau_1}^{\tau_2}
        \sqrt{1-\norm{\EulerQ\cdot(\xQ-\zQ)}^2}
        \; \fm\big(\norm{\xQ-\zQ}\big)\,\delta\big(\uQ\cdot
        (\xQ-\zQ)\big)\,\dd \tau\, .
        \label{eq:BARElagrangianDENS}
\end{equation}

        \subsection{Invariants for the action functional}

        Our action functional is invariant under several symmetry
transformations of base and target spaces, associated with which 
are invariants that give the physical conservation laws.
        These invariants are listed below.
	We consider only ordered histories, so that the domain of 
integration in the conservation laws can be written $\Rset^{1,3}$ 
instead of $\Xi$.

\subsubsection{Invariance under base space transformation}

        Invariance of the action functional under translations of
spacetime implies conservation of the \emph{total Minkowski momentum} 
four-vector $\qv{P}$ of the field-particle system, the constant time 
component of which expresses energy conservation, the space components 
the conservation of linear three-momentum. 
        The Minkowski momentum of the field-particle system
can be written in the form 
\begin{equation}
        \qv{P}
        =
        \Mfp (\tau) \cdot \uQ(\tau) 
\, ,
\end{equation}
where 
\begin{equation}
        \Mfp (\tau) 
 =
	\rotM(\norm{\eulerQ(\tau)})\, \gQ  
+  	\Mnodv (\tau)       
+ 	\int_{\Rset^{1,3}}
		 \tenseur{T}(\xQ)
        \,\delta\Big(\uQ(\tau)\cdot\big(\xQ-\zQ(\tau)\big)\Big)\, \dd^4\xQ 
\, , \label{eq:sMmt}
\end{equation}
is the symmetric field-particle tensor mass, itself composed of the following
tensor masses: the first term on its r.h.s. is just the bare
gyrational mass of the particle written as diagonal tensor with the
help of the metric tensor  $\gQ$; the second term on its r.h.s is
the symmetric \emph{Nodvik tensor mass},
\begin{equation}
        \Mnodv (\tau) 
=
	- \int_{\Rset^{1,3}}
        	\left[\xQ\otimes\xQ, 
			\left[ \FQ(\zQ(\tau)  +\xQ),\EulerQ(\tau) \right]_+ 
		\right]_+ \fe\big(\norm{\xQ}\big)\,
        	\delta(\uQ(\tau) \cdot \xQ)\,\dd^4\xQ
\, , \label{eq:spinORBITmass}
\end{equation}
extracted (see Appendix A.1) from the Minkowski momentum four-vector of
electromagnetic spin-orbit coupling given in~\cite{nodvik}; 
the last term on its r.h.s. finally is the electromagnetic field mass tensor,
with 
\begin{equation}
        \tenseur{T}(\xQ)
        =
        \frac{1}{4\pi}
        \left(
        \FQ(\xQ)\cdot \FQ(\xQ) -  \frac{1}{4}
        \tr \bigl(\FQ(\xQ)\cdot \FQ(\xQ) \bigr) \gQ
      \right)
      \label{eq:stressenergy}
\end{equation}
the conventional traceless tensor of the electromagnetic 
energy-momentum-stress density, a symmetric tensor field 
of rank two.
	It's four-divergence, given by
\begin{equation}
        \nabQ\cdot \tenseur{T}(\xQ)
        =
        \FQ(\xQ)\cdot\JQ(\xQ)
\, ,
\label{eq:DIVstressenergy}
\end{equation}
does in general not vanish on the support of the particle
densities $\fm$ and $\fe$; in a nutshell, this is the origin for 
the need of the Fermi prescription of integration over space slices of
simultaneity (cf.~\cite{rohrlichAJPold, rohrlichBOOK}).

        Invariance of the action functional under 
spacetime rotations implies conservation of the anti-symmetric tensor of 
\emph{total Minkowski angular momentum} $\LQ$ of the field-particle
system,
\begin{equation}
         \LQ
        = 
        \SQb (\tau) 
        +
        \zQ(\tau) \wedge \qv{P}
        +
        \int_{\Rset^{1,3}}
        \xQ\wedge\Big(\tenseur{T}\big(\xQ+\zQ(\tau)\big)\cdot \uQ(\tau)\Big)
        \,\delta\big(\uQ(\tau) \cdot \xQ\big)\, \dd^4\xQ \, ,
\end{equation}
where $\qv{P}$ and $\tenseur{T}$ are given above, and where
\begin{equation}
        \SQb(\tau)
=
        -\int_{\Rset^{1,3}}
\xQ\wedge \frac{ \EulerQ(\tau)\cdot \xQ}
        {\sqrt{\displaystyle 1-\norm{\EulerQ(\tau)\cdot\xQ}^2}}
        \fm\big(\norm{\xQ}\big)\,\delta\big(\uQ(\tau)\cdot\xQ\big)\,
        \dd^4 \xQ
\label{eq:barespinTENSOR}
\end{equation}
is the anti-symmetric tensor of \emph{bare Minkowski spin (about $\zQ$)} 
of the particle.
        The tensor  $\SQb$ is dual to the 
bare spin four-vector \refeq{eq:bareSPINvect}, 
hence it is of space-space type w.r.t. $\uQ$, viz.
\begin{equation}
        \SQb\cdot\uQ = 0\, .
\end{equation}
        Since $\LQ$ is anti-symmetric, its constancy
expresses six individual physical conservation laws.
        The constancy of the three space-space components 
expresses the conservation of the total angular momentum, 
and the constancy of the three time-space components 
correspond to the statement that the center of mass
of the field-particle system moves with constant three-velocity
along a straight line in the three-space slice of the Lorentz
frame.

\subsubsection{Invariance under target space transformation}

        Invariance of the action functional under gauge transformations 
$\AQ\to\AQ+\nabQ\psi$ implies charge conservation locally, i.e. 
\refeq{eq:Jnodvik} satisfies the continuity equation
\begin{equation}
        \nabQ\cdot\JQ(\xQ) 
= 
        0
\, ,\label{eq:continuityEQ}
\end{equation}
which is demanded by the Maxwell--Lorentz equations and
readily verified to hold also by direct computation with 
\refeq{eq:Jnodvik}, cf.~\cite{nodvik}.
	Of course, \refeq{eq:continuityEQ} implies the
conservation of the total \emph{charge} 
\begin{equation}
	Q
=       
	-\int_{\Rset^{1,3}}  \qv{e}_0 \cdot\JQ(\xQ)
        \,\delta\big(\qv{e}_0 \cdot\xQ - t_0\big)\, \dd^4\xQ 
\,, \label{eq:charge}
\end{equation}
i.e. $Q=-e$ in our model of the electron, for all $t_0$.

        In addition, there is a symmetry associated with the 
spherical rest symmetry of the particle, giving 
invariance of the action functional under shift of the Euler angle
around the instantaneous axis of gyration.
        (For the definition of the Euler angles, see Appendix A.1.)
	This corresponds to the conservation of 
\begin{equation}
        s^2     
=
        - \frac12 \tr \left(\SQ^\perp(\tau) \cdot\SQ^\perp(\tau)\right)
\, ,
\end{equation}
where
\begin{equation}
        \SQ^\perp 
=       
        \SQ + \big[ \uQ \otimes\uQ\, , \SQ\big]_+ 
\end{equation}
is the space-space part w.r.t. $\uQ$ of the anti-symmetric
tensor of \emph{total Minkowski spin}, 
\begin{equation}
        \SQ (\tau) 
\defeg  
        \SQe(\tau)  +  \SQb(\tau) \, ,
\end{equation}
with
\begin{equation}
        \SQe (\tau) 
\defeg
        \int_{\Rset^{1,3}}
\big(\xQ-\zQ(\tau)\big)\, \wedge\, \AQ(\xQ)
        \fe\big(\norm{\xQ-\zQ(\tau)}\big)\,
        \delta\Big(\uQ(\tau) \cdot\big(\xQ-\zQ(\tau) \big)\Big)\,\dd^4\xQ  
\label{eq:sqe}
\end{equation}
the tensor of the particle's 
\emph{electromagnetic Minkowski spin} about $\zQ(\tau)$.

        We remark that despite its appearance, $\SQe$ is invariant 
under a gauge transformation $\AQ\to\AQ+\nabQ\psi$.

        The law for $s^2$ means that the dual four-vector to $\SQ^\perp$ 
evolves on the spacelike genus-one hyperboloid. 

\subsection{The principle of `least' action}

	Our action principle is formulated for 
a fixed four-domain $\Xi$, with the constraints $\delta \uQ(\tau)=0$ 
for $\tau=\tau_1$ and $\tau=\tau_2$. 
	Likewise, the variations of the world-line $\delta\zQ(\tau)$ 
and of the Euler angles (see Appendix A.1) vanish 
at $\tau=\tau_1$ and $\tau=\tau_2$.  
	Following standard physics terminology we refer to this
Hamiltonian principle of stationary action as `principle of least action,'
although we do not investigate the second variation here.

        The particle world-line satisfies the variational  principle
\begin{equation}
        \delta( \Ab + \Abf )
= 
        0
\end{equation}
under an independent deformation of the world-line $\tau\mapsto \zQ(\tau)$ 
(with $\tau\in [\tau_1,\tau_2]$) into a world-line 
$\tau\mapsto \zQ(\tau)+\delta\zQ(\tau)$ (with $\tau$ in the same range), 
with Euler variables $\thetaQ$ and four-potential $\AQ$ fixed. 

        The gyrograph satisfies the variational principle
\begin{equation}
  \delta( \Ab + \Abf ) = 0
\end{equation}
under an independent variation of the gyrograph of the particle, with
fixed world-line $\tau\mapsto\zQ$ and four-potential $\AQ$.  

        The fields satisfy the standard Schwarzschild 
principle~\cite{barutBOOKa, misnerthornewheeler},
\begin{equation}
  \delta( \Af + \Abf ) = 0
\end{equation}
for prescribed world-line $\tau\mapsto\zQ$ and gyrograph 
$\tau\mapsto\EulerQ$.

        \section{The dynamical equations of massive LED}

        The dynamical equations are the Euler-Lagrange equations 
for the variational principle 
\begin{equation}
        \delta {\scr A} = 0 ,
\end{equation}
with $\zQ,\AQ$ and the Euler angles $\thetaQ$ 
treated as independent variables.          
        We will state the relativistic dynamical equations in a format 
closely reminiscent of the semi-relativistic equations of
the massive Abraham theory described in Appendix A.3.

	\subsection{Field equations}

        Recalling that $\FQ = \nabQ\wedge \qv{A}$ by definition,
which already implies the homogeneous Maxwell--Lorentz equations
\refeq{eq:homMLeq}, the principle of least action together with 
$\nabQ\, \cdot \qv{A} = 0$ now yields the equivalent potential 
form of the inhomogeneous Maxwell--Lorentz equations, i.e. the 
inhomogeneous wave equation 
\begin{equation}
        \wop \qv{A} (\xQ)
= 
        4\pi \JQ(\xQ)\, ,
\end{equation}
with $\JQ$ Nodvik's four-vector of charge-current density 
given in \refeq{eq:Jnodvik}.

\subsection{World-line equations}

	The world-line equations read
\begin{equation}
        \frac{\dd }{\dd \tau}\pQ
=
	\fQ
\, ,\label{eq:wlEQ}
\end{equation}
where
\begin{equation}
        \pQ(\tau) 
=
        \Mmink(\tau) \cdot \uQ(\tau) 
\, \label{eq:MinkEnMom}
\end{equation}
is the \emph{Minkowski momentum} of the particle, 
with $\Mmink = \rotM\gQ + \Mnodv$ (see \refeq{eq:sMmt}), and where
\begin{equation}
	\fQ(\tau)
= 
      \int_{{\mathbb R}^{1,3}} \FQ(\xQ)\cdot
\qv{U}(\xQ;\tau)
        \fe(\norm{\xQ- \zQ(\tau)})
	\,\delta\Big(\uQ(\tau)\cdot\big(\xQ-\zQ(\tau)\big)\Big)\,\dd^4\xQ, 
\label{eq:MinkALforce}
\end{equation}
is Nodvik's~\cite{nodvik} Abraham--Lorentz type \emph{Minkowski force}.

\subsection{Gyrograph equations}

	The equations for the gyrograph read
\begin{equation}
	\dds {\SQb}  +\, \big[\ThomasQ ,\SQb \big]_- 
= 
	\tQT
\, ,  \label{eq:ggEQ}
\end{equation}
where
\begin{equation}
	\tQT(\tau) 
=
	\int_{{\mathbb R}^{1,3}}
		\big(\xQ-\zQ(\tau)\big)\wedge 
		\bigl(\FQ(\xQ)\cdot \qv{U}(\xQ,\tau)\bigr)^\perp
	\fe\big(\norm{\xQ-\zQ(\tau)}\big)\,
	\delta\Big(\uQ(\tau)\cdot\big(\xQ-\zQ(\tau)\big)\Big)
	\,\dd^4\xQ
\label{eq:MinkALtorque}
\end{equation}
is the anti-symmetric tensor of the Abraham--Lorentz type
\emph{Minkowski torque}, with $\aQ^\perp\defeg \bigl(\gQ +
\uQ\otimes\uQ\bigr)\cdot\aQ$ the  projection of the four-vector $\aQ$
onto the space slice w.r.t. $\uQ$. 
	The space projector $\gQ + \uQ\otimes\uQ$ under the integral 
guarantees that \refeq{eq:ggEQ} preserves the intrinsic space-space 
character of $\SQb$. 
	The commutator with the Fermi--Walker tensor accounts for the 
four-precession of the Fermi--Walker-transported, natural co-moving 
frame with respect to which the inertial gyrations are defined.  

         \section{Evolution of the state in massive LED}

	In the concise component-free Minkowski space notation
the similarity of the world-line equation \refeq{eq:wlEQ}, 
supplemented with \refeq{eq:MinkEnMom} and \refeq{eq:MinkALforce}, 
to the Newtonian type evolution equation \refeq{eq:ALeqTmb} of the 
semi-relativistic massive LED with spin is clearly recognizable, and
so is the similarity of the gyrograph equation \refeq{eq:ggEQ}, 
supplemented with \refeq{eq:MinkALtorque}, to the Eulerian evolution equation
\refeq{eq:ALeqRmb}.
	The almost familiar appearance of the dynamical equations of 
our relativistic massive LED is, however, deceptive.
	As we will now see, the physically natural interpretation of
these dynamical equations as a set of evolution equations for the 
\emph{dynamical state} of the theory unveils some radically new
elements of physical reality (as defined by the theory).
 
	The characteristic of a physical state, loosely speaking, 
is that its specification at any particular instant completely
determines, via first-order evolution equations, the time derivative 
of the state at that instant and, hence, allows to continue uniquely 
to the state at the next instant.
	To find out what precisely constitutes a `state at any particular
instant' in our massive LED, we have to analyze for which dynamical 
variables the equations pose a first-order initial value problem 
(Cauchy problem) in an orderly manner.
	This is done in the following two subsections.
	We then address the problem of nearly free global evolutions. 
	Since the evolution equations of our massive LED 
are implicit in a rather unfamiliar manner, we  first rewrite them
into a quasi-explicit format.
	We then propose a direct generalization of the familiar Picard 
iteration to solve the equations --- in principle.
	We finally remark on the question of global well-posedness
of the Cauchy problem for arbitrary admissible states.
	This issue is only partially settled. 
	Various possible `mechanisms' that may cause some sort of 
singularity to occur at a finite time have to be controlled.
	However, for the special case of purely gyrational particle
dynamics coupled to the Maxwell--Lorentz fields we succeeded in 
proving, by using our generalized Picard iteration, a global existence 
and uniqueness result. 
	This proof is going to appear in~\cite{appelkiessling}.

        \subsection{The states in massive LED}

	We first need to clarify the phrase \emph{at an instant of time}.
	In special relativity, this is a quite flexible notion.
	Generally it refers to the collection of all the events on a 
three-dimensional, simply connected, spacelike hypersurface. 
	We assume without loss of generality that such a hypersurface
is defined as the level set $\Sigma_t =\{\xQ: T(\xQ) =t\}$ of a 
differentiable function $T$ on spacetime. 
	Here, a three-dimensional hypersurface $\Sigma_t$ is called 
spacelike if $\nabQ T(\xQ)$ is timelike at each $\xQ$ in $\Sigma_t$. 
	If the intersection of any two such spacelike hypersurfaces
$\Sigma_{t_1}$ and $\Sigma_{t_2}$, with $t_1\neq t_2$, 
is empty, and if 
furthermore the four-dimensional set $\cup_{t\in{\mathbb R}}\Sigma_t$
covers spacetime, $\cup_{t\in{\mathbb R}}\Sigma_t$ is called a 
\emph{foliation} of spacetime (generated by $T$), and each 
three-dimensional hypersurface $\Sigma_t$ is called a \emph{leaf} of 
that foliation.  
	For example, the standard foliation of a Lorentz frame 
$\LoF=\{\eQ_0,\eQ_1,\eQ_2,\eQ_3\}$ into affine space slices of the form
$\{\xQ : x^0 = t\}$ is generated by the function 
$T_{\LoF}(\xQ)\defeg -\eQ_0\cdot\xQ$, which has a constant 
timelike four-gradient, $\nabQ T_{\LoF}(\xQ) = -\eQ_0$.
	There are uncountably many foliations of spacetime among which
one has to find any one that is \emph{compatible} with the structure 
of the dynamical equations, preferably a convenient one.
	Unfortunately, as we shall see in a moment, the standard
foliation is not compatible with the dynamical equations of massive LED.
	In any event, a `state at an instant $t$' thus means the 
specification of the complete set of dynamical variables on the leaf 
with index $t$ of a compatible foliation of spacetime.
	
	To find out what constitutes such a complete set of dynamical 
variables, we notice that our dynamical equations are of first order 
in the proper-time derivative of each of the kinematical particle variables
$\zQ$, $\uQ$, $\EulerQ$, and of first order in the Lorentz time derivative 
of the fields $\FQ$. 
	A dynamical state in our massive LED will therefore be a set
of the form $\{\zQ(\tau), \uQ(\tau), \EulerQ(\tau), \FQ(\ .\ )\}$, 
specified on a leaf $\Sigma_t$ of a compatible foliation of spacetime
indexed by $t$, satisfying
various consistency conditions that constrain the values of
$\zQ, \uQ, \EulerQ$ at $\tau$ and the field $\FQ(\ .\ )$ on the leaf. 
	In particular, the four-velocity has to satisfy $\norm{\uQ}^2= -1$, 
and the anti-symmetric rank-two Euler tensor has to satisfy 
$\EulerQ(\tau)\cdot\uQ(\tau) = \qv{0}$. 
	Furthermore, the Maxwell--Lorentz equations impose the constraint 
that $\FQ$ satisfy the three-divergence part of the Maxwell--Lorentz 
equations, which fixes the longitudinal part of $\FQ$ as a passive 
attribute of the charged particle.

	These are all \emph{local constraints} imposed by relativity,
which occur also in the traditional limiting theories, i.e. the 
relativistic point charge mechanics in given external force fields, 
the mechanics of a point spin in given external torque fields, and 
the Maxwell electrodynamics with given source terms, in all of which
cases the constraints are known to propagate in time if they are 
satisfied initially.
	This will be true also for our massive Lorentz electrodynamics. 

	However, our covariant first-order dynamical equations are also
mildly \emph{non-local}, and this is going to be reflected in the 
compatibility conditions to be satisfied by  the foliation of spacetime.
	Namely, since the first-order proper-time derivatives of $\uQ$ and
$\EulerQ$ at proper-time $\tau$ are given (implicitly) in terms of
integrals of the fields over the particle's support in its instantaneous 
space slice of simultaneity, the transversal part (i.e. the dynamical 
degrees of freedom) of $\FQ(\xQ)$ must be known for all 
$\xQ \in \supp \big(\fe(\| .\| )\big)$ satisfying 
$\uQ(\tau)\cdot(\xQ - \zQ(\tau)) =0$ (recall that the passive 
longitudinal part of the field $\FQ(\xQ)$ is not a dynamical 
degree of freedom).
	Hence, we say that a foliation of spacetime is \emph{compatible} 
with a world-line $\tau \mapsto \zQ (\tau)$ of an extended Lorentz
electron (i.e., a particle history) if each leaf of the foliation contains 
precisely one of the sets 
$\{\uQ(\tau)\cdot(\xQ - \zQ(\tau)) =0\} \cap\supp \big(\fe(\| .\| )\big)$
(for simplicity, we assume that $\supp(\fm) = \supp(\fe)$).
	A compatible foliation sets the spacetime parameterization w.r.t. 
which the Maxwell--Lorentz equations must be solved in order to enable the
evaluation of the Minkowski force and torque.

	A compatible foliation is of course not known explicitly
until the particle world-line is known; it has to be determined 
along with the particle world-line and the fields by solving the
second variation equations for the extrinsic curvature of the 
leaf $\Sigma_t$ coupled with the constraint equations of 
Gauss--Codacci and Gauss in the complement of the sets 
$\xQ \in \supp \big(\fe(\| .\| )\big)$ satisfying 
$\uQ(\tau)\cdot(\xQ - \zQ(\tau)) =0$, which is a non-linear 
free boundary problem.  
	Interestingly, these circumstances inject technical 
elements from general relativity into this special relativistic theory. 

\emph{Remark:}
	No foliation is compatible with a particle history that is not 
ordered in the sense described earlier (see Fig.1 and 2).

	The condition that the particle history be ordered
imposes size restrictions on the accelerating fields.
	Indeed, we easily find locally consistent data for which 
the particle's three-acceleration in its instantaneous Lorentz
rest frame exceeds the value $1/R$, where $R$ is the radius of 
the particle.
	For such  accelerations, part of the support of 
the particle in its instantaneous space slice of simultaneity 
will be launched \emph{backward in Lorentz time} rather than 
forward, and this motion does not generate an ordered history.
	This implies the following bound on the particle's initial
three-acceleration in the Lorentz frame in which the particle 
is instantaneously at rest,
\begin{equation}
	\abs{\ddot{\qV}}R < 1
\, ,\label{eq:initACCELbound}
\end{equation}
which has to be translated into bounds on the accelerating  field strengths
in the instantaneous particle rest frame.
	We will soon see that these bounds on the field strengths 
are about $100$ times the particle's self-fields at it's surface which, 
happily, is way outside the physically reasonable range of data for a 
classical electrodynamics.
	Last not least, the condition that the gyrational speed for 
points on the particle's equator must be subluminal 
imposes a size restriction on $\EulerQ$, viz. (recall that $c=1$) 
\begin{equation}
	\norm{\eulerQ} R < 1
\,. \label{eq:gyroSUBc}
\end{equation}

        \subsection{The Cauchy problem for the state}

	We now formulate the initial value problem for the states.
	We begin by specifying the generic initial data on a
convenient initial leaf.
	The compatible foliation itself is characterized subsequently
by dynamical equations which together with the equations of 
massive LED (rewritten w.r.t. the compatible foliation) 
pose a nonlinear, implicit Cauchy problem for the joint evolution 
of the state and the leaf of the foliation on which the state is defined.

        \subsubsection{Cauchy data and the initial leaf}

	Since by a suitable spacetime translation we can
always transform to a Lorentz frame in which the event of the 
particle's initial spacetime position is at the origin of that 
Lorentz frame, and since by a suitable Lorentz boost we can 
furthermore transform to another Lorentz frame with the same
origin but in which the particle's  initial three-velocity vanishes, 
and since finally we can choose the proper-time of the particle equal 
to zero at that initial event, the Cauchy data for the world-line 
$\tau\mapsto \zQ(\tau)$ can be chosen conveniently as
\begin{align}
         \zQ(0) &  = \qv{0} 
\, , \label{eq:EventInitCond} 
\\
         \uQ(0) &  = \qv{e}_0
\, . \label{eq:HodoInitCond} 
\end{align}
        Notice that $\qv{e}_0$ satisfies the 
constraint $\norm{\qv{e}_0}^2 = -1$ required of a four-velocity.
	Moreover, by at most a space rotation of this 
Lorentz frame we can achieve that $\eulerQ(0)$, the dual vector to
$\EulerQ(0)$, has only a single non-vanishing component, say the 
$3$ component, $\eulerQ(0) = \euler_0\qv{e}_3$, with $\euler_0$
satisfying \refeq{eq:gyroSUBc} but otherwise at our disposal. 
        The Cauchy data for the particle's gyrograph 
$\tau\mapsto \EulerQ(\tau)$ are then of the form
\begin{equation}
        \EulerQ(0) 
= 	
	\euler_0\, \qv{e}_1\wedge\qv{e}_2
\, . \label{eq:GyroInitCond} 
\end{equation}
	Clearly, $\EulerQ(0)$ is anti-symmetric, and 
since $(\qv{e}_1\wedge\qv{e}_2)\cdot\qv{e}_0 = \qv{0}$, also of 
space-space type w.r.t. $\uQ(0)$.

	The Lorentz frame in which 
\refeq{eq:EventInitCond}, \refeq{eq:HodoInitCond}, and \refeq{eq:GyroInitCond} 
holds will be referred to as the \emph{initial rest frame} ${\LoF}_0$.
	The initial leaf $\Sigma_0$ of all of our foliations is now chosen
to be the particle's affine space slice of simultaneity in the initial rest 
frame ${\LoF}_0$, i.e. the set $\{\qv{e}_0\cdot \xQ = 0\}$ on which the 
Lorentz time $t\, (=x^0)=0$.

	Accordingly, the Cauchy data for $\FQ(\xQ)$ are 
specified on $\Sigma_0$,
\begin{equation}
	\FQ(\xQ)|_{t=0} 
=
	\FQ_0(\xV)
\, ,
\end{equation}
where $\FQ_0(\xV)$ satisfies the three-divergence equations in the
Maxwell--Lorentz equations restricted to $\Sigma_0$.
	To make this somewhat implicit characterization of $\FQ_0$
more explicit, we decompose $\FQ_0$ into its electric and magnetic
components w.r.t. ${\LoF}_0$ and group them together 
into a complex electromagnetic three-vector field on $\Sigma_0$,
\begin{equation}
	\GV_0
\defeg 
	\EV_0 + i\BV_0
\, ,
\end{equation}
whose real and imaginary part are, respectively, the electric
(i.e. time-space) and magnetic (i.e. space-space) components 
of the field tensor $\FQ_0$ in $\Sigma_0$.
	The three-divergence equations in $\Sigma_0$ now read
\begin{equation}
	\nab\cdot\GV_0 (\xV)
= 
	4\pi\rho_0(\xV)
\, , \label{eq:MLdivEQqexplINIT}
\end{equation}
where $\rho_0$ is the time component of $\JQ(0)$, uniquely determined
in terms of the initial values of the other state variables 
$\zQ_0, \uQ_0$ and $\EulerQ_0$. 

	\emph{Remark:} Through the presence of $\qdot{\uQ}$ in the
right-hand side of \refeq{eq:Jnodvik} the (initial) constraint 
appears to be implicit.
	However, by direct computation it is readily verified that 
the $\qdot{\uQ}$ contribution to $\JQ$ drops out initially as a 
consequence of the sphericity of the charge distribution in the particle's 
instantaneous rest frame, which initially coincides with the Lorentz
frame ${\LoF}_0$.

 	The initial fields are chosen in accordance with the
asymptotic condition that $\GV_0(\xV)\to\vect{0}$ as 
$|\xV| \to \infty$, the real part as 
$\EV_0 (\xV) \sim - e \xV/|\xV|^3 +o(|\xV|^{-2})$,
the imaginary part sufficiently fast for surface integrals at
infinity involving $\BV_0$ to vanish.

    \subsubsection{Evolution equations for state and compatible foliation}

	To each differentiable particle world-line starting with
\refeq{eq:EventInitCond} and \refeq{eq:HodoInitCond} and generating
an ordered particle history (referred to by an index $_h$), we have
to associate a convenient,  compatible foliation of spacetime which
contains $\Sigma_0$ as initial leaf.
	Unfortunately, the standard foliation is generally not  
a compatible foliation, for notice that any world-line is 
the soul of the tubular set of events  
$\cup_{\tau\geq 0}\supp (\fe(\norm{\xQ-\zQ(\tau)}))$,
called the future \emph{world-tube} of the particle, 
inside of which the foliation is necessarily given by the 
$\tau$-indexed affine segments $\{\xQ | \uQ(\tau)\cdot(\xQ -\zQ(\tau))=0 \}$
which intersect with the standard leaf of Lorentz time $t$ but are 
generally strict subsets of that leaf.
	However, one expects that it is possible to select a  compatible 
foliation whose leafs, sufficiently far away from the world-line,
will arbitrarily closely approximate the leafs of the standard foliation 
of the initial Lorentz frame, i.e. $\Sigma_0$ parallelly transported in 
Lorentz time. 
	By further choosing the $t$-parameterization  of the
compatible foliation to be Lorentz time at spatial infinity, 
by selecting the corresponding value of $\tau$ through the identification 
$q^0(\tau) = t$, and  by parameterizing points in $\Sigma_t$ by the 
points $\xV\in\Sigma_0$ through following the integral curves of $\nabQ T_h$, 
we obtain a foliation which nearly coincides with the standard foliation 
in most of spacetime, except for a small ``wiggle'' accompanying the 
world-line of the particle.
	Notice that the world-line is one such integral curve, so that
in this parameterization the segment of the world-tube that belongs to
$\Sigma_t$ is always centered at the origin of $\Sigma_0$. 
	Equipped with such a global parameterization of a compatible
foliation, spacetime is diffeomorphic to the product manifold 
${\mathbb R}\times \Sigma_0$, i.e. ${\mathbb R}\times {\mathbb R}^3$,
with spacetime metric
\begin{equation}
	\dd s^2 
= 
	- \psi^2(\xV,t)\, \dd t^2 
	+ \sum_{1\leq k,l\leq 3}\gamma^{kl}(\xV,t) \dd x^k \dd x^l
\end{equation}
where $\psi(\xV,t) = \norm{-\nabQ T_h(\xQ)}^{-1}$ is the 
\emph{lapse function}, and the $\gamma^{kl}(\xV,t)$ are the components of the 
\emph{first fundamental form} $\gamma$ of the foliation induced on 
$\Sigma_t$ from the Minkowski metric. 

	Such an extension of the family of affine segments of the future 
world-tube to the outside of the world-tube is found by solving the 
structure equations for a foliation of Minkowski spacetime, supplemented
with standard foliation asymptotics at spatial infinity, and with
appropriate boundary conditions along the perimeter of each segment
in the tube which guarantee a differentiable transition at
the segment's boundary. 
	The boundary  is determined by $\zQ(\tau)$ and 
$\uQ(\tau)$ (given $\fe$, which as a model parameter
does not vary during the dynamics). 
	Since we have to determine the world-line along with the
foliation outside the world-tube, our problem belongs to the category
of the dynamical free-boundary problems.
	Notice that even if the world-line is analytic, the foliation
is surely not analytic unless the particle is stationary, in which case
one can choose the standard foliation of spacetime.

	By following~\cite{hawkingellisBOOK, christodoulouklainermanBOOK},
we conclude that outside the world-tube the foliation is a solution of the
system of evolution equations
\begin{equation}
\partial_t \gamma = - 2 \psi K
\label{eq:Gevol}
\end{equation}
\begin{equation}
	\partial_t K 
= 
	- \nab_\gamma\tens\nab_\gamma \psi 
	+ \psi\big(Ric + K \tr K - 2 K\cdot K\big)
\label{eq:Kevol}
\end{equation}
where $K$ is the \emph{second fundamental form} of the leaf of the foliation, 
i.e. the extrinsic curvature (actually defined by \refeq{eq:Gevol} 
if $T_h$ were known), $Ric$ is the Ricci tensor of $\Sigma_t$, and 
$\nab_\gamma$ denotes covariant gradient w.r.t. $\gamma$.  
	Equation \refeq{eq:Kevol} is known as the second variation formula.
	In addition, $\gamma$, $\psi$ and $K$ have to satisfy the constraints 
given by the Gauss--Codacci equation
\begin{equation}
	\nab_\gamma\cdot K - \nab_\gamma \tr K = 0
\end{equation}
and the Gauss equation
\begin{equation}
	Ric -  K\cdot K + K\tr K = 0
\end{equation}
which propagate in time if satisfied on the initial leaf. 
	This set of equations has to be supplemented by boundary conditions
along the perimeter of the world-tube segment for $\psi$ and for $K$;
and by asymptotic conditions as $|\xV|\to\infty$. 
	We demand that the foliation be \emph{normal}, i.e. that 
$\psi(\xV,t)\to 1$ as $|\xV|\to\infty$, which means that at 
spatial infinity the foliation tends to the standard foliation of spacetime.
	Furthermore, a ``gauge'' condition is at our disposal, for 
instance the maximality condition 
\begin{equation}
	\tr K = 0
\, .
\end{equation}

	The Maxwell--Lorentz equations for the field tensor $\FQ$ have to
be rewritten w.r.t. the foliation in terms of the associated decomposition 
of $\FQ$ into electric and magnetic components (cf. \cite{thorneetal}),
here conveniently grouped together as a complex electromagnetic three-vector 
field,
\begin{equation}
	\GV
\defeg 
	\EV + i\BV
\, , \label{eq:EBfoli}
\end{equation}
the initial decomposition being defined already earlier.
	The Maxwell--Lorentz evolution equations for $\GV$
then read~\cite{thorneetal}
\begin{equation}
	\partial_t\GV (\xV,t)
= 
	-i\nab_\gamma\times\big( \psi \GV\big)(\xV,t) 
	- 4\pi\big(\psi\jV\big)(\xV,t)
\, , \label{eq:MLevolvEQfoli}
\end{equation}
where $\partial_t$ means first-order foliation time derivative 
(which equals Lorentz time derivative far away from the particle), 
$\nabla_\gamma\times$ is the curl w.r.t. $\gamma^{k,l}$, and 
the electrical current density three-vector $\jV$ is the $\Sigma_t$
projection of $\JQ$.
	Clearly, \refeq{eq:MLevolvEQfoli} is an explicit equation for
the foliation time derivative of $\GV$; however, note that $\jV(\xV,t)$ 
is a (vector) functional not only of all the other state variables 
$\EulerQ, \zQ, \uQ$ but in addition also of $\qdot{\uQ}$.

	Beside the evolution equation \refeq{eq:MLevolvEQfoli}, $\GV$
has to satisfy the three-divergence equations 
\begin{equation}
	\nab_\gamma\cdot\GV (\xV,t)
= 
	4\pi\rho(\xV,t)
\, , \label{eq:MLdivEQfoli}
\end{equation}
where $\rho$, the time component of $\JQ$, is likewise 
a functional of all the other state variables $\EulerQ, \zQ, \uQ$ and 
in addition also of $\qdot{\uQ}$.
	However, \refeq{eq:MLdivEQfoli} is already included in our set of 
conditions to be satisfied by a state, for notice that by taking the 
(three-)divergence of \refeq{eq:MLevolvEQfoli} it follows that 
a solution $\GV(\xV,t)$ of \refeq{eq:MLevolvEQfoli} for $\JQ(\xQ)$ 
given in terms of consistent $\zQ(\tau)$ and $\EulerQ(\tau)$, 
automatically satisfies \refeq{eq:MLdivEQfoli} for all $t>0$ if 
it satisfies \refeq{eq:MLdivEQfoli} initially. 
 
	The Maxwell--Lorentz equations are supplemented by the
asymptotic condition that $\GV(\xV)\to\vect{0}$ as 
$|\xV| \to \infty$, the real part as 
$\EV(\xV) \sim - e \xV/|\xV|^3 +o(|\xV|^{-2})$,
the imaginary part sufficiently fast for surface integrals at
infinity involving $\BV$ to vanish.

	Finally, following~\cite{thorneetal} one converts $\GV(\xV,t)$
into $\FQ$ on a given leaf, which inserted into the world-line and
gyrograph equations \refeq{eq:wlEQ} and \refeq{eq:ggEQ} closes
the system of evolution equations for the state in massive LED.

	In summary, the Cauchy problem for the state evolution turns
out to bear a surprising resemblance to the general relativistic
evolution problem of a black hole interacting with electromagnetic 
radiation as described in~\cite{thorneetal}, and to the
evolution problem of the vacuum Einstein
equations~\cite{christodoulouklainermanBOOK}.
	This gives the equations of massive LED an unexpected additional
appeal as a somewhat simpler ``toy problem'' for these more
formidable type of problems of general relativity.
	At the same time, the enormous progress recently made in the 
rigorous treatment of the Einstein equations~\cite{christodoulouklainermanBOOK}
gives hope that the complete control of the Cauchy problem for the
state in massive LED can be achieved along similar lines of analysis. 

	We shall have recourse to the notion of the state in massive 
LED in our section about scattering. 
	In particular, the invariants stated earlier have to be 
converted into conservation laws for the evolution w.r.t. a compatible
foliation.
	In the remainder of this section now we will discuss an alternate,
global way of looking at massive LED, including an algorithm for the actual
computation of solutions to any degree of precision.

	\subsection{Nearly free global evolutions}

	If, instead of evolving the state locally in time, we 
look at the putatively harder problem of global particle histories 
and field decorations  of spacetime, one gains however the 
advantage that one can work with the standard foliation. 
	The price to pay, of having to solve the global problem, 
will be somewhat lessened by noting that for all practical purposes 
of computation, one will need only an approximate solution.
	We restrict ourselves to nearly free global future
histories and decorations, with data prescribed at $t=\tau=0$
on $\Sigma_0$, as before.

	Nearly free evolutions start from Cauchy data
which are sufficiently close to a stationary state. 
	For such evolutions, the system of fully 
implicit massive LED evolution equations can be converted into an equivalent, 
regular \emph{quasi-explicit} first-order system.
	We then explain that this quasi-explicit system of equations
can be treated via a generalized Picard iteration.\footnote{ 
	The terminology of ``explicit, quasi-explicit, and fully implicit 
	first-order system of equations,'' is readily explained at hand 
	of just two coupled dynamical variables $\xi(t), \zeta(t)$. 
	(A) a pair of equations 
	$\dot\xi = f_1(\xi,\zeta)$, 
	$\dot\zeta = f_2(\xi,\zeta)$ 
	is called an explicit first-order system.
	(B) a pair
	$\dot\xi = g_1(\xi,\zeta,\dot\zeta)$, 
	$\dot\zeta = g_2(\xi,\dot\xi,\zeta)$ 
	which is not also of the form (A) is called a 
	regular quasi-explicit first-order system. Thus, in a regular
	quasi-explicit system each dynamical variable satisfies an 
	explicit equation \emph{conditioned on the histories of the 
	other variables}.
	(C) a pair
	$F_1(\xi,\dot\xi,\zeta,\dot\zeta)=0$, 
	$F_2(\xi,\dot\xi,\zeta,\dot\zeta)=0$ 
	which is not also of the form (B) or (A) is called a fully implicit 
	first-order system. The class of fully implicit systems
	contains the singular quasi-explicit systems, which are of the form 
	$h_1(\xi,\zeta,\dot\zeta)\dot\xi = g_1(\xi,\zeta,\dot\zeta)$, 
	$h_2(\xi,\dot\xi,\zeta)\dot\zeta = g_2(\xi,\dot\xi,\zeta)$
	with coefficient functions $h_1$ and $h_2$ that vanish
	somewhere in the range of their arguments.}

        \subsubsection{Maxwell--Lorentz equations revisited}
 
	We now work with the standard foliation, so that \refeq{eq:EBfoli}
is the decomposition of $\FQ$ into electric and magnetic components
w.r.t. this standard foliation. 
	The Maxwell--Lorentz evolution equations for $\GV$
now read simply as
\begin{equation}
	\partial_t\GV (\xV,t)
= 
	-i\nab\times \GV(\xV,t) 
	- 4\pi \jV(\xV,t)
\, , \label{eq:MLevolvEQqexpl}
\end{equation}
where $\partial_t$ means first-order Lorentz time derivative,
$\nabla \times$ is the conventional curl, and the electrical 
current density three-vector $\jV$ is the space component of $\JQ$.
	Clearly, \refeq{eq:MLevolvEQqexpl} is an explicit equation for
the Lorentz time derivative of $\GV$. 
	Once again, $\jV(\xV,t)$ is a (vector) functional not only of 
all the other state variables $\EulerQ, \zQ, \uQ$ but in addition also 
of $\qdot{\uQ}$.

	Beside the evolution equation \refeq{eq:MLevolvEQqexpl}, $\GV$
has to satisfy the three-divergence equations 
\begin{equation}
	\nab\cdot\GV (\xV,t)
= 
	4\pi\rho(\xV,t)
\, , \label{eq:MLdivEQqexpl}
\end{equation}
where $\rho$, the time component of $\JQ$, is likewise 
a functional of all the other state variables $\EulerQ, \zQ, \uQ$ and 
in addition also of $\qdot{\uQ}$.
	As explained earlier, \refeq{eq:MLdivEQqexpl} will be
satisfied if it is satisfied by a initial state.
 
	The Maxwell--Lorentz equations are supplemented by the
asymptotic condition that $\GV(\xV)\to\vect{0}$ as 
$|\xV| \to \infty$, the real part as 
$\EV(\xV) \sim - e \xV/|\xV|^3 +o(|\xV|^{-2})$,
the imaginary part sufficiently fast for surface integrals at
infinity involving $\BV$ to vanish.

        \subsubsection{Gyrograph equations revisited}

	The gyrograph equation~\refeq{eq:ggEQ}, when viewed as an
equation for $\EulerQ$, does not have the desired format in the 
first-order proper-time derivative of $\EulerQ$ because $\qdot{\SQb}$ 
involves $\qdot{\EulerQ}$ in a rather complicated manner. 
	However, it is possible and also convenient to 
view \refeq{eq:ggEQ} as an equation for $\SQb$. 
        To see that  \refeq{eq:ggEQ} can indeed be viewed as
an equation for $\SQb$, we recall that we required $\fm$ to
generate a strictly positive moment of inertia ${\cal I}_{\text{b}}$.
	As a consequence, the map 
$\norm{\eulerQ}\mapsto{\cal I}_{\text{b}}(\norm{\eulerQ})$
is {strictly positive}, {increasing}, and {strictly convex} 
for $\norm{\eulerQ}\in[0,1/R)$.
	This implies that we can invert 
$\SQb = {\cal I}_{\text{b}}(\norm{\eulerQ})\EulerQ$
to get the Euler angular velocity tensor $\EulerQ$
uniquely in terms of the bare spin tensor $\SQb$, viz.
$\EulerQ = {{\cal I}^{-1}_{\text{b}}}^*(\|\SQb\|)\SQb$, 
where ${{\cal I}^{-1}_{\text{b}}}^* (\xi) = 
\xi^{-1}({\cal I}_{\text{b}}\, {\rm id})^{-1}(\xi)$.
	With $\EulerQ$ understood in this way as unique (vector) 
functional of $\SQb$, the Minkowski torque $\tQT$ given in 
\refeq{eq:MinkALtorque} is now a (tensor) functional of 
$\SQb,\FQ,\zQ,\uQ,\qdot{\uQ}$, so that \refeq{eq:ggEQ}, 
as equation  for $\SQb$, 
\begin{equation}
	\frac{\dd\phantom{\tau}}{\dd\tau}\SQb
= 
	 -\, \big[\ThomasQ ,\SQb \big]_-\, +	\tQT
\, ,\label{eq:gyroEQqexpl}
\end{equation}
is already in the desired format regarding $\qdot{\SQb}$. 
	Notice that not only $\tQT$ but also $\ThomasQ$ still 
depends on $\qdot{\uQ}$. 

        \subsubsection{World-line equations revisited}

	We now come to the world-line equation~\refeq{eq:wlEQ},
which is of second order in the proper-time derivatives of $\zQ$.
	Since we can treat $\zQ$ and $\uQ$ as independent
degrees of freedom, this second-order equation is equivalent to two
first-order equations; namely, \refeq{eq:wlEQ}, now viewed as 
equation for ${\uQ}$, together with
\begin{equation}
        \frac{\dd\zQ}{\dd\tau} 
        =
        \uQ
\, \label{eq:qEQqexpl}
\end{equation}
for $\zQ$.
	While \refeq{eq:qEQqexpl} already is in the desired format,
\refeq{eq:wlEQ} is not.
	In fact,  $\qdot{\uQ}$ enters through  $\qdot{\pQ}$ in
the l.h.s. of \refeq{eq:wlEQ} and through the tensor 
$\ThomasQ =  \qdot{\uQ}\, \wedge\,\uQ$ in the r.h.s. of \refeq{eq:wlEQ}, 
so that \refeq{eq:wlEQ} is fully implicit first-order 
no matter whether viewed as equation for $\uQ$ or for $\pQ$. 
	Switching to $\pQ$ as dynamical variable therefore does not 
seem to offer an advantage in the conversion of \refeq{eq:wlEQ} 
into the desired format. 

	Continuing thus to handle \refeq{eq:wlEQ} as an equation for ${\uQ}$,
we first sort out all terms involving $\qdot{\uQ}$. 
	Beginning with the r.h.s. of the world-line equation~\refeq{eq:wlEQ}, 
we notice that $\fQ$ can be decomposed into a sum of two terms, one of which 
is independent of $\qdot{\uQ}$ and the other one linear in $\qdot{\uQ}$,
\begin{align}
        \fQ
 = 
& 	\int_{{\mathbb R}^{1,3}}
                 \FQ(\zQ+\xQ)\cdot\big(\uQ-\EulerQ\cdot\xQ\big) 
         \fe(\norm{\xQ})\delta(\uQ\cdot\xQ)
        \dd^4\xQ 
\notag \\
& \qquad + \int_{{\mathbb R}^{1,3}} 
	\FQ(\zQ+\xQ)\cdot\uQ\, (\qdot{\uQ}\cdot\xQ)
        \fe(\norm{\xQ})\delta(\uQ\cdot\xQ) \dd^4\xQ
\, .
\end{align}
	The l.h.s. of \refeq{eq:wlEQ} can be decomposed in a similar fashion.
	For $\pQ=\Mmink\cdot\uQ$ we have
\begin{equation}
        \qdot\pQ
        = 
        \qdot \Mmink\, \cdot\,\uQ + \Mmink\,\cdot\qdot{\uQ}
\, ,\end{equation}
with $\Mmink$ given in ~\refeq{eq:gyroMASS}, and with
\begin{align}
        \qdot\Mmink 
= {} 
	\qdot{\rotM}\tenseur{g} 
& 	
 	 - \int_{{\mathbb R}^{1,3}}
	\Big[\xQ\tens\xQ,
		\big[\FQ(\zQ+\xQ),\qdot{\EulerQ}\big]_+
	\Big]_+ 
        \fe(\norm{\xQ})\delta(\uQ\cdot\xQ)\dd^4\xQ
\notag \\
&
	- \int_{{\mathbb R}^{1,3}} 
        	\Big[\xQ\tens\xQ,\big[\uQ\cdot\nabQ\FQ(\zQ+\xQ),
\EulerQ\big]_+\Big]_+
        \fe(\norm{\xQ})\delta(\uQ\cdot\xQ)\dd^4\xQ 
\notag \\
&
	+\int_{{\mathbb R}^{1,3}} 
	\Big[\xQ\tens\xQ,
		\big[\FQ(\zQ+\xQ),{\EulerQ}\big]_+
	\Big]_+
        \fe(\norm{\xQ})(\qdot{\uQ}\cdot\xQ)\uQ\cdot\nabQ\delta(\uQ\cdot\xQ)
	\dd^4\xQ
\, .
\end{align}
	The tensor $\Mmink$ is independent of $\qdot{\uQ}$, and so
are the first three terms of $\qdot{\Mmink}$, while the last term
in $\qdot{\Mmink}$ is linear in~$\qdot{\uQ}$.
	In conclusion, the world-line equation can therefore be
written as
\begin{equation}
        \widetilde{\tenseur{M}}\,\cdot\qdot{\uQ} 
= 
	\widetilde{\fQ}
\, ,\label{eq:wlEQqexpl}
\end{equation}
where the four-force $\widetilde{\fQ}$ depends only on~$\zQ$, $\uQ$,
$\EulerQ$, $\qdot{\EulerQ}$, $\FQ$, and  $\partial_t\FQ$, thus
\begin{align}
        \widetilde{\fQ} 
= 
	-	\qdot{\rotM}\uQ      
& +
 	\int_{{\mathbb R}^{1,3}} 
		\FQ(\zQ+\xQ) \cdot\big(\uQ- \EulerQ\cdot\xQ\big)  
	\fe(\norm{\xQ})\delta(\uQ\cdot\xQ) \dd^4\xQ 
\nonumber \\
& +
 	\int_{{\mathbb R}^{1,3}} 
\xQ\big(\xQ\cdot\EulerQ\cdot\big(\uQ\cdot\nabQ\FQ(\zQ+\xQ)\big)\cdot\uQ\big)
        \fe(\norm{\xQ})\delta(\uQ\cdot\xQ)\dd^4\xQ
\nonumber \\
& +
 	\int_{{\mathbb R}^{1,3}}  
        \xQ\big(\xQ\cdot 
		\big[\FQ(\zQ+\xQ),\qdot{\EulerQ}\big]_+\cdot\uQ	\big)
        \fe(\norm{\xQ})\delta(\uQ\cdot\xQ)\dd^4\xQ
\, ,
\end{align}
and the pseudo-inertia tensor $\widetilde{\tenseur{M}}$ only on~$\zQ$, $\uQ$,
$\EulerQ$, $\FQ$, and  $\partial_t\FQ$, thus
\begin{align}
        \widetilde{\tenseur{M}} 
=
\  	\rotM\tenseur{g} 
&       -\int_{{\mathbb R}^{1,3}} 
		\Big[\xQ\tens\xQ,
			\big[ \FQ(\zQ+\xQ),\EulerQ\big]_+
		\Big]_+ 
	\fe\big(\norm{\xQ}\big) \delta(\uQ\cdot\xQ)\,\dd^4\xQ
\notag \\
  &     - \int_{{\mathbb R}^{1,3}}
\xQ\tens\xQ \big(\xQ\cdot\EulerQ\cdot \FQ(\zQ+\xQ)\cdot\uQ\big)
	\fe\big(\norm{\xQ}\big)\uQ\cdot\nabQ \delta(\uQ\cdot\xQ)\,\dd^4\xQ
\notag \\ 
  &     - \int_{{\mathbb R}^{1,3}}
                \FQ(\zQ+\xQ)\cdot\uQ\tens\xQ  \fe\big(\norm{\xQ}\big) 
        \delta(\uQ\cdot\xQ)\,\dd^4\xQ
\, .\label{eq:tildeMtensor}
\end{align}

	\emph{Remark:} The first three terms in \refeq{eq:tildeMtensor} 
are manifestly symmetric, but the fourth one is not and, as far as we
can see, also cannot be symmetrized by adding a term whose action on
$\qdot{\uQ}$ yields zero.
	Although this brings some technical inconveniences for concrete
computations, it does not interfere with the remaining steps in
our analysis. 
	
	At this point, the system of evolution equations 
\refeq{eq:MLevolvEQqexpl}, \refeq{eq:gyroEQqexpl}, \refeq{eq:qEQqexpl}, 
and \refeq{eq:wlEQqexpl} is already a quasi-explicit first-order
system; however the tensor $\widetilde{\tenseur{M}}$ could be 
singular (as an operator acting by matrix multiplication (w.r.t. the
Minkowski metric) on four-vectors). 
	Our next step is to show that in typical situations
$\widetilde{\tenseur{M}}$ is not singular.
	We will establish the invertibility of $\widetilde{\tenseur{M}}$ 
for admissible initial data in a large neighborhood of those data that 
correspond to the stationary solutions with empirical electron
characteristics that covers essentially all physically sensible 
situations.

	We begin with the observation that, since we have chosen a
Lorentz frame for the initial value problem which coincides with the 
particle's initial rest frame, we can in particular choose initial 
data such that the particle and the fields remain stationary for all time. 
	The stationary electric Coulomb field and the
stationary gyro-magnetic field field, denoted here by an index $_{st}$,  
are defined by the equations
\begin{equation}
	{-i}{\nab}\times{\GV}_{st}({\xV}) 
= 
	4\pi \fe\big(|{\xV}|\big)\,\eulerV\times\xV
\, , \label{eq:MLevolvEQcomove}
\end{equation}
with $\eulerV=\eulerV_0$ constant, and 
\begin{equation}
\quad	
	{\nab}\cdot{\GV}_{st} ({\xV})
= 
	4\pi	\fe\big(|{\xV}|\big)\,
\, , \label{eq:MLdivEQcomove}
\end{equation}
together with the condition that 
$\GV_{st}({\xV})\to \vect{0}$ as $|{\xV}|\to \infty$.
	For  a stationary situation in which the charge, 
gyro-magnetic moment, and renormalized  mass are matched to those of 
the empirical electron, with $\mz>0$ but small (details are given in 
section 10.1), the first and third field integrals in \refeq{eq:tildeMtensor},
evaluated with the corresponding $\FQ_{st}$, are of magnitude $\alpha$ 
(Sommerfeld's fine structure constant) relative to the gyrational 
bare term, while the second field integral in \refeq{eq:tildeMtensor}, 
evaluated with the corresponding $\FQ_{st}$, vanishes. 
	The field integrals in \refeq{eq:tildeMtensor} are therefore
only small perturbations (in the mathematical sense of operator theory)
to the gyrational bare term $\rotM\gQ$ in \refeq{eq:tildeMtensor},
which in turn is invertible because it acts just as multiplication
by the strictly positive coefficient $\rotM$ ($\approx m_{\text{e}}$, 
the empirical electron mass) --- recall that the metric tensor~$\gQ$ acts 
as identity on four-vectors. 
	Hence, we conclude that for a stationary, empirical data-honoring 
situation, $\widetilde{\tenseur{M}}$ is invertible. 

	Having established the invertibility of such a  stationary 
$\widetilde{\tenseur{M}}$, we now turn to the discussion of
initial data which do not launch a stationary solution. 
	Since $\alpha < 10^{-2}$, in such a nonstationary
situation the tensor $\widetilde{\tenseur{M}}$ will remain 
invertible roughly up to the point where the field integrals 
in \refeq{eq:tildeMtensor} become comparable to the gyrational bare term. 
	Inspection of the field integrals reveals that this limit is 
reached (a) for applied initial field strengths about 
$100$ times the surface field strength of the stationary electron, 
i.e. about $10^{8} \rm{V/m}$, or (b) for values of 
${-i}{\nab}\times{\GV}_0({\xV}) - 
	4\pi \fe\big(|{\xV}|\big)\,\eulerV_0\times\xV$ 
about $10^2$ times the surface field strength of the stationary 
electron per radial size of the electron.
	Interestingly, field configurations that strong are precisely
those that saturate the admissibility condition 
\refeq{eq:initACCELbound}. 
	While therefore it is to be expected that a proof, if at all possible,
of the general invertibility of $\widetilde{\tenseur{M}}$ for all \apriori\
admissible initial data will be a technically hard problem, 
from a physical perspective it is idle to worry 
about field strengths that strong within the Lorentz model.
	These critical field strengths are way outside
the reasonable range of applied field strengths for which one can hope 
that the Lorentz model makes physically sensible predictions; whence, 
we will not address the issue of the critical field strengths any
further and simply conclude with the observation that for essentially all 
physically acceptable initial data the tensor $\widetilde{\tenseur{M}}$  
is a small perturbation of $\rotM\gQ$ and therefore invertible.

	We may now multiply \refeq{eq:wlEQqexpl} by the inverse of
$\widetilde{\tenseur{M}}$ to obtain the first-order evolution equation
\begin{equation}
         \frac{\dd\uQ}{\dd\tau} 
=
        \widetilde{\tenseur{M}}^{-1}\cdot\widetilde{\fQ}
\, , \label{eq:hgEQ}
\end{equation}
valid surely initially and, if a continuously differentiable solution $\uQ$
of \refeq{eq:wlEQqexpl} exists, valid also for at least 
a short ensuing time interval during which $\widetilde{\tenseur{M}}$ 
will continue to be invertible, then, literally by continuity. 
	Conversely, if \refeq{eq:hgEQ} has a continuously differentiable 
solution $\uQ$, then as long as $\widetilde{\tenseur{M}}$ remains invertible 
this solution will also solve \refeq{eq:wlEQqexpl}. 
	Equation \refeq{eq:hgEQ} is called the \emph{hodograph} equation for 
$\uQ$.
	It is of the desired format in $\qdot{\uQ}$.

        \subsubsection{Consistency}

	Before we address the question of how to solve our 
quasi-explicit system of evolution equations, a remark is in order
regarding the following subtle point. 
	Since $\uQ\cdot\qdot\uQ =0$, by taking the inner product of 
\refeq{eq:hgEQ} we see that we must have
$\uQ\cdot \widetilde{\tenseur{M}}^{-1}\cdot\widetilde{\fQ}=0$. 
	Unfortunately, this last equation is not an automatic identity
(as is, for instance, $\uQ\cdot\FQ\cdot\uQ = 0$, which follows just by 
the anti-symmetry of $\FQ$, irrespective of whether $\uQ$ and $\FQ$
are dynamically related or not), and this may cause some concern.
	However, it is only necessary that 
$\uQ\cdot \widetilde{\tenseur{M}}^{-1}\cdot\widetilde{\fQ}$ 
vanishes identically for solutions to the dynamical equations,
and this can be shown to be the case. 
	In fact, it holds for any solution of \refeq{eq:hgEQ} with 
given fields $\FQ$ and gyrograph $\EulerQ$ \emph{conditioned} on 
$\tau\mapsto \EulerQ(\tau)$ being a solution of \refeq{eq:ggEQ}.
	In brief, the apparent difficulty traces back to the fact
that Nodvik's Minkowski force $\fQ$ given in \refeq{eq:MinkALforce} is 
generally not four-orthogonal to $\uQ$~\cite{nodvik} because it 
couples the spin-generated electromagnetic (dipole and higher
multipole) moments to the inhomogeneities of the electromagnetic 
fields over the size of the particle. 
	We have
\begin{equation}
	\fQ\cdot\uQ
= 
 -   \int_{{\mathbb R}^{1,3}} (\xQ-\zQ)\cdot\EulerQ\cdot\FQ(\xQ)
        \fe(\norm{\xQ- \zQ})
	\,\delta\Big(\uQ\cdot\big(\xQ-\zQ\big)\Big)\,\dd^4\xQ \cdot\uQ
\,,\label{eq:MinkALforceCDOTu}
\end{equation}
which vanishes identically (irrespective of $\FQ$) only for a so-called 
`particle without spin' for which $\EulerQ\equiv \tenseur{0}$ (more on 
particle models without spin in Appendices A.2 and A.3).
	This non-orthogonality of four-velocity and Minkowski force is
countered on the left side of \refeq{eq:wlEQ} by the spin-orbit coupling 
term $\Mnodv$ in $\Mmink$, which creates a generally anisotropic
translational inertia of the gyrating charged particle so that the 
Minkowski momentum $\pQ$ and the four-velocity $\uQ$ are not anymore 
parallel to each other but related by the genuine
tensor proportionality \refeq{eq:MinkEnMom}. 
	By adapting the corresponding computations of Nodvik~\cite{nodvik} 
to the present model, it can now be shown that the equation obtained 
by taking the inner product of \refeq{eq:wlEQ} with $\uQ$ is identically 
satisfied conditioned on $\tau\mapsto \EulerQ(\tau)$ being a solution of the 
gyrograph equations \refeq{eq:ggEQ}, \refeq{eq:MinkALtorque}.
	The correspondingly conditioned identity
$\uQ\cdot \widetilde{\tenseur{M}}^{-1}\cdot\widetilde{\fQ}=0$ now
follows from the fact that \refeq{eq:hgEQ} is equivalent to
\refeq{eq:wlEQ}.

        \subsubsection{Iterative treatment}

	Having succeeded in writing the whole evolution problem
as a regular quasi-explicit first-order system~\refeq{eq:MLevolvEQqexpl},
\refeq{eq:gyroEQqexpl}, \refeq{eq:qEQqexpl}, and \refeq{eq:hgEQ}, in
the final step in our local analysis of the  initial value problem 
we now explain that this regular quasi-explicit first-order system 
can be treated by a direct generalization of the familiar Picard iteration 
scheme.
	This then verifies that a unique regular solution exists and,
above all, shows how it can actually be computed. 

	Thus, we consider the coupled sequences of approximate
world-lines $\{\zQ^{(n)}(\tau)\}_{n=0}^\infty$, $\tau \geq 0$;
approximate hodographs
 $\{\uQ^{(n)}(\tau)\}_{n=0}^\infty$,  $\tau \geq 0$;
approximate gyrographs
$\{\EulerQ^{(n)}(\tau)\}_{n=0}^\infty$,  $\tau \geq 0$; 
and approximate electromagnetic fields 
$\{ \GV^{(n)}(\xV,t)\}_{n=0}^\infty$, with 
$\xV\in {\mathbb R}^{3}$ and $t\geq 0$, jointly defined by 
\begin{equation}
           \frac{\dd}{\dd\tau}  \zQ^{(n+1)}
=
	\uQ^{(n)}
\, , \label{eq:zEQitera}
\end{equation}
\begin{equation}
         \frac{\dd}{\dd\tau}  \uQ^{(n+1)}  \,
=
        \big(\widetilde{\tenseur{M}}^{-1}\big)^{(n)}\cdot\widetilde{\fQ}^{(n)}
\, , \label{eq:hgEQitera}
\end{equation}
\begin{equation}
	\frac{\dd\phantom{\tau}}{\dd\tau}\SQb^{(n+1)}
= 
	 -\, \big[\ThomasQ^{(n)} ,\SQb^{(n)} \big]_-\, +\tQT^{(n)}
\, ,\label{eq:gyroEQitera}
\end{equation}
and
\begin{equation}
	\partial_t\GV^{(n+1)} (\xV,t)
= 
   {-i}\nab_\gamma\times(\psi\GV)^{(n)}(\xV,t) - 4\pi(\psi\jV)^{(n)}(\xV,t)
\, , \label{eq:MLevolvEQitera}
\end{equation}
together with the initial-time constraint
\begin{equation}
\quad	
	{\nab}\cdot{\GV}^{(n+1)} (\xV,0)
= 
	4\pi	\rho^{(n)}({\xV},0)
\, , \label{eq:MLdivEQitera}
\end{equation}
starting the iteration with functions that are constant in time,  
\begin{align}
\GV^{(0)}(\xQ) &= \GV_0(\xV) \\
\zQ^{(0)} &= \qv{0} \\
\uQ^{(0)} &= \qv{e}_0 \\
\SQb^{(0)}&={\cal I}_{\text{b}}(|\eulerV_0|)\,\euler_0\,\qv{e}_1\wedge\qv{e}_2
\, ,\label{eq:initDATAitera}
\end{align}
and with correspondingly vanishing time derivatives of the starting data, 
\begin{align}
\partial_t\GV^{(0)}(\xQ) & = \vect{0} \\
\qdot\zQ{}^{(0)} & = \qv{0} \\
\qdot\uQ{}^{(0)} & = \qv{0} \\
\qdot{\SQ}_{\text{b}}{}^{(0)} & = \tenseur{0}
\, .\label{eq:DOTinitDATAitera}
\end{align}
	This set of equations determines the first-order time derivatives
of the maps $t\mapsto \FQ^{(n+1)}$, $\tau\mapsto\zQ^{(n+1)}$,
$\tau\mapsto \uQ^{(n+1)}$, and $\tau\mapsto \SQb^{(n+1)}$, in terms 
of the  order $n$ maps 
$t\mapsto \GV^{(n)}$, $\tau\mapsto\zQ^{(n)}$,
$\tau\mapsto \uQ^{(n)}$, and $\tau\mapsto \SQb^{(n)}$, \emph{and} the
first-order time derivatives of the  order $n$ maps.
	The  maps $t\mapsto \FQ^{(n+1)}$, $\tau\mapsto\zQ^{(n+1)}$,
$\tau\mapsto \uQ^{(n+1)}$, and $\tau\mapsto \SQb^{(n+1)}$ themselves
are now obtained by simply integrating the whole system with
respect to the respective time variables. 
	This is done for the Maxwell--Lorentz equations 
\refeq{eq:MLevolvEQitera}, \refeq{eq:MLdivEQitera} by 
the standard Laplace transform technique, thus picking up the
consistent initial data $\GV_0$ (i.e. $\FQ_0$) posed at $t=0$
through the familiar integration by parts on $\partial_t\GV$.
	Equations  \refeq{eq:zEQitera}, \refeq{eq:hgEQitera},
and \refeq{eq:gyroEQitera} are simply integrated w.r.t. the proper-time 
from $\tau =0$ to $\tau$, thereby picking up the initial data 
$\zQ(0)$, $\uQ(0)$, and $\SQb(0)$ (viz. $\EulerQ(0)$) from the integration 
of the explicit proper-time derivatives in the respective left-hand sides. 
	This completes the $n+1^{st}$ iteration cycle.
	Starting with the prescribed order zero
maps \refeq{eq:initDATAitera} and their derivatives
\refeq{eq:DOTinitDATAitera}, one can thus systematically proceed 
to higher order iterates.
	The upshot is an algorithm for the practical computation of the
solution, provided the iteration converges. 

	\emph{Remark:} It is important to notice that ``practical
computation'' means to be able to obtain an approximate solution to 
any degree of precision in some finite region of spacetime.
	To reach this goal it is \emph{not} necessary to actually
compute the global iterands in spacetime (which would be impossible
practically).
	A finite number of iterations on a nested, decreasing sequence
of spacetime regions, starting with a sufficiently big initial region,
will suffice. 

	\subsubsection{Some remarks on rigorous results}

	The convergence of our algorithm is equivalent
to a contraction principle for the fixed-point map in the global
future particle histories starting with the specified Cauchy
data, their time derivatives (treated as independent
functions in the iteration and in the fixed point map), and
the field decorations of spacetime in the future of $\Sigma_0$,
starting with the initial field configuration on $\Sigma_0$.
	Proving the contraction principle establishes the convergence 
of the iteration to a unique global solution in the future (and under
similar conditions also the past) of the initial time for those 
admissible Cauchy data for which  the tensor $\widetilde{\tenseur{M}}$  
remains a small perturbation of $\rotM\gQ$ and therefore invertible.
		
	So far, for the model in which the bare rest mass and 
the charge are both concentrated on the same sphere, and provided 
the ratio of electrostatic to bare rest mass is smaller than $1$,
we proved global existence and uniqueness for the iteration algorithm
in the special case that only the particle's spin degrees of freedom 
participate in the particle + field dynamics.	
        In the further specialization that the axis of rotation
remains fixed, we were able to prove exponentially fast approach
to a stationary gyration state for the field-particle system,
convergence understood on families of nested compact sets,
superposed on which is an outgoing field of electromagnetic radiation.
	For the proofs of these results, see~\cite{appelkiessling}.

	For more general evolutions, convergence has not yet been
established. 
	The dynamical equations of our massive LED may spoil the
global existence of the dynamics in several ways.
	First, the equatorial rotation speed has to remain subluminal 
for a particle with finite bare rest mass $\mz$. (As we will see later on,
the situation is different in the limit $\mz\to 0$ where the rotation
speed can reach the speed of light, or actually is locked in at that 
speed). 
	Second, the acceleration of the particle has to remain 
below a critical threshold (mentioned above) to guarantee an
ordered particle history. 
	Third, but only relevant to the multi-particle systems not
considered here, any two particles have to have a minimum distance 
between them to avoid overlapping cutoffs.
	We now comment on the first two problems.

        The first problem can be controlled by an appropriate
\apriori\ choice of the bare rest mass distribution.
	It suffices to discuss two representative choices, 
volume and surface inertia. 
	They differ in their limiting behavior when $\|\eulerQ\|\to 1/R$.
        The volume choice for $\fm$ is representative for the larger class 
of continuous functions $\fm \in C^0([0,R])$, in all of which
cases $\lim_{\|\eulerQ\| \to 1/R}\rotM(\|\eulerQ\|)<\infty$.  
        In such a case it takes only a finite amount of gyrational 
energy to bring an equatorial point to the speed of light. 
        This signals that the choice of a continuous $\fm$
will create a mathematically delicate problem when it comes to the 
question whether the gyrational dynamics remains subluminal, and so 
one may expect that this choice may easily lead to a singularity in 
finite time.
	In contrast, for the surface inertia choice for $\fm$, the 
function $\rotM$ is not only increasing and convex, but satisfies 
$\rotM(\norm{\eulerQ})\to \infty$ as $\|\eulerQ\| \to 1/R$.
        In this case it would require an infinite amount of gyrational 
energy to bring an equatorial point to the speed of light,
so that it is to be expected that the dynamics will
not lead to a singularity in finite time.
        These considerations show that surface inertia is preferable,
although we do not claim that this requirement is  necessary for 
global or even local existence.

	The second problem is much harder to control \apriori.
	For a moving particle one now has to rule out 
that constructive interference of  electromagnetic radiation fields 
leads to a violation of the acceleration limit (in our Lorentz frame)
\begin{equation}
  \abs{\ddot{\vect{q}}}<\frac{1}{R\gamma^3}
\, ,\label{eq:condONaccel}
\end{equation}
where $R$ is the radius of the particle and 
$\gamma=1/\sqrt{1-\dot{\vect{q}}^2}$. 
	At the initial instant $t=\tau =0$, the particle is at rest
and the estimate \refeq{eq:condONaccel} reduces to \refeq{eq:initACCELbound}.

	Notice finally that the smallness condition that we need in
our proof in~\cite{appelkiessling} is surely going to be violated if 
one lets the bare rest mass flow to zero. 
	The control of this situation apparently requires  
completely different techniques. 

	Due to all these technical difficulties, the global existence 
and uniqueness of regular solutions, at least for the physically
relevant parameter range of the model, still awaits its verification.
	In the following sections we proceed under the assumption that 
the dynamics of the field-particle system is globally well-defined.
	Put differently, whatever is concluded in those sections applies 
only to those situations for which a unique solution exists for all time.

        \section{Scattering}

	Having established that the dynamical state in our 
single-particle massive LED in spacetime ${\mathbb R}^{1,3}$ is given by 
the transversal part of $\GV$ (viz., $\FQ$) on a spacelike hypersurface 
of spacetime containing the particle support, together with the particle 
variables $\zQ$, $\uQ$, and $\EulerQ$ (or $\SQb$) coincidental with
the particle support in the spacelike hypersurface, we now turn the 
important question which of these states are scattering states, and 
in which sense.
	Loosely speaking, a \emph{scattering state} is a state whose
backward evolution tends asymptotically in the infinite past
to a superposition of free evolutions of propagating (in space) 
field and particle degrees of freedom, and similarly its forward 
evolution tends asymptotically in the infinite future
to a(nother) superposition of free evolutions of propagating
field and particle degrees of freedom.
        The \emph{scattering problem} is concerned with establishing the 
existence of the scattering states, with their unique identification, 
and further with their classification according as to how their 
asymptotically free evolution in the infinite past is connected 
with their asymptotically free evolution in the infinite future. 

	We will limit our discussion to situations in which 
the conserved quantities $\PQ$, $\LQ$, and $s^2$ are finite. 
	Moreover, we take a foliation in which $\PQ = (M,\vect{0})$,
i.e. a center of mass frame for the field-particle system.
	We assume the center of mass is at the origin of the initial  leaf. 

	We expect that in our massive LED there are basically three
categories of states, namely the non-propagating (a.k.a. bound) states, 
the scattering states, and the `catastrophic states'. 
	For the catastrophic states the 
evolution either terminates at a finite time, or blows up in
infinity time, or suffers some other pathological behavior like
loss of uniqueness. 
	As explained in the previous section, the global existence
and uniqueness problem is largely unsettled so that we will not be 
able to rule out catastrophic states nor characterize them as 
physically irrelevant (although we expect the latter to be true).
	However, the special global existence and uniqueness result
in~\cite{appelkiessling} shows that for a certain class of initial 
conditions the evolution does exist globally, and these states are 
in fact scattering states. 
	We shall show in this section that the scattering states have
the remarkable, and in fact physically indispensable character of 
soliton dynamics for the renormalized particle degrees of freedom. 
	As for the bound states, we readily characterize the 
stationary states, but we have no control yet on the question of 
whether periodic (or perhaps quasi-periodic) non-propagating states
exist as well.

	This section on scattering  is mainly included for the 
purpose of demonstrating the important solitonic character of the 
renormalized particle dynamics in scattering situations. 
	As such, we will necessarily be brief on the other scattering
issues. 

	As a general technical reference to scattering theory we recommend
\cite{reedsimonBOOKiii}, though most of the emphasis there is on 
quantum mechanical potential scattering; in this context, see 
also~\cite{sigalsofferA, sigalsofferB} for 
subsequent technical breakthroughs, and \cite{duerretal} for a very clear 
exposition of the conceptual issues involved and further references.
	Various classical scattering results in electrodynamics are discussed 
in~\cite{jacksonBOOK}.

        \subsection{The bound states} 

	Massive LED describes the interacting dynamics of two 
\emph{a priori} subsystems, the charged bare particle and the 
electromagnetic field, each of which is characterized by its
own  dynamical degrees of freedom. 
	With only a single charged bare particle interacting with the
electromagnetic field in otherwise empty space, it could seem that 
there are no bound states.
	However, some of the particle's spin degrees of freedom 
couple strongly to some of the dynamical degrees of freedom of the 
electromagnetic field, as a result of which bound states exist.
	The most elementary ones give rise to the notion of the 
\emph{renormalized particle}.
	In addition, more subtle scenarios may possibly lead to dynamical
bound states, on which we comment only briefly. 

        \subsubsection{The stationary bound states} 

	The most obvious bound states are stationary solutions 
with nonvanishing particle spin. 
	Clearly we can choose the standard foliation of spacetime
for these states, i.e. a Lorentz frame~$\LoF$ in which the particle is
at rest at the origin of the space slice $\Sigma_0\sim\Sigma_t$; i.e. 
$\zQ(\tau)=\qv{0}$ for all $\tau$ so that $\uQ(\tau) = \qv{e}_0$.
	The stationary states compatible with our conventions made
in the Cauchy problem have been defined earlier; all other stationary
bound states that satisfy our convention about the center of mass
are now obtained by at most a rotation from these stationary states.
        
	For later convenience, we here list the solutions explicitly.
	The stationary particle in the center of mass frame
is spinning with constant angular velocity tensor $\EulerQ$.
	The stationary charge-current density four-vector thus reads
\begin{equation}
	\JQ_{st}(\xQ)
=
	\Big(\qv{e}_0 - \EulerQ\cdot{\xQ}\Big)\fe\big(\norm{\xQ}\big).
\end{equation}
	We seek a stationary solution of the vector wave equation
$\wop \AQ_{st} = \JQ_{st}$ satisfying the Lorentz gauge condition and
vanishing at spatial infinity.
 	Recalling that $- \EulerQ\cdot{\xQ} = (0, \eulerV\times \xV)$,
and introducing the time-space decompositions for the current
density $\JQ=(\rho,\jV)$ and electromagnetic potential four-vector
$\AQ=(\phi,\AV)$, for the stationary  Coulomb potential $\phi_{st}(\xV)$
we find
\begin{equation}
	\phi_{st}(\xV) 
= 
  \int_{\Rset^3}\frac{1}{\abs{\xV-\xV'}}\fe\big(\abs{\xV'}\big)\,\dd^3x'
\end{equation}
and for the stationary vector potential $\AV_{st}(\xV)$,
\begin{equation}
	\AV_{st}(\xV)
=
	\eulerV\times \int_{\Rset^3}\frac{\xV'}
	  {\abs{\xV-\xV'}}\fe\big(\abs{\xV'}\big)\,\dd^3 \xV'
\, .\label{eq:statA}
\end{equation}
        The electric and magnetic fields are obtained in the usual manner 
as $\EV_{st}(\xV)=-\nab\phi_{st}(\xV)$ and 
$\BV_{st}(\xV)=\nab\times\AV_{st}(\xV)$, 
so that
\begin{equation}
  \GV_{st}(\xV,t)  =-\nab \phi_{st}(\xV) + i \nab\times\AV_{st}(\xV).
\end{equation}

	Since we chose a spherical distribution~$\fe(|\ .\ |)$ with
compact support in ${\mathbb R}^3$, the Coulomb and the vector potential 
take a universal form outside the support of the particle, given by
\begin{equation}
 \left.
  \begin{aligned}
	\phi_{st}(\xV) 
& = 
	-e\frac{1}{\abs{\xV}} 
\\ \\
	\AV_{st}(\xV) 
& = 
	\vect{\mu}\times\frac{\xV}{\abs{\xV}^3} \\
  \end{aligned}
 \right\}
		\qquad\text{for $r>R$}
\, , 
\end{equation}
where $\vect{\mu}$ is the magnetic moment for the stationary $\eulerV$,
\begin{equation}
        \vect{\mu} 
= 
	\frac{1}{ 2} \int_{\mathbb{R}^3}
	\xV\times(\eulerV\times\xV)\fe(|\xV|)\dd^3x
\, .\label{eq:magmom}
\end{equation}
	Clearly, away from its support the stationary charged particle's
electromagnetic signature is that of an electric point charge and a 
magnetic point dipole, as was to be anticipated.

	Finally, we remark that the static limit $|\eulerV|\to 0$ 
for the stationary bound state is no longer a bound state of the 
field-particle system, in the sense that only the passive Coulomb 
field remains attached to the charged bare particle.

        \subsubsection{The non-stationary bound states} 

	In various approximate, linearized versions of (massive) LED, 
time-periodic non-propagating field-particle solutions have been found.
	Time-periodic solutions do not exist if the
\emph{Wiener} condition is satisfied, viz. the Fourier transform of 
$\fe$ be strictly positive. 
	However, notice that neither the surface density nor the 
volume density discussed in this paper satisfy the Wiener condition.
	In principle one should therefore be prepared for the 
possibility of periodic solutions in our massive LED. 
	On the other hand, rigorous studies of simpler semi-relativistic 
models of a particle interacting with a scalar wave field show that
time-periodic bound states of their linearized version are structurally 
unstable to the nonlinear  perturbation of the linear dynamics~\cite{kunze}; 
see also~\cite{sofferweinstein}.
	No such study has yet been carried out for our massive LED.

        \subsection{Scattering states}

	In our center of mass initial leaf, we can divide the 
scattering states into two categories:
(a) scattering of electromagnetic radiation off a non-moving particle,
and (b) scattering of radiation and moving particle in an encounter.  

        \subsubsection{Scattering with a non-moving particle}

	Scattering with a non-moving particle is considerably
simpler because the family of space slices of simultaneity for
the particle are just the standard foliation of spacetime, as 
for a stationary particle. 
	The only dynamically active degree of freedom of the particle 
is its Euler tensor, or in the stationary Lorentz frame of our
foliation, the angular velocity vector $\eulerV(t)$.
	The dynamical degrees of freedom of the electromagnetic field
are restricted to be compatible with the condition that the particle
sit still at the origin of space. 
	For instance, large classes of axisymmetric field decorations
of spacetime will be fine. 
	This case has been studied in by us in rigor; 
	details will be given in~\cite{appelkiessling}.
	Here we summarize the main results, obtained under the 
condition that the ratio of the particle's electrostatic Coulomb
energy to the bare rest mass is smaller than $1$.

	The best understood situation is when the particle's axis of
rotation  does not change during the evolution, say $\eulerV =\euler
\vect{e}_3$, so that $\euler$ is the only remaining dynamical particle 
degree of freedom. 
	For all admissible initial conditions corresponding to these
restrictions, we know that on families of nested compact sets the 
field-particle system converges exponentially fast to a stationary 
state which is a bound state if the conserved $s^2\neq 0$ initially
and the static state if $s^2 = 0$, superposed on which in 
either case is an outgoing field of electromagnetic radiation. 
	The evolution of the electromagnetic field is thus
given by the scattering formula
\begin{equation}
	\GV(\xV,t) 
\quad{\stackrel{t \to +\infty}{\longrightarrow }}\quad
	\GV_{st}(\xV) 
	+ e^{-it\nabla\times}\ \GV^{\text{out}}_{\text{rad}}(\xV) 
\, ,
\end{equation}
where $\GV^{\text{out}}_{\text{rad}}$ is a  divergence-free
electromagnetic field, orthogonal to all bound-state fields, uniquely
determined by the initial state, characterizing the outgoing radiation.
	Similarly
\begin{equation}
	\GV^*(\xV,t) 
\quad{\stackrel{ t \to -\infty}{\longrightarrow }}\quad
	\GV^*_{st}(\xV) 
	+ e^{it\nabla\times}\ \GV^{\text{in}}_{\text{rad}}{^*}(\xV) 
\end{equation}
where $^*$ denotes complex conjugate.
	We remark that $\GV_{st}$ is the {same} stationary
bound state in both formulas, which is a simple  illustration
of the soliton dynamics of the renormalized particle in this special 
situation.

	In the general non-moving particle case where the axis of
$\eulerV$ is not fixed during the evolution, we know that every 
admissible initial state compatible with the constraints evolves 
uniquely into the future (respectively, past), but we do not yet 
know that on families of nested compact sets the field-particle 
system converges (exponentially fast) to a stationary state. 
	However, we do conjecture that this will be the case.
	Assuming this to be true we were able to show that for large
times the evolution of the electromagnetic field satisfies 
the scattering formulas
\begin{equation}
	\GV(\xV,t) 
\quad{\stackrel{t \to +\infty}{\longrightarrow }}\quad
	\GV_{st}^{\text{out}}(\xV) 
	+ e^{-it\nabla\times}\ \GV^{\text{out}}_{\text{rad}}(\xV) 
\, ,
\end{equation}
and
\begin{equation}
	\GV^*(\xV,t) 
\quad{\stackrel{ t \to -\infty}{\longrightarrow }}\quad
	\GV_{st}^{\text{in}}{^*}(\xV) 
	+ e^{it\nabla\times}\ \GV_{\text{rad}}^{\text{in}}{^*}(\xV) 
\, ,
\end{equation}
where now $\GV_{st}^{\text{in}}$ and $\GV_{st}^{\text{out}}$ 
are generally \emph{not the same} stationary bound state;
however, they differ by at most a space rotation, and this 
once again illustrates the soliton dynamics of the renormalized particle.
	
	In both situations, the explicit characterization of the 
scattering operator from the ``in'' states to the ``out'' states
has not yet been worked out.
	
        \subsubsection{Scattering with a moving particle}

	Scattering with a moving particle can mean any scattering process
in which the particle moves for at least a finite interval of time.
	In that case we need to work with a nonstandard foliation,
as explained in the section on the Cauchy problem for the state. 
	Needless to say, so far this is the least explored case. 
	The explicit characterization of the scattering states, 
and of the scattering operator from the ``in'' states to the ``out'' 
states still have to be worked out. 

	However, although it is rigorously unknown which initial states
are scattering states for which the particle moves, \emph{a priori} 
speaking there are several types of scattering processes to be distinguished: 
	(A) the particle moves only a finite distance, associated with
which are bound states in the distant past and future --- this 
is a perturbation of the previous 
``scattering of radiation off a bound state;''
	(B) the particle is at rest only in the past (respectively,
only in the future) but moves freely in the asymptotic
future (respectively, past);
	(C) the particle moves freely in the asymptotic past \emph{and}
future.
	Furthermore, one can give the following partial 
characterization of the ``in'' and ``out'' states. 

	In the situations (A) and the parts of (B) associated with a 
stationary bound state, the scattering asymptotics differs from 
the one discussed in the previous subsection by at most a translation 
of the stationary bound state. 

	In the situations (C) and the parts of (B) not associated with 
a stationary bound state, the scattering asymptotics in the specified
foliation is given as follows. 
	As for the asymptotic particle motion, all the ``in'' states
$\{\zQ^{\text{in}}, \uQ^{\text{in}}, \EulerQ^{\text{in}}\}$, as well as
all the ``out'' states
$\{\zQ^{\text{out}}, \uQ^{\text{out}}, \EulerQ^{\text{out}}\}$,
are obtained from the stationary particle states considered above
by the action of an element of the Poincar\'e group, more explicitly a 
combination of a translation, a rotation, and a Lorentz boost.
	The free evolution of a particle is obtained by applying
the Lorentz-time propagation operator to these states, or expressed
in terms of proper-time $\tau$ (recall that $q^0(\tau)=t$),
\begin{equation}
	\zQ(\tau) 
\quad{\stackrel{ \tau \to +\infty}{\longrightarrow }}\quad
	\zQ^{\text{out}}
	+ \uQ^{\text{out}}\tau
\end{equation}
\begin{equation}
	\uQ(\tau) 
\quad{\stackrel{ \tau \to +\infty}{\longrightarrow }}\quad
	\uQ^{\text{out}}
\end{equation}
\begin{equation}
	\EulerQ(\tau) 
\quad{\stackrel{ \tau \to +\infty}{\longrightarrow }}\quad
	\EulerQ^{\text{out}}
\end{equation}
and correspondingly for $\tau\to -\infty$ using the ``in'' states.
	Associated with these free asymptotic evolution of the particle 
are the co-moving electromagnetic fields.
	Sufficiently far away from the particle these are just co-moving 
in the standard foliation to which the compatible foliation is asymptotically 
close except in a neighborhood of the world-tube.
	Their electromagnetic ``in'' and ``out'' states are parameterized
by the space part $\xV^{\text{out}}$ of $\zQ^{\text{out}}$, the constant 
velocity $\vV^{\text{out}}$ in 
$\uQ^{\text{out}}=(\gamma,\gamma\vV)^{\text{out}}$, and the 
angular velocity vector $\eulerV^{\text{out}}$ corresponding to
$\EulerQ^{\text{out}}$. 
	To have a handy abbreviation for this parameter set, we choose
the index $_{\text{com}}$ for `co-moving.'
	We set
\begin{align}
	\phi^{\text{out}}_{\text{com}}(\xV) 
& \defeg 
	\frac{1}{\abs{y_\sssp
	      \hat{\vV}+\sqrt{1-\vV^2}\yV_\perp}} +
	(1-\vV^2) 
	\frac{\vV\cdot(\vect{\mu}\times\yV)}
		{\abs{y_\sssp \hat{\vV}+\sqrt{1-\vV^2}\yV_\perp}^3} 
\\ 
	\AV^{\text{out}}_{\text{com}}(\xV) 
& \defeg
	  \frac{\vV}{\abs{y_\sssp
	      \hat{\vV}+\sqrt{1-\vV^2}\yV_\perp}} +
	(1-\vV^2)\frac{\vect{\mu}\times(\gamma
	y_\sssp
\hat{\vV}+\yV_\perp)}
  {\abs{y_\sssp
\hat{\vV}+\sqrt{1-\vV^2}\yV_\perp}^3} 
\\ 
  \intertext{with} \hat{\vV}& \defeg \vV/\abs{\vV}, \qquad
	y_\sssp
\defeg
	(\xV-\xV^{\text{out}})\cdot\hat{\vV},  
\qquad
	\yV_\perp  
=
	\xV - \xV^{\text{out}}- y_\sssp\hat{\vV}  
\, ,
\end{align}
and where $\vV = \vV^{\text{out}}$ and $\vect{\mu}= \vect{\mu}^{\text{out}}$.
	After these preparations, we can give the asymptotic future
evolution of the electromagnetic field for such a scattering state as
\begin{align}
	\GV(\xV,t) 
\quad{\stackrel{t \to +\infty}{\longrightarrow }}\quad
&	-\nabla \Big(
	e^{-t\vV^{\text{out}}\cdot\nabla}\phi_{\text{com}}^{\text{out}}(\xV) 
		\Big)
	-(\partial_t - i\nabla\times) 
		\Big(
	e^{-t\vV^{\text{out}}\cdot\nabla}\AV_{\text{com}}^{\text{out}}(\xV) 
		\Big)
\nonumber \\
&	+ e^{-it\nabla\times}\ \GV^{\text{out}}_{\text{rad}}(\xV) 
\end{align}
	Similar formulas hold for the asymptotic past evolution in
terms of the ``in'' potentials.

        \subsection{Soliton dynamics}

	While the rigorous treatment of the scattering problem of 
massive LED is still in its infancy, in all instances of scattering
a  remarkable conclusion about the gyrational particle degrees of 
freedom can be made. 
	Na\"{\i}vely one would expect that generally the orientation
\emph{and} the norm  of the angular velocity of the particle 
would change during the scattering process---and with it the rotational 
and renormalized mass of the particle, its spin and magnetic moment.
	However, precisely this does not happen.
	Any scattering process connects two 
freely evolving particle states with identical renormalized mass
and identical magnitudes of spin  and magnetic moment.
        The relativistic Lorentz electron, equipped with non-vanishing
positive bare inertia, thus has the  dynamical characteristics 
of an elementary particle in the best sense one could have hoped for
in a classical theory. 

        By adapting the terminology of Spohn~\cite{spohn} and 
collaborators, who considered charged particles without spin, we 
call the renormalized particle a \emph{spinning charge soliton.} 
	As remarkable as this soliton dynamics itself, is the fact
that it is a simple joint consequence of the conservation of 
$s^2 = - (1/2)\tr \SQ^\perp\cdot\SQ^\perp$ together with
the fact that $\SQ^\perp = \SQ$ for a stationary bound state, 
and the invertibility of the map $\EulerQ\mapsto \SQ$ for such a state.
	Indeed, we need only consider a stationary bound state, for we know
that the asymptotic future (past) evolution of the scattered 
particle dressed with its co-moving fields (the renormalized particle)
is just a translated and boosted such state.

	Explicitly, for a stationary bound state we have, in the
Lorentz frame and in dual vector notation,
\begin{equation}
	\sV
=  
        \int_{{\mathbb R}^3} 
        \frac{\xV\times(\eulerV\times\xV)}
             {\sqrt{1-|\eulerV\times\xV|^2}}
                {\fm}(|\xV|)\,\dd^3x 
+ 
	 \int_{\mathbb{R}^3} \xV \times \AV_{st} (\xV)
	f_e(|\xV|)\dd^3x
\, ,\label{eq:totalSPINvect}
\end{equation}
the first term being the bare spin, the second the electromagnetic
spin of the particle.
	For the bare spin $\sVb$ we already saw, in our section on the Cauchy
problem, that it is parallel to $\eulerV$ and the map $\sVb\mapsto \eulerV$
invertible.
	Using our formula \refeq{eq:statA} for the stationary vector
potential of a bound state, we see by explicit computation that
also the total spin $\sV$ is parallel to $\eulerV$, and  
the map $\sV\mapsto \eulerV$ is invertible.
	We have  $\sV= I(|\eulerV|)\eulerV$, with $I>0$
and map $x\mapsto xI(x)$ increasing,  vanishing for $x=0$. 
	Now, $\abs{\sV}=\abs{\eulerV}I(|\eulerV|)$, so that
from the invertibility of $x\mapsto xI(x)$ we conclude that
$|\sV|$ determines $|\eulerV|$ uniquely in a stationary bound state.
	Thus, from the constancy of $s^2$ we conclude that $\abs{\eulerV}$ 
in the asymptotic future state has the same value as in the asymptotic
past state. 
	The solitonic character of the renormalized particle is proved.

	\section{The limit of vanishing bare rest mass}

        We carry out the renormalization group analysis 
for a simple example where the various dynamical functions 
can be computed explicitly.
        Specifically, we here consider `densities' $\fm$ and $\fe$ 
which are Dirac measures concentrated  on the surface of a two-sphere
of radius $R$, viz.
\begin{equation} 
        \frac{1}{\mz} \fm(|\xV|) 
= 
        \frac{1}{-e}   \fe(|\xV|) 
=
        \frac{1}{4\pi R^2} \delta(|\xV| - R)
\, .\label{eq:SHELL}
\end{equation} 
	The elementary charge $e$  will not be tempered with and therefore 
not displayed as variable in the arguments of the functions below.
        Beside having set the charge $-e$ of the model particle equal 
to that of the real electron, we also demand that its total model mass 
$m$ matches the empirical electron mass, $m = m_{\text{e}}$, and 
that its magnetic moment $|\mu|$ matches the empirical electron
magnetic moment, $|\mu| = \mu_{\text{e}}$.

	\emph{Remark:} In this and the next section we resort to units 
in which the speed of light $c$ is displayed explicitly, to facilitate 
the comparison with the formulas used in physical data books.

        The renormalized particle spin will be {\it derived} in our 
model in the limit of vanishing bare rest mass, whence is \emph{not}
to be matched to a `corresponding empirical value,' but rather compared.
        Na\"{\i}vely speaking it is to be expected that in our 
classical theory with ultraviolet cutoff the spin--to--magnetic-moment 
relation 
\begin{equation}
|\mu| = g \frac{e}{2m_{\text{e}}c} s
\label{eq:mugspin}
\end{equation} 
with Land\'e factor $g \approx  g_{\text{\tiny Cl}} = 1$ should hold.
	Interestingly, we will find $g\approx 2/3$. 

        We remark that, without further elaboration,
there remains a certain ambiguity in matching our stationary 
model data to empirical electron data inferred  with 
the help of quantum mechanics from data of \emph{dynamical} experiments,
viz. $m_{\text{e}}$ in scattering experiments, and $ \mu_{\text{e}}$
in resonance frequency experiments in a Penning trap.
        The ultimate justification for these choices has to wait for
some future dynamical analysis of our model.

	\subsection{The limit $\mz\to 0^+$ for the stationary electron}

        By performing at most a space translation, we can assume that
the stationary particle, and its self-fields, are centered the 
at the origin of the space slice $\mathbb{R}^3$ 
of the Lorentz frame in which the system is stationary.
        Since the time-component of the spin four-vector and 
the angular velocity four-vector vanish in our stationary frame, 
after at most a space rotation we can furthermore assume that the 
spin four-vector of the stationary bare particle, $\sQb$, and of 
its self-fields, $\sQe$, points along the $\qv{e}_3$ direction. 
        Since $\qv{e}_3 = (0,\vect{e}_3)$, we can canonically 
identify these spin four-vectors with the familiar bare and field 
spin three-vectors $\sVb$, $\sVe$,  respectively the angular velocity
four-vector with the
angular velocity three-vector $\eulerV$, which moreover can 
be written as  $\sVb =  s_{\text{b}}\vect{e}_3$, 
$\sVe =  s_{\text{f}}\vect{e}_3$, and $\eulerV = \euler\vect{e}_3$, 
with $\euler >0$.
        With these conventions, and with \refeq{eq:SHELL}, 
the relevant formulae for the renormalization analysis are the following.

        Beside the electric monopole moment (i.e. the charge $-e$), 
the magnetic dipole moment of the empirical electron is the second 
electromagnetic characteristic measurable in a classical manner. 
        The magnetic moment three-vector $\muV$ given in 
\refeq{eq:magmom} (and re-installing a factor $1/c$) evaluates to 
\begin{align}
	\vect{\mu}
&= 
	-\frac{e}{2c} \frac{1}{4\pi R^2}  \int_{\mathbb{R}^3}
	\big(|\xV|^2 \mathrm{Id} - \xV\otimes\xV\big)\delta(|\xV| - R)
	\,\dd^3x\, \cdot \eulerV 
 \\
& =
 -\frac{1}{ 3} \frac{e}{ c} R^2 \euler\, \vect{e}_3 \, .
\label{eq:XXXII}
\end{align}
\label{eq:XXV}\\
        Since it points in the $\vect{e}_3$ direction as well,  
we will write $\vect{\mu}= \mu \vect{e}_3$, with $\mu$ 
read off from \refeq{eq:XXXII}.

        The  mass $m$ of the renormalized particle is defined by 
\begin{equation}
m{c^2} = W_{\text{b}} + W_{\text{f} }\, .
\label{eq:XXIX}
\end{equation}
	Here, $W_{\text{b}}$ is the gyrational energy of the bare particle,
defined earlier in \refeq{eq:gyroMASS} and evaluated for general
spherical $\fm$ in \refeq{eq:gyroMASSeval}. 
	Re-installing a factor $c^2$ and inserting now 
\refeq{eq:SHELL} for $\fm$ in \refeq{eq:gyroMASSeval}, we obtain
\begin{align}
	W_{\text{b}} 
&= 
	\mz c^2 \frac{1}{4\pi R^2} \int_{\mathbb{R}^3}
	 \frac{1}{ \sqrt{1- |\eulerV\times \xV|^2/c^2}} \delta(|\xV| - R)
	\dd^3x
\label{eq:XXIII}\\
&= 
     \mz c^2 \frac{c}{ \euler R} \Artanh \left( \frac{\euler R}{c}\right)
\, .\label{eq:XXVI}
\end{align}
	Furthermore, $W_{\text{f}}$ is the electromagnetic field energy 
of the gyrating charged particle, 
\begin{align}
	W_{\text{f}}
&= 
	\frac{1}{8\pi} \int_{\mathbb{R}^3}
	\Big(\big|\EV_{st}(\xV)\big|^2 + \big|\BV_{st}(\xV)\big|^2\Big)\dd^3x
\label{eq:XXVII}\\
&= 
	\frac{1}{2}\frac{e^2}{R}
	\left(1 + \frac{2}{9}\frac{\euler^2 R^2}{c^2}\right)
\label{eq:XXVIII} \, . 
\end{align}

        Our system of equations for the stationary renormalization flow 
is given by
\begin{align}
        m_{\text{e}}
&=      m(R,\euler,\mz) 
\, ,\label{eq:renormM}\\
        \mu_{\text{e}}
& =     |\mu|(R,\euler) 
\,,\label{eq:renormMU}
\end{align}
with the function $m(R,\euler,\mz)$ given through \refeq{eq:XXIX},
\refeq{eq:XXVI}, \refeq{eq:XXVIII}, and with the function
 $|\mu|(R,\euler)$ given through \refeq{eq:XXXII}.
        Explicitly, we have 
\begin{align}
        m_{\text{e}}
&=      \mz \frac{c}{ \euler R} \Artanh \left( \frac{\euler R}{c}\right)
        +
        \frac{1}{2}\frac{e^2}{c^2}\frac{1}{R}
        \left(1 + \frac{2}{9}\frac{\euler^2 R^2}{ c^2}\right)\, ,
\label{eq:renormMexpl}\\
        \mu_{\text{e}}
& =     \frac{1}{ 3} \frac{e}{ c} R^2 \euler\, .      
\label{eq:renormMUexpl}
\end{align}
        These two coupled equations determine a  curve 
in $(R,\euler,\mz)$-space for which we will study the limit $\mz\to 0$. 

        Equation \refeq{eq:renormMUexpl} can immediately be solved 
for $\euler$ in terms of the radius~$R$, 
\begin{equation*}
	\euler 
= 
	3\frac{\mu_{\text{e}}c}{e}\frac{1}{R^2}
\, .\label{eq:renormEuR}
\end{equation*}
	Inserted into \refeq{eq:renormMexpl}, this gives
\begin{equation}
        m_{\text{e}}  
= 
        \mz R \frac{1}{3} \frac{e}{\mu_{\text{e}}}
 \Artanh 
        \left( 3\frac{\mu_{\text{e}}}{e} \frac{1}{ R }\right)
+
        \frac{1}{2}\frac{e^2}{ c^2}\frac{1}{R}
\left(1+ {2}\frac{\mu^2_{\text{e}}}{e^2}\frac{1}{R^2}\right)
\, ,\label{eq:LXIV}
\end{equation}
which rewrites into a formula for the bare rest mass as function of $R$, 
\begin{equation}
        \mz 
=
        m_{\text{e}}
{3} \frac{\mu_{\text{e}}}{e}\frac{1}{R}
\frac{\displaystyle 1 - 
\frac{1}{2}\frac{e^2}{ m_{\text{e}} c^2}\frac{1}{R}
                \left(1 + 
{2} \frac{\mu_{\text{e}}^2}{e^2}\frac{1}{ R^2 }\right)}
{\displaystyle 
        \Artanh \left(
        {3} \frac{\mu_{\text{e}}}{e}\frac{1}{R}\right)}
\, .\label{eq:LXV}
\end{equation}
        We are interested in the inverse function $R(\mz)$.
        In particular, we want to know whether $\lim_{\mz\to 0} R$ 
can be taken without encountering superluminal equatorial rotation speeds.

        We readily read off from \refeq{eq:LXV} that the right side
is well-defined only on the $R$-interval 
$({3}{\mu_{\text{e}}}/{e},\infty)$.
        From \refeq{eq:renormMUexpl} and \refeq{eq:ComptonL} it
furthermore follows that $|\euler| R < c$ for  
$R\in ({3} {\mu_{\text{e}}}/{e},\infty)$, 
viz., no superluminal rotation speeds occur for $R$ in the
domain of definition for the r.h.s. of \refeq{eq:LXV}. 
        Furthermore, $\mz(R)$ is a monotonically increasing
function, so that the limit $R\searrow {3} {\mu_{\text{e}}}/{e}$
for $\mz$ will answer the question whether one can take the limit 
$\mz\to 0$ for $R$. 
        A simple inspection of \refeq{eq:LXV} reveals that 
$\mz\to m_{\text{e}}$ as $R\to \infty$ and that
$\mz\to 0$ as $R\to {3}{\mu_{\text{e}}}/{e}$.
        Hence, by monotonicity there is a unique limit of 
vanishing bare rest mass for the radius of the
renormalized particle, given by
\begin{equation}
        \lim_{\mz\to 0} R 
= 
        {3} \frac{\mu_{\text{e}}}{e} 
\, .\label{eq:Rlim}
\end{equation}
        It also follows from our discussion that the equatorial
rotation speed in this limit is precisely equal to the speed of light $c$. 
        
        The renormalization flow analysis reveals that 
the  Lorentz program of a  ``purely electromagnetic electron'' 
can be completed successfully, in the stationary setting at least.
	The resolution of the puzzlement is that beside the static 
electromagnetic field energy contributions considered by Lorentz,  
the mass of the properly renormalized purely electromagnetic particle 
consists of an additional contribution which we call \emph{photonic}, 
for it is the mass of an object without rest mass which rotates with  
equatorial speed equal to the speed of light. 
        This contribution was missed by Lorentz who calculated
without bare rest mass from the outset.

        It is instructive now to make use of the 
empirical relation for the electron magnetic moment, 
$\mu_{\text{e}}=(1+a) \mu_{\textsc{b}}$, where 
\begin{equation}
 \mu_{\textsc{b}}
 =\frac{1}{2}\frac{\hbar e}{ m_{\text{e}} c} 
\label{eq:BOHRmagneton}
\end{equation}
is the \emph{Bohr magneton} and $a = 0.001159652...$ the anomaly factor
(see, e.g., \cite{cottingham}). 
        Up to corrections of $O(a)$, equation \refeq{eq:Rlim} then 
reveals that the radius of the renormalized purely electromagnetic
electron is
\begin{equation}
        {3} \frac{\mu_{\text{e}}}{e} 
 = 
	\frac{3}{2} \RC 
\, ,  \label{eq:Rlim2}
\end{equation}
where
\begin{equation}
        \RC = \frac{\hbar}{m_{\text{e}} c}\, 
\label{eq:ComptonL}
\end{equation}
is the electron's \emph{Compton length}.
	This result is quite remarkable in many respects. 
        It shows that our Lorentz-covariant LED in fact contains 
the correct ultraviolet cutoff length where heuristic
discussions traditionally place the limit of applicability of a classical
field theory, viz. at the order of the electron's  Compton length $\RC$,
rather than at the (lamentably so-called) \emph{classical electron radius},
given by
\begin{equation}
R_{\text{\tiny Cl}} = \frac{ e^2 }{ m_{\text{e}} c^2} \, ,
\label{eq:Rclass}
\end{equation}
which gives the size of a `purely electrostatic classical electron.' 
        Recalling \emph{Sommerfeld's fine structure constant}, given by
\begin{equation}
\alpha = \frac{e^2}{ \hbar c} \approx \frac{1}{ 137.036}\, ,
\label{eq:SOMMERFELDalpha}
\end{equation}
we have the relation
\begin{equation}
R_{\text{\tiny Cl}} = \alpha \RC 
\end{equation}
so that the true Lorentz electron is actually about $(3/2\alpha)\approx 200$
times larger in diameter than predicted by a purely electrostatic calculation.

	We conclude this subsection with a discussion of the total spin, 
given by
\begin{equation}
	\sV 
=  
	\sVb +  \sVe 
\, .\label{eq:totSPINvect}
\end{equation}
        Here, $\sVb$ is the spin three-vector of the bare particle,
which for $\mz > 0$ evaluates to
\begin{align}
	\sVb 
& =     
	\mz  \frac{1}{4\pi R^2} \int_{\mathbb{R}^3}
        \frac{\xV\times(\eulerV\times\xV)}
             {\sqrt{1-|\eulerV\times\xV|^2/c^2}}
	\delta(|\xV| - R)\,\dd^3x 
\label{eq:inenB}\\ 
&=
	\mz c R\left(\frac{1+ c^2/\euler^2R^2}{2}
	\Artanh\left(\frac{\euler R}{c}\right)- \frac{c}{2\euler R}\right)
	\,\vect{e}_3
\, ,\label{eq:XXX}
\end{align}
which, by inserting \refeq{eq:renormMUexpl} into \refeq{eq:XXX}, 
with \refeq{eq:BOHRmagneton} for $\mu_{\text{e}}$, becomes
\begin{equation}
	s_{\text{b}} 
= 
	\hbar \,\frac{\mz}{ m_{\text{e}}} \frac{ R }{ \RC}
\left(\frac{1 + (2R/3\RC)^2}{2}
	\Artanh \left(\frac{3}{2}\frac{\RC}{ R }\right)
	- \frac{1}{3}\frac{R}{\RC}
\right)
\,.
\end{equation}
	Using now \refeq{eq:LXV}, we get
\begin{equation}
        s_{\text{b}} 
= 
      	\frac{3}{2}\hbar \,\frac{1 +  (2R/3\RC)^2}{2}
        \left( 1 - \frac{1}{ 2}\frac{R_{Cl}}{ R}
                \left(1 + \frac{1}{ 2} \frac{\RC^2}{ R^2 }\right)
        \right) - \hbar\frac{1}{3}\frac{R^2}{\RC^2}\frac{\mz}{m_{\text{e}}}
\, .\label{eq:spinBofR}
\end{equation}
	The second spin term, $\sVe$, is the stationary field spin 
three-vector associated with the rotating charged particle, which 
evaluates to
\begin{align}
	\sVe 
&= 
	-\frac{e}{c} \frac{1}{4\pi R^2} 
	\int_{\mathbb{R}^3}
		\xV\times\AV_{st}(\xV)\delta(|\xV| - R)
	\,\dd^3x 
\label{eq:XXIVa}\\
&=
	\frac{1}{4\pi c} \int_{\mathbb{R}^3} \xV \times 
                \big(\EV_{st}(\xV) \times \BV_{st}(\xV)\big)\dd^3x
\label{eq:XXIVb}\\
& = 
	\frac{2}{9} \frac{e^2 }{ c^2} \euler R \, \vect{e}_3
\, .\label{eq:XXXI}
\end{align}
        We write $\sV = (s_{\text{b}} + s_{\text{f}} ) \vect{e}_3$, with
$s_{\text{b}}$ and $s_{\text{f}}$ given in \refeq{eq:spinBofR} 
and \refeq{eq:XXXI}.

	We consider the spin coefficients $s_{\text{b}}$ and $s_{\text{f}}$ 
as function of $R$.
        Letting $R\to 1.5 \RC$, i.e. taking $\mz\to 0$,  for the
bare spin we get,
\begin{equation}
        \lim_{\mz\to 0} s_{\text{b}} 
= 
	\frac{3}{2} \hbar \, \left( 1 - \frac{11}{27} \alpha \right)
\, ,\label{eq:SPINbLIM}
\end{equation}
while for the field spin, recalling that $\euler R\to c$
as $\mz\to 0$, we immediately obtain from \refeq{eq:XXXI} that
\begin{equation}
        \lim_{\mz\to 0} s_{\text{f}}  
= 
         \frac{2}{9}\frac{e^2}{c}
=
         \frac{2}{9}\alpha\hbar
\,. \label{eq:SPINfLIM}
\end{equation}  
        Finally, adding $s_{\text{f}}$ and $s_{\text{b}}$ and 
taking the limit equals the sum of the individual limits, whence
the renormalized limit spin magnitude is given by
\begin{equation}
        s_{\text{\tiny ren}}
\defeg  
        \lim_{\mz\to 0}s
= 
        \frac{3}{2}\hbar \, \left(1 - \frac{7}{27}\alpha\right)
\, .\label{eq:SPINlim}
\end{equation}  
	Put differently, from \refeq{eq:SPINlim} we read off that 
the spin magnitude $s_{\text{\tiny ren}}$ of the renormalized electron
satisfies
\begin{equation}
        s_{\text{\tiny ren}} \approx \frac{3}{2}\hbar  
\label{eq:spinLIMhbar}
\end{equation}  
to high precision, with relative corrections of order $\alpha$,
compared to our na\"{\i}ve expectation $s\approx \hbar$, and
compared to $s\approx\hbar/2$ in a quantum mechanical framework.

	Curiously, furthermore, a closer look at the corrections reveals 
that in our LED an expansion of the $g$-factor in the Land\'e relation 
\refeq{eq:mugspin} in terms of powers of $\alpha$ holds.
	We set $g_0 \defeg 2/3$, then 
\begin{equation}
        g_{\textsc{led}}
 = 
	g_0
        \left( 1 +
                \frac{7}{27}\alpha + 
                \left(\frac{7}{27}\right)^2 \alpha^2+
                O\left(\alpha^3\right)
        \right)
\label{eq:gLED}
\end{equation}  
which is surprisingly similar to the familiar expansion in 
QED~\cite{cottingham, glimmjaffeBOOK},
\begin{equation}
        g_{\textsc{qed}}
 = 
        g_{\textsc{qm}}
        \left( 1 +
                \frac{1}{2\pi} \alpha -
        \frac{0.328...}{\pi^2}\alpha^2 +
                O\left(\alpha^3\right)
        \right)
\,.\label{eq:gQED}
\end{equation}  

		\subsection{The limit $\mz\to 0^+$ for the dynamical electron}

	We finally come the dynamical renormalization problem.
	In contrast to the stationary problem, no rigorous results 
concerning the existence of this limit are available yet.
	However, it is of interest to inquire into the features 
expected for a renormalized dynamical limit of vanishing 
bare rest mass of massive LED. 

	First of all, the matching of empirical electron data in 
the dynamical renormalization problem  with scattering states has to
be carried out in the asymptotic past (or future), for we want to 
``send in'' the particle with the correct physical characteristic 
particle data.
	The incoming-data matching in the asymptotic past will fix 
$R$ and $\mz$  for the rest of the motion, since as parameters of 
the model these do not vary during the motion. 

	This matching in the asymptotic past will also fix the 
value of $|\eulerV|$ in the asymptotic future, which is due to the 
soliton dynamics of the renormalized particle. 
	During the motion, of course, $|\eulerV|$ will in general vary.
	However, our results on the scattering with a non-moving 
particle indicate that $|\eulerV|$ approaches its asymptotic value
extremely fast. 
	Extrapolating from this observation, we expect that  
the soliton dynamics of the particle in massive LED in fact will
try to keep the angular frequency $|\eulerV|$ of the particle
as close as possible to the asymptotic value in the free motions.
	By the same token the gyrational mass (hence, also the renormalized
mass, which is a small perturbation of the former in the stationary state) 
of the particle should remain (in the instantaneous rest frame)
very close to the empirical mass during the motion.

	An important conclusion can then be drawn if one
can show that the gyrational bare mass of the particle with 
$\mz >0$ stays \emph{uniformly close} to the empirical mass
as $\mz\to 0$ for an $\mz$-indexed sequence of motions.
	Namely, in that case the equatorial gyration speed of
the renormalized purely electromagnetic particle must be 
locked at the speed of light during the whole motion!
	For assuming to the contrary that the gyrational speed would
remain bounded away from the speed of light for at least a finite
fraction of the evolution (of course still close to it in the
uniform sense stipulated), it would then
follow that the bare gyrational mass converges
to zero, while the electromagnetic mass would  contribute at best an
amount of order $\alpha$ of the empirical mass.
	This is in contradiction to its hypothesized uniform proximity 
to the empirical mass. 

	We infer that in the limit  $\mz \to 0$ the bare gyrational mass 
$\rotM$ is independent of $\eulerV$, hence a \emph{constant of the motion}
identical to what we have called the {photonic mass} of the stationary
particle; explicitly,
\begin{equation}
	m_{\text{ph}} 
= 
	m_{\text{e}}\left(1 - \frac{11}{27} \alpha\right)
\, .
\end{equation}
        By the same reasoning,  in the limit of vanishing
bare rest mass the bare spin vector becomes the  \emph{photonic spin},
a four-vector with only two independent variables. 
        It's dual tensor is given by
\begin{equation}
        \SQ_{\text{ph}}  
= 
        \frac{3}{2}\hbar \, \left(1 - \frac{11}{27}\alpha\right)\SigmaQ
\, ,\label{eq:SPINph}
\end{equation}
where $\SigmaQ$ is anti-symmetric, of space-space type, has norm 1, 
satisfying $\SigmaQ\cdot \uQ =0$.
        The inertial gyro-frequency $\EulerQ$ and the spin tensor
 $\SigmaQ$ are related by a simple proportionality,
\begin{equation}
        \EulerQ 
= 
         \frac{m_e c^2}{ \hbar}\SigmaQ
 \, ,\label{eq:SPINpropOMEGA}
\end{equation}
emphasizing the characteristic Euler angular frequency $ m_e c^2/ \hbar$
explicitly.
	Similarly, $\EulerQ = \kappa\SQ_{\text{ph}}$, with 
\begin{equation}
	\kappa 
=        
	\frac{m_e c^2}{\hbar^2}\frac{1}{1 - \frac{11}{27}\alpha}
\, .
\end{equation}

	After these preparations, we can write down the 
putative dynamical equations of renormalized LED. 
        The world-line equation of renormalized LED reads
\begin{equation}
	\frac{\dd }{\dd \tau}\pQ
 = 
	  \int_{{\mathbb R}^{1,3}} \FQ(\zQ +\xQ)\cdot
	  \Big(\uQ - (\ThomasQ + \kappa \SQ_{\text{ph}}) \cdot\xQ\Big)
	  \fe(\norm{\xQ})\,\delta(\uQ\cdot\xQ)\,\dd^4\xQ 
\,, \label{eq:worldEQnull}
\end{equation}
where
\begin{equation}
	\pQ(\tau) 
 =
	\tenseur{M}_{\text{ren}}(\tau) \cdot \uQ(\tau) 
\end{equation}
is the {energy-momentum} four-vector of the renormalized purely
electromagnetic electron, with $\tenseur{M}_{\text{ren}}$ given by
\begin{equation}
	\tenseur{M}_{\text{ren}}
=
	  m_{\text{ph}}\, \gQ   - \int_{{\mathbb R}^{1,3}}
	  \left[\xQ\otimes\xQ, \left[
	      \FQ(\zQ +\xQ), \kappa\SQ_{\text{ph}} \right]_+ \right]_+
	  \fe\big(\norm{\xQ}\big)\, \delta(\uQ\cdot \xQ)\,\dd^4\xQ
\label{eq:sMmtnull}
\end{equation}
now the renormalization limit of the tensor mass $\Mmink$.
        The gyrograph equation of renormalized LED reads
\begin{align}
	\dds {\SQ_{\text{ph}}}
&  +\, 
	\big[\ThomasQ ,\SQ_{\text{ph}} \big]_- 
= \nonumber \\
&
	\int_{{\mathbb R}^{1,3}} 
	  \xQ\wedge\bigl(\gQ + \uQ\otimes\uQ\bigr)\cdot \FQ(\zQ +\xQ)\cdot
	\Big(\uQ - (\ThomasQ + \kappa \SQ_{\text{ph}})\cdot\xQ\Big)
	 \fe(\norm{\xQ})\,\delta(\uQ\cdot\xQ)\,\dd^4\xQ
\, .  \label{eq:spinEQnull}
\end{align}
	The Maxwell--Lorentz equations for renormalized LED 
have their familiar structure, but the charge-current density 
is now given by 
\begin{equation}
        \JQ(\xQ)
=
        \int_{-\infty}^{+\infty}
	 \Big(\uQ - (\ThomasQ + \kappa\SQ_{\text{ph}})\cdot (\xQ-\zQ)\Big)
        \;{\fe}\big(\norm{\xQ-\zQ}\big)\,\delta\big(\uQ\cdot (\xQ-\zQ)\big)
	\,\dd \tau
\,.\label{eq:currentdensBare}
\end{equation}

        The state in renormalized LED differs from that of massive LED 
in that only the orientation but not the norm of $\SQ_{\text{ph}}$ is
a dynamical degree of freedom. 
	This imposes some consistency conditions also on the initial
conditions.
	We remark that \refeq{eq:spinEQnull} does \emph{not} automatically 
preserve $\norm{\SQ_{\text{ph}}}$ for an arbitrary electromagnetic 
tensor field $\FQ$ and world-line $\tau\mapsto\zQ$. 
	However, it is only mandatory that the norm of $\SQ_{\text{ph}}$ 
be conserved for consistent joint solutions of the world-line, gyrograph, 
and field equations; the situation is similar to the one related to the
conservation of $\norm{\uQ}^2$ by the world-line equation.

        \section{Summary and Outlook}

	In the previous sections we have constructed the first 
viable relativistic Lorentz electrodynamics with positive 
bare rest mass (massive LED). 
	We studied its renormalization flow limit to 
vanishing bare rest mass (renormalized LED).
	This  modern completion of Lorentz' program 
of a purely electromagnetic theory avoids the superluminal 
pitfalls of Lorentz' purely electromagnetic calculations.
	In the limit we obtain a particle with correct empirical
mass, charge and magnetic moment, though with spin $3\hbar/2$.
	Although the bare rest mass vanishes, the particle retains
a ``photonic mass'' resulting from luminal equatorial motions, which
actually accounts for most of the particle's mass while the
traditional electromagnetic self-energies contribute merely a 
correction of order $\alpha$.
	The size of the limit particle is of the order of the
electron's Compton length, about $200$  classical electron radii.

	The finite size of this ultraviolet cutoff in the 
`purely electromagnetic' limit endows our LED with a mildly 
\emph{a-causal} appearance.
	However, the ultraviolet cutoff is itself
an individual but \emph{non-local event} which cannot be 
subdivided into a collection of individual point events.
	The cutoff moves as a non-local unit in a Lorentz-covariant 
way, in fact accelerated in a \emph{causal} manner --- in the sense that 
fields outside the cutoff event do not influence its motion --- 
by the field integrated over the cutoff at any given instant of 
its proper-time.  
        It is perhaps surprising that such an instantaneous,
non-local action is compatible with Lorentz covariance and causality, 
yet it is.

         The  existence of this physically viable properly 
renormalized LED  reinforces some hopes that LED may ultimately 
have some bearing on the quest for a consistent QED, with LED or some
improved version thereof (LED$^+$ say) as its (semi-)classical limit.
	Recall that the perturbative series for QED~\cite{dysonA}
is most likely divergent~\cite{dysonB} and hence does not give a
mathematically acceptable definition of QED.
        Many experts nowadays are inclined to  believe that the constructive 
quantum field theory approach to QED~\cite{glimmjaffeBOOK} will not lead to 
a mathematically consistent theory either~\cite{balaban},
though the failure of this approach to QED has not yet been conclusively 
demonstrated~\cite{wightman}.
        There is also a strong anticipation that a physically correct
and mathematically well-defined relativistic quantum theory must
involve also the strong and weak nuclear interactions, as well as 
gravity, and all this supersymmetrically~\cite{wilczek, witten, connes}.
        It should however not be necessary to unify physics at the 
Planck scale in order to obtain a consistent working theory for
electromagnetic phenomena above the electron's Compton scale.
	Inspired by Kramer's program of `non-relativistic 
QED'~\cite{schweberBOOK, griesemerliebloss}, which aims at the 
quantization of LED's semi-relativistic predecessor (see Appendix A.3), 
and which is expected to describe the low energy limit of 
QED~\cite{griesemerliebloss}, one may want to investigate the possibility 
of deforming our LED into a `quantized LED', perhaps along the ideas of 
Kontsevich~\cite{kontsevich}, and it would be interesting to see
whether that would teach us something about QED proper.

	Such fundamental hopes aside, our LED is interesting in itself 
as a nonlinear  relativistic microscopic theory which couples particle 
and field degrees  of freedom in  a self-consistent manner. 
	While the rigorous control of its semi-relativistic
predecessor has made significant advances in recent years due
to the efforts of Spohn and collaborators and others (discussed in the
Appendix), the extension of this rigorous control to our 
relativistic LED has just started~\cite{appelkiessling}. 
	The most surprising new dynamical challenge is the need to 
construct a compatible foliation of spacetime along the evolution
of the state in LED by also solving the vacuum Einstein equations as 
a free-boundary problem coupled to the evolution of the state. 
	Unexpectedly, massive and renormalized LED may therefore 
serve as playground for the harder problem of black hole dynamics
in general relativity.
	
	Among the future projects, one is certainly the rigorous 
extraction of effective equations of motion for the particle in 
the adiabatic regime, where to lowest order one expects equations of 
the type discussed in~\cite{nyborg, benguria}, with radiation 
reaction-corrections showing in the next order of approximation.

	Another interesting project is the  many particle problem, and in 
particular the rigorous microscopic foundation of relativistic Vlasov--Maxwell
theory~\cite{glasseyschaeffer, glasseystraussA, glasseystraussB, vlasovBOOK},
which so far could not even be attempted for lack of a consistent 
microscopic theory of electromagnetic particles and fields.

\appendix
                \section{Appendices}

        Appendix A.1 contains the more technical derivations of the
Euler-Lagrange equations for our massive LED, Appendix A.2 discusses
the singular point mass limit in which our massive LED reduces to Nodvik's
massive LED, and Appendix A.3 provides supplementary material about the 
semi-relativistic LED.

     \subsection{The Euler-Lagrange equations of massive LED}
     \label{sec:appendixELeqs}

        In this subsection, we first collect various formulas needed to
carry out the variations of the action functional.
        Next we apply these formulas to derive the Euler-Lagrange
equations for the action principle.

	\subsubsection{Independent variation of the world-line}

To a given world-line $\tau\mapsto \zQ(\tau)$ we may add a small
perturbation $\tau\mapsto \delta\zQ(\tau)$ to get a deformed particle
world-line $\check\zQ(\tau) = (\zQ + \delta \zQ)(\tau)$.  
        Here, $\tau$ is the proper-time of the unperturbed world-line.  
        The proper-time along the
perturbed world-line at the point $\check\zQ(\tau)$ will in general
\emph{not} coincide with the value $\tau$, i.e. the proper-time of the
unperturbed world-line.  To first order, the difference in the proper-times
along the deformed and original world-line reads
\begin{equation}
        \delta (\dd \tau) 
        =       
        - \uQ\,\cdot \qdot {\delta \zQ}\;\dd\tau \, .
\label{eq:deltadtau}
\end{equation}
	Accordingly, the derivative of ${\delta \zQ}$ with respect to
$\tau$, viz. $\qdot {\delta \zQ}$, is generally \emph{not} the first-order 
variation in the four-velocity, $\delta\uQ$, which is defined by 
taking the derivative of $\check\zQ$ w.r.t. its proper-time, then 
computing the difference $\check\uQ - \uQ$ to first order.  
	As a function of $\tau$, the first-order
variation in the four-velocity is given by the space projection (w.r.t.
$\uQ$) of $\qdot {\delta \zQ}$,
\begin{equation}
        \delta \uQ
        =
        (\gQ + \uQ\otimes\uQ )\,\cdot \qdot {\delta \zQ}\, ,
\label{eq:deltaU}
\end{equation}
where $\gQ$ is the metric tensor of Minkowski space, defined in
\refeq{eq:gT}. 
	Taking the inner product of \refeq{eq:deltaU} with $\uQ$,
and recalling that $\gQ$ acts as identity on four-vectors, we find that
$\uQ\cdot\delta\uQ = 0$, as required by
$(\uQ+\delta\uQ)\cdot(\uQ+\delta\uQ) =-1$ in first order.  
	The variation of $\ThomasQ$ under an independent change 
$\delta \zQ$, the gyrograph being kept fixed, is given by~\cite{nodvik} :
\begin{equation}
        \delta \ThomasQ 
        =      
        - \uQ\wedge\,{\qdot{\delta\zQ}}
        + \ThomasQ\, \uQ\,\cdot {\qdot{\delta\zQ}}\, .
\label{eq:varthom1}
\end{equation}
	Defining quasi-coordinates
\begin{equation}
        \delta\varthetaQ 
        \defeg  
        - \uQ\,\wedge\delta \uQ\, ,
\end{equation}
equation~\refeq{eq:varthom1} can be rewritten into
\begin{equation}
        \delta\ThomasQ
        =       
        \dds \delta\varthetaQ + [\ThomasQ,\delta\varthetaQ]_-
        + \ThomasQ \uQ\,\cdot\qdot{\delta \zQ}\, ,
\end{equation}
which proves convenient later on. 

        Moreover, although the gyrograph is fixed during 
an independent variation of the  world-line, the Fermi--Walker frame changes,
and with it the tensor of the Euler gyration, for the gyrograph
 is defined w.r.t. the Fermi--Walker frame.
        The variation in $\EulerQ$ under an independent change $\delta \zQ$
is given by~\cite{nodvik}
\begin{equation}
        \delta\EulerQ 
        =       
        - \uQ\wedge\,(\EulerQ\,\cdot\delta\uQ) 
        + \EulerQ\, \uQ\,\cdot {\qdot{\delta\zQ}} \, ,
\label{eq:vareuler1}
\end{equation}
 whence 
\begin{equation} 
        \delta\OmegaQ 
        =
        - \uQ\wedge\,(\OmegaQ\,\cdot\delta\uQ) 
        - \uQ\wedge\,{\qdot{\delta\zQ}}
        + \OmegaQ\, \uQ\, \cdot {\qdot{\delta\zQ}}\, .
\label{eq:varOm1}
\end{equation}

 \subsubsection{Independent variation of the gyrograph}

	The orientation of the set of three spacelike vectors 
$\{\tilde{\eQ}_1,\tilde{\eQ}_2,\tilde{\eQ}_3\}$ relative to the 
spacelike vectors $\{\bar{\eQ}_1,\bar{\eQ}_2,\bar{\eQ}_3\}$
can be specified by Euler angles $\theta$, $\phi$, $\psi$.
	Independent changes $\delta\theta$, $\delta\phi$, $\delta\psi$ of
the particle's history of Euler angles, i.e. keeping its 
world-line fixed, give rise to the variation~\cite{nodvik}
\begin{equation}
        \delta\EulerQ 
        =
        \qdot{\delta\thetaQ} + [\OmegaQ ,\delta\thetaQ ]_- \, ,
\label{eq:varEUL2}
\end{equation}
where the anti-symmetric tensor $\delta\thetaQ$ is given by 
\begin{equation}
        \delta\thetaQ 
   =
        \sum_{0\leq \mu,\nu\leq 3}
        \delta\bar\theta^{\mu\nu}\, {\bar\eQ}_{\mu}\otimes
	{\bar\eQ}_{\nu}\, ,
\end{equation}
with $\{\bar\eQ_\mu\}$ the tetrad of the frame~$\FeWaF$ 
and $\delta\bar\theta_\mu{}^\nu$ given by
\begin{equation}
  \left\{
     \begin{array}{rll}
        \delta\bar\theta_1{}^2 & = \delta\phi +\cos\theta\;\delta\psi \\
        \delta\bar\theta_2{}^3 & = \cos\phi\;\delta\theta +
                \sin\theta\sin\phi\;\delta\psi \\
        \delta\bar\theta_3{}^1 & = \sin\phi\;\delta\theta - 
                \sin\theta\cos\phi\;\delta\psi \\
        \delta\bar\theta_0{}^i & = 0 & \text{for}\  i=1,2,3\, ,
     \end{array}
  \right.
\end{equation}
 cf. \cite{nodvik} for details. 

     \subsubsection{Euler-Lagrange equations for the world-line}

	We first turn to the action of the bare particle, which we rewrite as 
\begin{equation}
        \Ab
        =
        \int_{\tau_1}^{\tau_2} \Lb(\EulerQ,\uQ)\,\dd \tau\, ,
\end{equation}
where
\begin{equation}
        \Lb(\EulerQ,\uQ)
        \defeg  
        \, -\int_{\Xiq} \sqrt{1-\norm{\EulerQ\cdot\xQ}^2}
        \; {\fm}\big(\norm{\xQ}\big)\,
        \delta\big(\uQ\cdot \xQ\big)\,\dd^4\xQ
\label{eq:lagrangiantwo}
\end{equation}
is the Lagrangian for the bare particle.  
	We first define the volume $\Xi$ bounded by two space slices defined 
by $\uQ(\tau)\cdot\big(\xQ-\zQ(\tau)\big)=0$ at $\tau=\tau_1$ and
$\tau=\tau_2$. 
	Its translation by $\zQ$ is denoted by
$\Xiq=\{\xQ-\zQ(\tau)\;;\;\xQ\in\Xi\}$.  
	Under an independent variation $\zQ \mapsto \zQ +\delta \zQ$ 
the resulting first variation of $\Ab$ reads
\begin{equation}
        \delta \Ab 
        =
        \int_{\tau_1}^{\tau_2} 
        \left(
          \frac12\tr         
          \left(
            \frac{\delta\Lb }{\delta \EulerQ}\cdot\delta\EulerQ
          \right) 
          + \frac{\delta\Lb }{\delta \uQ}\cdot\delta\uQ\, 
          - \Lb\uQ\,\cdot\qdot{\delta\zQ}
        \right)
        \,\dd \tau \, ,
\end{equation}
where the last term under the integral results from the change in proper
time along the perturbed world-line, see equation~\refeq{eq:deltadtau}. 
	As for the variation of $\Lb$ w.r.t. $\uQ$, we get 
\begin{alignat}{1}
        \frac{\delta\Lb }{\delta \uQ}
        =
        &\, -\int_{\Xiq} \sqrt{1-\norm{\EulerQ\cdot\xQ}^2}
        \; {\fm}\big(\norm{\xQ}\big)\,
        {\partial_\uQ}\,
        \delta\big(\uQ\cdot \xQ\big)\,\dd^4\xQ
        \\
        =
        &\, -   \uQ\int_{\Xiq} \sqrt{1-\norm{\EulerQ\cdot\xQ}^2}
        \; {\fm}\big(\norm{\xQ}\big)
        \,\delta\big(\uQ\cdot \xQ\big)
        \,\dd^4\xQ,
\label{eq:deltaLbu}
\end{alignat}
where we used 
\begin{equation}
        {\partial_\uQ}\, \delta(\uQ\cdot\xQ) 
        =
        - \xQ(\uQ\cdot{\nabQ})\, \delta(\uQ\cdot\xQ) \, ,
\label{eq:dudelta}
\end{equation} 
then integrated by parts, carried out the differentiations, 
and noticed that terms proportional to $\uQ\cdot\xQ$ in the 
integrand vanish upon integration against $\delta(\uQ\cdot \xQ)$. 
	But with the help of the four-orthogonality of $\uQ$ and 
$\delta\uQ$ we see that
\begin{equation}
        \frac{\delta\Lb }{\delta \uQ}\cdot\delta\uQ
        =
        0\, .
\end{equation}
	As for the variation of $\Lb$ w.r.t. $\EulerQ$, we find that
\begin{equation}
        \frac{\delta \Lb}{\delta \EulerQ} 
        =
        \SQb\, ,
\end{equation}
 where $\SQb$ is the spin tensor of the bare particle, 
 defined earlier in \refeq{eq:barespinTENSOR}.
Thus, we have
\begin{equation}
        \delta \Ab 
        =
        \int_{\tau_1}^{\tau_2} 
        \left(
          \frac12\tr
          \left(
            \SQb\cdot\delta\EulerQ
          \right) 
          - \Lb\uQ\,\cdot\qdot{\delta\zQ}
        \right)
        \,\dd \tau \, .
\end{equation}
	Inserting now \refeq{eq:vareuler1} for $\delta\EulerQ$, with
\refeq{eq:deltaU} in place for $\delta\uQ$, then using $\EulerQ\cdot\uQ
= \qv{0}$ once, next integrating by parts terms containing
$\qdot{\delta\zQ}$, recalling that $\delta \zQ =0$ at the boundaries of
integration, then using again $\EulerQ\cdot\uQ = \qv{0}$ and also
noticing once again that terms proportional to $\uQ\cdot\xQ$ in the
integrand vanish upon integration against $\delta(\uQ\cdot \xQ)$, 
we obtain
\begin{equation}
         \delta\Ab
         =
         -\int_{\tau_1}^{\tau_2}
         \qdot{\pQ}_{\text{b}}(\tau)\cdot\delta \zQ \;\dd \tau \, .
         \label{eq:bareVAR}
\end{equation}
	Here, $\pQ_{\text{b}}$ is the Minkowski momentum four-vector of
the bare particle, which for later convenience we write in the form
\begin{equation}
        \pQ_{\text{b}}
        =
        \tenseur{M}_{\text{b}}\cdot\uQ 
\, ,\label{eq:baremom}
\end{equation}
where 
\begin{equation}
        \Mbare
        =
	{\cal M}_{\text{b}}\, \gQ
\label{eq:bareMtensor}
\end{equation}
is the diagonal bare mass tensor, with 
\begin{equation}
	{\cal M}_{\text{b}}(\norm{\eulerQ})
=
	\int_{\Xiq} 
        \frac{1}{\sqrt{\displaystyle 1-\norm{\EulerQ\cdot\xQ}^2}}
        \, \fm\big(\norm{\xQ}\big)\delta(\uQ\cdot \xQ)
        \dd^4\xQ
\label{eq:bareM}
\end{equation}
the gyrational bare mass of the particle.

	Turning next to the action of field-particle coupling, we rewrite it as
\begin{equation}
        \Abf 
        =
        \int_{\tau_1}^{\tau_2} \Lc(\EulerQ,\zQ,\uQ,\qdot\uQ)\,\dd \tau\, ,
\end{equation}
where
\begin{equation}
        \Lc(\EulerQ,\zQ,\uQ,\qdot\uQ)
        = 
        \int_{\Xiq}
         \big( \uQ-\OmegaQ\cdot
        \xQ\big)\cdot \AQ(\zQ+\xQ)  \fe\big(\norm{\xQ}\big)
        \delta\big(\uQ\cdot\xQ\big)\,\dd^4\xQ 
\end{equation}
is the Lagrangian for the coupling of the particle to the electromagnetic 
fields.
	Since $\EulerQ$ and $\qdot\uQ$ occur in $\Lc$ only in the
combination $\EulerQ\, +\qdot\uQ\wedge\,\uQ = \OmegaQ$, we 
can treat $\Lc$ as a function of $\OmegaQ,\zQ,\uQ$. 
	Thus, abusing notation a little bit, we write 
$\Lc(\OmegaQ,\zQ,\uQ)$ and obtain
\begin{equation}
        \delta\Abf 
   = 
        \int_{\tau_1}^{\tau_2} 
        \Bigl(
        \frac12 \tr 
        \Bigl(
        \frac{\delta \Lc}{\delta \OmegaQ} \cdot \delta\OmegaQ
        \Bigr)
     +  \frac{\delta \Lc}{\delta \zQ} \cdot \delta\zQ 
     +  \frac{\delta \Lc}{\delta \uQ} \cdot \delta\uQ 
     -  \Lc \uQ\,\cdot \qdot {\delta \zQ}
        \Bigr) 
        \dd\tau 
\end{equation}
under independent variation of the world-line.
	As for the variation of $\Lc$ w.r.t. $\uQ$, we have
\begin{alignat}{1}
        \frac{\delta \Lc}{\delta \uQ} 
        =
        & \int_{\Xiq} \AQ(\zQ+\xQ)
        \fe\big(\norm{\xQ}\big)\,
        \delta(\uQ\cdot\xQ) \,\dd^4\xQ 
        \nonumber
        \\
        & + \int_{\Xiq} (\uQ-\OmegaQ\cdot\xQ\big)\cdot\AQ(\zQ+\xQ)
        \fe\big(\norm{\xQ}\big)\,
        {\partial_\uQ}\, \delta(\uQ\cdot\xQ)\,\dd^4\xQ.
\end{alignat}
	Using once again \refeq{eq:dudelta}, then integrating by parts, 
carrying out the differentiations,  and noticing once again that terms 
proportional to $\uQ\cdot\xQ$ in the integrand vanish upon integration 
against $\delta(\uQ\cdot \xQ)$, we obtain
\begin{alignat}{1}
        \frac{\delta \Lc}{\delta \uQ} 
        =
        &       \int_{\Xiq} \AQ(\zQ+\xQ)\,\delta(\uQ\cdot\xQ) 
        \fe\big(\norm{\xQ}\big)\,\dd^4\xQ 
        \nonumber
        \\
        &       +\uQ \int_{\Xiq}
        (\uQ-\OmegaQ\cdot\xQ\big)\cdot\AQ(\zQ+\xQ)\fe\big(\norm{\xQ}\big)\,
        \delta(\uQ\cdot\xQ) \,\dd^4\xQ
        \nonumber
        \\
        &       +\int_{\Xiq}\xQ
        \qdot\uQ\cdot\AQ(\zQ+\xQ)\fe\big(\norm{\xQ}\big)\,
        \delta(\uQ\cdot\xQ) \,\dd^4\xQ
        \nonumber
        \\
        &       +\int_{\Xiq}
        \xQ(\uQ-\OmegaQ\cdot\xQ\big)\cdot
        \bigl(\uQ\cdot{\nabQ}\AQ(\zQ+\xQ)\bigr)
        \fe\big(\norm{\xQ}\big)\,
        \delta(\uQ\cdot\xQ) \,\dd^4\xQ.
\end{alignat}
Using the four-orthogonality of $\uQ$ and $\delta\uQ$, we thus have
\begin{alignat}{1}
        \frac{\delta \Lc}{\delta \uQ} \cdot \delta\uQ
        =
        &       \int_{\Xiq} \delta\uQ\cdot\AQ(\zQ+\xQ)\,\delta(\uQ\cdot\xQ) 
        \fe\big(\norm{\xQ}\big)\,\dd^4\xQ 
        \nonumber
        \\
        &       +\int_{\Xiq}  \delta\uQ \cdot\xQ
        \qdot\uQ\cdot\AQ(\zQ+\xQ)\fe\big(\norm{\xQ}\big)\,
        \delta(\uQ\cdot\xQ) \,\dd^4\xQ 
        \nonumber
        \\
        &       +\int_{\Xiq}\delta\uQ \cdot
        \xQ (\uQ-\OmegaQ\cdot\xQ\big)\cdot
        \bigl(\uQ\cdot{\nabQ}\AQ(\zQ+\xQ)\bigr)
        \fe\big(\norm{\xQ}\big)\,
        \delta(\uQ\cdot\xQ) \,\dd^4\xQ.
\end{alignat}
	As for the variation of $\Lc$ w.r.t. $\zQ$, we have
\begin{equation}
        \frac{\delta \Lc}{\delta \zQ} \cdot\delta\zQ
        =
        \int_{\Xiq}
        (\uQ-\OmegaQ\cdot\xQ\big)\cdot
        \left(\frac{\partial\AQ(\zQ+\xQ)}{\partial\zQ}\cdot\delta\zQ\right)
        \fe\big(\norm{\xQ}\big)\,
        \delta\big(\uQ\cdot\xQ\big) \,\dd^4\xQ,
\end{equation}
while for the variation of $\Lc$ w.r.t. $\OmegaQ$ we simply have
\begin{equation}
        \frac12 \tr \Bigl(
        \frac{\delta \Lc}{\delta \OmegaQ} \cdot \delta\OmegaQ
        \Bigr)        
        =
        -\int_{\Xiq}
        (\delta\OmegaQ\cdot\xQ)\cdot\AQ(\zQ+\xQ)\,
        \fe\big(\norm{\xQ}\big)\,
        \delta\big(\uQ\cdot\xQ\big) \,\dd^4\xQ.
\end{equation}
        We use \refeq{eq:varOm1} for the variation of the total 
gyration tensor $\OmegaQ$ along the independently 
perturbed particle world-line, and find
\begin{multline}
        \delta\Abf 
=
        \int_{\tau_1}^{\tau_2}\dds\Bigg\{
        -(\tenseur{g} +\uQ\,\tens\uQ)\, \cdot \dds
        \int_{\Xiq}\xQ\, \uQ\cdot\AQ(\zQ+\xQ)
        \fe\big(\norm{\xQ}\big)
        \,\delta\big(\uQ\cdot\xQ\big) \,\dd^4\xQ 
 \\
\qquad\qquad\qquad
 +   \int_{\Xiq} \left[\AQ(\zQ+\xQ)-\big(\uQ\cdot\AQ(\zQ+\xQ)\big)
        \EulerQ\cdot\xQ\right] \fe\big(\norm{\xQ}\big)
        \,\delta\big(\uQ\cdot\xQ\big)
        \,\dd^4\xQ 
\\
\qquad\qquad\qquad 
+  \int_{\Xiq} \big[ \xQ+ (\uQ\cdot\xQ)\uQ\big]
        \big(\uQ-\OmegaQ\cdot\xQ\big)\cdot\AQ(\zQ+\xQ) 
         \fe\big(\norm{\xQ}\big)\,\delta'\big(\uQ\cdot\xQ\big)
        \,\dd^4\xQ \Bigg\}\delta\zQ\,\dd\tau 
\\
        - \int_{\tau_1}^{\tau_2} 
		\int_{\Xiq} \nabQ\tens\AQ(\zQ+\xQ)\cdot\big[\uQ
        -\OmegaQ\cdot\xQ\big] \fe\big(\norm{\xQ}\big)
        \,\delta\big(\uQ\cdot\xQ\big) \,\dd^4\xQ\,\delta\zQ\,\dd\tau\, .
        \label{eq:deltaSmf}
\end{multline}
	We next simplify the previous equation with the help of the
following useful identity.
	Performing an integration by parts, one can easily check that
\begin{multline}
        \int_{\Xiq}\big[\uQ -\OmegaQ\cdot\xQ\big] \cdot
        \nabQ\tens\AQ(\zQ+\xQ) \fe\big(\norm{\xQ}\big)
        \,\delta\big(\uQ\cdot\xQ\big) \,\dd^4\xQ  
\\
=        \int_{{\mathbb R}^{1,3}}
                \AQ(\zQ+\xQ) (\tr\OmegaQ) \fe\big(\norm{\xQ}\big)
        \,\delta\big(\uQ\cdot\xQ\big) \,\dd^4\xQ 
        \\ \qquad
        - \int_{\Xiq} \AQ(\zQ+\xQ)\big[
		\uQ\cdot\xQ-\xQ\cdot\OmegaQ\cdot\xQ\big]
        \fe'\big(\norm{\xQ}\big)/\norm{\xQ}
        \,\delta\big(\uQ\cdot\xQ\big) \,\dd^4\xQ 
        \\
        - \int_{\Xiq} \AQ(\zQ+\xQ)\,\uQ\cdot\big[ \uQ-\OmegaQ\cdot\xQ\big]
        \fe\big(\norm{\xQ}\big)
        \,\delta'\big(\uQ\cdot\xQ\big) \,\dd^4\xQ .
\label{eq:intermedSTEPa}
\end{multline} 
	By the anti-symmetry of $\OmegaQ$ we have
$\tr\OmegaQ =0$, so  that the first integral of the r.h.s. of 
\refeq{eq:intermedSTEPa} vanishes.
	Again by the anti-symmetry of $\OmegaQ$, we also have that
$\xQ\cdot\OmegaQ\cdot\xQ =0$, and furthermore we have
$\uQ\cdot\xQ\delta (\uQ\cdot\xQ)=0$.  
	Therefore, the second integral of the r.h.s. of 
\refeq{eq:intermedSTEPa} vanishes, too.
	As for the last integral of the r.h.s. of \refeq{eq:intermedSTEPa}, 
we use that $\OmegaQ=\EulerQ + \ThomasQ$ with
$\ThomasQ=\qdot{\uQ}\wedge\uQ$, and moreover that $\uQ\cdot\EulerQ=0$, to
rewrite it into 
\begin{align}
	- &\int_{\Xiq} \AQ(\zQ+\xQ)\,\uQ\cdot\big[ \uQ-\OmegaQ\cdot\xQ\big]
        \fe\big(\norm{\xQ}\big)
        \,\delta'\big(\uQ\cdot\xQ\big) \,\dd^4\xQ 
\notag \\
=        &\qquad\qquad\qquad\int_{\Xiq} \AQ(\zQ+\xQ)
                \big[1+\qdot{\uQ}\cdot\xQ\big] \fe\big(\norm{\xQ}\big)
        \,\delta'\big(\uQ\cdot\xQ\big) \,\dd^4\xQ \, .
\label{eq:intermedSTEPb}
\end{align} 
	Finally, the r.h.s. of \refeq{eq:intermedSTEPb} is integrated by
parts once, and we arrive at the  final identity 
\begin{align} 
& 	\int_{\Xiq}\big[\uQ -\OmegaQ\cdot\xQ\big]  \cdot
        \nabQ\tens\AQ(\zQ+\xQ) \fe\big(\norm{\xQ}\big)
        \,\delta\big(\uQ\cdot\xQ\big) \,\dd^4\xQ  
 \notag \\
&\qquad\qquad\qquad
=	\dds \int_{\Xiq}\AQ(\zQ+\xQ) \fe\big(\norm{\xQ}\big)
        \,\delta\big(\uQ\cdot\xQ\big) \,\dd^4\xQ.  
\end{align}
	This identity is now used to anti-symmetrize the last integral
in~\refeq{eq:deltaSmf}.
        Noticing next that $\nabQ\tens\AQ\cdot \vQ - \vQ 
\cdot\nabQ \tens \AQ = \nabQ\wedge\AQ\cdot\vQ$ for an arbitrary four-vector 
$\vQ$, and recalling that $\nabQ\wedge\AQ = \FQ$, the variation 
\refeq{eq:deltaSmf} of $\Abf$ takes the form
\begin{equation} 
        \delta\Abf = \int_{\tau_1}^{\tau_2}\Big[\qdot{\pQ}_{\textsc{N}}
        -\qv{f}\Big]\, \delta\zQ\,\dd\tau,
        \label{eq:deltaSmfnew}
\end{equation}
where 
\begin{equation}
        \qv{f}
=
        \int_{\Xiq}\FQ(\zQ +\xQ)\cdot
        \Big(\uQ - \OmegaQ \cdot\xQ\Big)
        \fe(\norm{\xQ})\,\delta(\uQ\cdot\xQ)\,\dd^4\xQ 
\label{eq:minkoF}
\end{equation}
is the Abraham--Lorentz type \emph{Minkowski force}, and
\begin{equation}
        \pQ_{\textsc{N}}
=
        -\int_{\Xiq}
        \xQ\Big[\xQ\cdot\EulerQ\cdot\FQ(\zQ +\xQ)\cdot\uQ\Big]
        \fe\big(\norm{\xQ}\big)\,
        \delta(\uQ\cdot \xQ)\,\dd^4\xQ
\label{eq:NODVIKpSPINORBIT}
\end{equation}
is Nodvik's Minkowski momentum for the electromagnetic 
spin-orbit coupling. 
	For later convenience, we notice that the charged particle's 
co-moving electromagnetic field, viz. its Coulomb field and its 
instantaneous gyro-magnetic field (both defined w.r.t. the Fermi--Walker 
frame) do not contribute to $\pQ_{\textsc{N}}$, so that $\FQ$ in 
\refeq{eq:NODVIKpSPINORBIT} can be replaced by the 
\emph{reduced field tensor} $\FQred$, which obtains 
from $\FQ$ by subtracting the particle's co-moving electromagnetic field.
	Furthermore, making use of the identity $\uQ\cdot\EulerQ=\qv{0}$,
and using the fact that terms proportional to $\xQ\cdot\uQ$ in 
the integrand vanish upon integration against $\delta(\uQ\cdot\xQ)$, we can 
rewrite the four-vector $\pQ_{\textsc{N}}$ in the more appealing format
\begin{equation}
        \pQ_{\textsc{N}}
        =
        \Mnodv\cdot \uQ 
\label{eq:OURpSPINORBIT}
\end{equation}
where we introduced the symmetric Nodvik tensor mass of spin-orbit coupling,
\begin{equation}
        \Mnodv
        = 
        - \int_{\Xiq}
        \Big[\xQ\tens\xQ,\big[\FQred(\zQ +\xQ),\EulerQ\big]_+\Big]_+ 
        \fe\big(\norm{\xQ}\big)\, \delta(\uQ\cdot \xQ)\,\dd^4\xQ
\, . \label{eq:Mspinorbit} 
\end{equation}

	The Euler-Lagrange equation for $\delta(\Ab + \Abf) =0$ under
the variation of $\zQ(\tau)$ for given $\FQ(\xQ)$ and $\EulerQ(\tau)$ 
can now be read off from \refeq{eq:deltaSmfnew} and \refeq{eq:bareVAR} as
\begin{equation}
        \frac{\dd \pQ}{\dd \tau}= \qv{f}
\, ,\label{eq:albertEQ}
\end{equation}
where 
\begin{equation}
        \pQ
        =
        \Mmink\cdot \uQ 
\, ,\label{eq:fourP}
\end{equation}
is the four-vector of the \emph{Minkowski momentum}, 
with $\Mmink = \Mbare + \Mnodv$.
	Equation~\refeq{eq:albertEQ} is the \emph{Minkowski equation 
of motion} with the expected Abraham--Lorentz type Minkowski force,
but with a somewhat unusual anisotropic tensor mass $\Mmink$.

          \subsubsection{Euler-Lagrange equations for the gyrograph}

	Using \refeq{eq:varEUL2} and integrating by parts, observing that 
the variations of the Euler variables vanish at the boundaries, for the 
first variation of  the action $\Ab$ of the bare particle we find  
\begin{equation}
        \delta\Ab 
        =
        \frac12 \int_{\tau_1}^{\tau_2} \tr \Bigl( \Bigl(
        \SQb - \bigl[ \SQb\, ,\OmegaQ \bigr]_- \Bigr)
        \cdot\delta\thetaQ \Bigr) \,\dd \tau ,
\end{equation}
and for the first variation of the coupling action we get
\begin{equation}
        \delta\Abf = \frac12 \int_{\tau_1}^{\tau_2} \tr \Bigl( \Bigl(
        \SQe - \bigl[ \SQe\, ,\OmegaQ \bigr]_- \Bigr)
        \cdot\delta\thetaQ \Bigr) \,\dd \tau
\end{equation}
where we introduced the tensor of
\emph{electromagnetic Minkowski spin (about $\zQ$)} of the charged particle,
\begin{equation}
        \SQe
        \defeg
        \int_\Xiq \xQ\, \wedge\, \AQ(\zQ +\xQ)
        \fe\big(\norm{\xQ}\big)\,
        \delta(\uQ\cdot \xQ)\,\dd^4\xQ.
        \tag{\ref{eq:sqe}}
\end{equation}
	We remark that $\SQe$ is invariant under a gauge transformation
$\AQ\to\AQ+\nabQ\Psi$.  
	Defining now the \emph{total spin} of the
particle as
\begin{equation}
        \SQ 
        \defeg  
        \SQe +  \SQb
\end{equation}
 and introducing the abbreviation
\begin{equation}
        \NQ
        \defeg 
        \dds \SQ + [\OmegaQ\, ,\SQ]_-,
\label{eq:Ndefine}
\end{equation}
the vanishing of the first variation of the action functional with respect
to all anti-symmetric $\delta\thetaQ$ that are of space-space type with
respect to $\uQ$ now simply reads
\begin{equation}
        0
        =
        \int_{\tau_1}^{\tau_2} \tr
        \left(
          \NQ \cdot\delta\thetaQ
        \right)
        \,\dd \tau 
\, .\label{eq:variaN}
\end{equation}

	At this point we recall that, given any tetrad, any
anti-symmetric tensor $\tenseur{A}$ can be uniquely decomposed into a sum 
of its space-space and time-space parts.
	With $\uQ$ as the timelike unit vector of the tetrad, we have
\begin{equation}
        \tenseur{A} 
        =
        \tenseur{A}^\perp + \tenseur{A}^\|
\, ,
\end{equation}
where
\begin{equation}
  \left\{
    \begin{array}{rl}
      \tenseur{A}^\perp 
      & \defeg
      \tenseur{A} + \big[ \uQ \otimes\uQ\, , \tenseur{A} \big]_+ 
         \text{ is the space-space part of $\tenseur{A}$, and}
      \\
      \tenseur{A}^\|
      & \defeg
      \uQ\, \wedge\,(\tenseur{A} \cdot \uQ) 
      \qquad\qquad\qquad \text{ its time-space part.}
    \end{array}
  \right.
\end{equation}

	Armed with the above decomposition, we conclude that
equation \refeq{eq:variaN} can only be satisfied if the space-space 
part w.r.t. $\uQ$ of the anti-symmetric tensor $\NQ$ vanishes, explicitly
\begin{equation}
	\NQ^\perp
        =
        \vect{0}
\, ,\label{eq:ELrot}
\end{equation}
which is the Euler-Lagrange equation for the rotational variations.

	To extract the evolution equation  for $\EulerQ$ from
\refeq{eq:ELrot}, we also decompose the anti-symmetric spin tensor $\SQ$ 
into its space-space and time-space parts with respect to $\uQ$,
\begin{equation}
        \SQ 
        =
        \SQ^\perp + \SQ^\|
\, .
\end{equation}
	Inserting the above decomposition of $\SQ$ into the definition 
\refeq{eq:Ndefine} of $\NQ$ simply gives 
\begin{equation}
        \NQ
        =
        \dds {\SQ^\perp} +\, \big[\OmegaQ ,\SQ^\perp \big]_- 
        + 
        \dds \SQ^\| + \big[\OmegaQ,\SQ^\| \big]_-
\, .
\end{equation}	
	With the help of the \emph{helicity} four-vector 
$\qv{H}\defeg\SQ\cdot\uQ$ we can write
\begin{equation}
  \left\{
    \begin{array}{rl}
      \tenseur{S}^\perp 
& =
      \tenseur{S} -  \qv{u}\, \wedge\,\qv{H}
\\
      \tenseur{S}^\|
& =
      \qv{u}\, \wedge\,\qv{H}
    \end{array}
  \right.
\, .
\end{equation}
	By straightforward calculation we furthermore find that
\begin{equation}
         \dds \SQ^\| + \big[\OmegaQ,\SQ^\| \big]_-
         =
         \uQ\wedge 
         \left(
           \qdot{\qv{H}} +\,  \EulerQ\cdot\qv{H}
         \right)
\end{equation}
and that 
\begin{equation}
        \left(\dds \SQ^\perp + \big[\OmegaQ,\SQ^\perp \big]_-\right)\cdot\uQ
        =
        \qv{0}
\, .
\end{equation}
	Recalling next that $\SQb \cdot \uQ = \qv{0}$
(i.e. $\SQb^\|=\qv{0}$, so that $\SQb=\SQb^\perp$) we find that 
$\SQ \cdot \uQ = \SQe\cdot\uQ$, viz.
\begin{equation}
        \qv{H}
	=
	\int_\Xiq \xQ\,\uQ\cdot \AQ(\zQ +\xQ)
        \fe\big(\norm{\xQ}\big)\,
        \delta(\uQ\cdot \xQ)\,\dd^4\xQ
\,,
\end{equation}
from where we read off that $\qv{H}\cdot\uQ = \qv{0}$, i.e.
$\qv{H}$ is itself spacelike for all $\tau$. 
	This implies that $\qdot{\qv{H}} +  \EulerQ\cdot\qv{H}$
is spacelike, and we conclude that
\begin{align}
      \dds \SQ^\|    + \big[\OmegaQ,\SQ^\|    \big]_-    
	& =     \NQ^\| \, ,
	\\
        \dds \SQ^\perp + \big[\OmegaQ,\SQ^\perp \big]_-
        & =  \NQ^\perp 
\, .
\end{align}
	Hence, the Euler-Lagrange equation~\refeq{eq:ELrot} reads
\begin{equation}
        \dds {\SQ^\perp} +\, \big[\OmegaQ ,\SQ^\perp \big]_- 
        =
        \tenseur{0}
\, . \label{eq:ELspinEQ}
\end{equation}

\emph{Remark:} The conservation law for $\norm{\SQ^\perp}$ 
follows immediately from \refeq{eq:ELspinEQ}.

	It remains to show that \refeq{eq:ELspinEQ} is identical
to \refeq{eq:ggEQ}. 
	We recall that $\SQ= \SQb +\SQe$, and notice that decomposition
into space-space and time-space parts commutes with addition, so that
$\SQ^\perp = \SQb^\perp +\SQe^\perp$.
	We also recall that $\SQb^\perp = \SQb$. 
	We furthermore notice that $\EulerQ \propto \SQb$, which
implies that $\big[\EulerQ ,\SQb \big]_- = \qv{0}$, so that 
$\big[\OmegaQ ,\SQb \big]_- = \big[\ThomasQ ,\SQb \big]_- $.
	Using all these pieces of information we see that 
\refeq{eq:ELspinEQ} states that
\begin{equation}
        \dds {\SQb} +\, \big[\ThomasQ ,\SQb \big]_- 
  =
 	- \dds {\SQe^\perp} -\, \big[\OmegaQ ,\SQe^\perp \big]_- 
\, . \label{eq:ELspinEQrewrite}
\end{equation}
	The l.h.s. of \refeq{eq:ELspinEQrewrite} is manifestly
identical to the l.h.s. of \refeq{eq:ggEQ}.
	To see that the r.h.s. of \refeq{eq:ELspinEQrewrite} 
is identical to the r.h.s. of \refeq{eq:ggEQ}, one can
rewrite the calculation in Appendix C of~\cite{nodvik}
into our notation, using the Nodvik Maxwell--Lorentz equations 
\refeq{eq:homMLeq}, \refeq{eq:inhMLeq}, \refeq{eq:Jnodvik},
and a few integrations by parts. 
	The details are not repeated here.

	\subsection{Point mass and point charge limits of massive LED}

	In the renormalization flow study of our massive LED with the choice
\refeq{eq:SHELL} for the mass and charge densities, the renormalized
particle reaches a finite smallest size $=1.5\, \RC$ in the `renormalized 
purely electromagnetic' limit of vanishing bare rest mass $\mz \to 0^+$. 
	This renormalization program for massive LED aside, it is of
interest to investigate which types of theories emerge in the point 
limits for mass and charge, either taken separately or jointly.

	\subsubsection{The point mass limit}

	In the point mass limit 
\begin{equation}
        \fm(\norm{\xQ}) 
\longrightarrow
	\mz\frac{1}{4\pi\norm{\xQ}^2}\,\delta(\norm{\xQ})
\defeg 
	\fm^{\textsc{Nodvik}}\big(\norm{\xQ}\big)
\, ,
\end{equation}
but with $\fe$ fixed and `regular' (not a point charge density),
and  with the condition that $\norm{\eulerQ}<C$,
our massive LED formally reduces to the massive LED of Nodvik~\cite{nodvik}. 
	In this limit, the square root terms in 
\refeq{eq:gyroMASS} and \refeq{eq:BARElagrangianDENS} 
become just factors of unity on the support of 
$\fm^{\textsc{Nodvik}}$, so that the bare particle's 
gyrational mass \refeq{eq:gyroMASS} degenerates identically into a constant, 
\begin{equation}
	\rotM(\|\eulerQ\|) \to \mz\, ,
\end{equation}
and our bare action reduces simply to the action  of a free mechanical
point particle with bare rest mass $\mz$, 
\begin{equation}
        \int_{\Xi} \Lm(\xQ)\,\dd^4\xQ
\longrightarrow
	- \mz \int_{\tau_1}^{\tau_2} \,\dd \tau
 =         {\scr A}_{\text{bare}}^{\textsc{Nodvik}}\, ,
        \label{eq:nodvikBAREaction}
\end{equation}
employed by Nodvik~\cite{nodvik}.
	The field and coupling actions remain the same, however.
	Accordingly, the electromagnetic part of the two models  
is identical, i.e. the Maxwell--Lorentz equations in Nodvik's massive 
LED are given by \refeq{eq:homMLeq}, \refeq{eq:inhMLeq}, and 
\refeq{eq:Jnodvik}. 
        Moreover, the world-line equation in Nodvik's massive LED 
can still be written as
\begin{equation}
        \frac{\dd }{\dd \tau}\pQ(\tau) 
= 
      \int_{{\mathbb R}^{1,3}} \FQ(\xQ)\cdot
\qv{U}(\xQ;\tau)
        \fe(\norm{\xQ- \zQ(\tau) })
\,\delta\Big(\uQ\cdot\big(\xQ-\zQ(\tau)\big)\Big)\,\dd^4\xQ, 
\label{eq:wlEQnodvik}
\end{equation}
but now one has
\begin{equation}
        \pQ(\tau) 
=
	\Mmink^{\text{Nodvik}}(\tau) \cdot \uQ(\tau) 
\label{eq:EnMomnodvik}
\end{equation}
with
\begin{equation}
	\Mmink^{\text{Nodvik}}(\tau) 
=        
	\mz \gQ  - \int_{{\mathbb R}^{1,3}}
        \Big[\xQ\tens\xQ,\big[\FQ(\zQ +\xQ),\EulerQ\big]_+\Big]_+ 
        \fe\big(\norm{\xQ}\big)\, \delta(\uQ\cdot \xQ)\,\dd^4\xQ
\, ,\label{eq:Minknodvik}
\end{equation}
which is a slightly simpler mass tensor than $\Mmink$.
	On the other hand, the point mass limit eliminates the moment of 
inertia of the bare particle while the norm of $\EulerQ$ remains bounded 
(by hypothesis) so that $\SQb\to \tenseur{0}$ identically; whence, the
gyrograph equation~\refeq{eq:ggEQ} degenerates into the statement
$\tenseur{t} =\tenseur{0}$, i.e., upon using \refeq{eq:MinkALtorque}, 
\begin{equation}
	\tenseur{0}  
= 
	\int_{{\mathbb R}^{1,3}}  
	(\xQ-\zQ)\wedge\bigl(\FQ(\xQ)\cdot\qv{U}(\xQ,\tau)\bigr)^\perp
	\fe(\norm{\xQ-\zQ})\,\delta(\uQ\cdot\big(\xQ-\zQ)\big)
	\,\dd^4\xQ
\, .  \label{eq:ggEQnodvik}
\end{equation}
	These are the equations of the massive LED of Nodvik~\cite{nodvik}.  
	We will now explain that this is a qualitatively quite different 
dynamical theory from our massive LED.

	First of all, since one loses a proper-time derivative for 
$\EulerQ$ by passing from \refeq{eq:ggEQ} to \refeq{eq:ggEQnodvik}, 
the point mass limit is  a \emph{singular limit} from the 
dynamical perspective of the Cauchy problem for $\EulerQ$.
	This fact in itself already means that it will no longer 
be possible to arbitrarily choose initial data for $\EulerQ$ 
from the set of admissible Cauchy data for \refeq{eq:ggEQ}.
	Indeed, choose (which we always may) a Lorentz frame in 
which initially (i.e. at $t = 0 = \tau$) we have $\zQ_0=\qv{0}$ and 
$\uQ_0=\qv{e}_0$, and pick the fields $\FQ_0$ on $\uQ_0\cdot\xQ=0$
at random from our set of admissible Cauchy data for $\FQ_0$.
	Then \refeq{eq:ggEQnodvik} at $\tau = 0$ has only a nontrivial
space part, which in three-space notation reads
\begin{equation}
	\tVE_0 + \eulerV_0\times \sigmaVB_0 
= 
	\vect{0}
\, ,\label{eq:initCONSTRAINT}
\end{equation}
where $\eulerV_0$ is the space part of the dual four-vector for
$\EulerQ_0$, and where
\begin{equation}
	\tVE_0 
\defeg 
	\int_{{\mathbb R}^3} \xV \times \EV_0(\xV) \fe(|\xV|)\dd^3 x
\label{eq:Etorque}
\end{equation}
is the initial electric torque on the particle and 
\begin{equation}
	\sigmaVB_0 
\defeg 
	\int_{{\mathbb R}^3} \xV \xV\cdot\BV_0(\xV)\fe(|\xV|)\dd^3 x
\label{eq:Bspin}
\end{equation}
its initial `magnetic spin.'
	Clearly, any attempt to also prescribe Cauchy data for
the angular velocity tensor $\EulerQ_0$, chosen at random from
our set of admissible data, will almost surely violate 
\refeq{eq:initCONSTRAINT}. 

	More specifically, \refeq{eq:initCONSTRAINT}, with \refeq{eq:Etorque} 
and \refeq{eq:Bspin}, imposes not just on the set of admissible initial data 
for $\EulerQ$ but also on the fields $\FQ_0$, in the sense that
for most choices of $\FQ_0$ \emph{no} choice for $\EulerQ_0$ will
satisfy \refeq{eq:initCONSTRAINT}. 
	Thus, one extracts directly from 
\refeq{eq:initCONSTRAINT}, \refeq{eq:Etorque}, \refeq{eq:Bspin}
that the originally admissible field data are now further 
restricted as follows: either $\BV_0$ satisfies $\sigmaVB_0 = \vect{0}$, 
and in this case also $\EV_0$ has to satisfy $\tVE_0 =\vect{0}$; or 
$\BV_0$ satisfies $\sigmaVB_0 \neq \vect{0}$, in which  case 
$\EV_0$ has to satisfy $\tVE_0\cdot\sigmaVB_0 =0$. 
	Only if these alternate constraints on the initial fields are
satisfied is it possible to satisfy \refeq{eq:initCONSTRAINT} by a
compatible class of initial data $\EulerQ_0$ (viz., $\eulerV_0$).
	Namely, if $\sigmaVB_0 = \vect{0}$ and $\tVE_0 =0$, then the 
admissible data for $\eulerV_0$ are the same as for our model. 
	If $\sigmaVB_0 \neq \vect{0}$ and $\tVE_0 \cdot\sigmaVB_0 =0$, 
however, then $\eulerV_0$ has to be of the form
\begin{equation}
	\eulerV_0
= 
	\frac{1}{|\sigmaVB_0|^2} \tVE_0\times \sigmaVB_0 + \alpha \sigmaVB_0
\, ,
\end{equation}
which leaves only one degree of freedom to choose from instead of three, 
i.e. the coefficient $\alpha$ in front of $\sigmaVB_0$ is at our disposal.
	This clarifies to which extent initial data for Nodvik's model
have to be compatible. 

	Compatibility  of a set of initial data in the sense just
explained does not, however, imply that all is well then, for not
only has one lost a proper-time derivative on $\EulerQ$ by passing
from \refeq{eq:ggEQ} to \refeq{eq:ggEQnodvik}, a proper-time derivative 
of $\EulerQ$ does not show any more at all in \refeq{eq:ggEQnodvik}. 
	Nodvik's \refeq{eq:ggEQnodvik} is therefore not an orderly evolution 
equation for $\EulerQ$, but rather a \emph{constraint} equation for 
the dynamical field variables $\FQ$ and the dynamical particle 
variables $\zQ$ and $\uQ$, which in case of actually being satisfied by
these variables for some time interval after the initial instant,  
still leaves a certain amount of freedom for choosing the value 
of the now purely kinematical variable $\EulerQ$.
	Therefore,  to the extent that compatible initial data for 
$\FQ$, $\zQ$, and $\uQ$, with a compatible initial choice of $\EulerQ$, 
can be continued at all according to  Nodvik's equations, such a 
continuation is presumably not unique.
	
	\emph{Remark:} We shall see that a similar assessment holds
for the semi-relativistic, purely electromagnetic theory of Abraham, 
which is discussed in Appendix A.3.

\subsubsection{Remarks on the spinless point particle limit}

	Ever since Dirac's surmise~\cite{diracA} that, as long as
we are willing to let the bare mass take the value $-\infty$ to
compensate for the $+\infty$ electromagnetic self-energies, we may
view the Abraham--Lorentz--Dirac (ALD) equation
\begin{equation}
	m\qddot\zQ
= 
	-e \FQ^{\text{\tiny ext}}\cdot{\qdot\zQ}
+	\frac{2}{3}\frac{e^2}{c^3}
	(\gQ +\qdot\zQ\tens\qdot\zQ)\cdot{\qdddot\zQ}
\, \label{eq:ALDeq}
\end{equation}
as the \emph{exact} equation for the radiation reaction-corrected motion of 
a spinless point particle with empirical electron mass $m$ and empirical 
electron charge $-e$ in external electromagnetic fields 
$\FQ^{\text{\tiny ext}}$, there has been a strong interest
in making sense out of the peculiar third-order proper-time derivative 
of $\zQ$ that figures in  \refeq{eq:ALDeq}, 
e.g.~\cite{rohrlichBOOK, diracA, caratigalganietal, gaftoilopezetal, luca, 
peierlsBOOK}.
	A rigorous clarification of Dirac's surmise is, however, not so
easy to come by. 
	Notable rigorous attempts to construct a consistent renormalized
electrodynamics with spinless point charges are 
\cite{gitteletal} and \cite{bambusinoja}; 
see also \cite{bambusigalgani} and \cite{nojaposilicanoA, nojaposilicanoB} 
for work on the classical Pauli--Fierz 
model, a semi-relativistic dipole approximation to the problem,
which also serves as point of departure for setting up a so-called 
non-relativistic QED~\cite{griesemerliebloss}.

	Naturally one may wonder whether \refeq{eq:ALDeq} 
emerges in some spinless point particle limit of our massive LED, or
to begin simpler, in the point particle limit of a spinless version
of our massive LED.
	This massive LED without spin obtains from our model 
by simply discarding \refeq{eq:ggEQ} and setting 
$\EulerQ\equiv \tenseur{0}$ in the remaining equations.
	Thus, the relativistic Maxwell--Lorentz equations are once again
\begin{equation}
        \nabQ\, \cdot\, {^\star\FQ} 
=
	\qv{0}
\, \label{eq:homMLeqSL}
\end{equation}
for the Hodge dual of $\FQ$, and 
\begin{equation}
        \nabQ\cdot \FQ (\xQ)
=
        4\pi\int_{-\infty}^{+\infty}
		\qv{U}(\qv{x};\tau)
        	\;{\fe}\big(\norm{\qv{x}-\zQ(\tau)}\big)
	        \,\delta\Big(\qv{u}(\tau)\cdot\big(\qv{x}-\zQ(\tau)\big)\Big)
	\,\dd \tau
\, \label{eq:inhMLeqSL}
\end{equation}
for $\FQ$ itself, but now with
\begin{equation}
        \qv{U}(\qv{x};\tau) 
=
	\qv{u}(\tau) -\,\big( \qdot{\uQ}(\tau)\, \wedge\,\uQ (\tau)\big)
	\cdot \big(\qv{x}-\zQ(\tau)\big)
\, .\label{eq:UeqSL}
\end{equation}
	The world-line equation still reads
\begin{equation}
        \frac{\dd }{\dd \tau}\pQ
=
      \int_{{\mathbb R}^{1,3}} \FQ(\xQ)\cdot \qv{U}(\xQ;\tau)
        \fe(\norm{\xQ- \zQ(\tau)})
	\,\delta\Big(\uQ(\tau)\cdot\big(\xQ-\zQ(\tau)\big)\Big)\,\dd^4\xQ 
\, ,\label{eq:wlEQSL}
\end{equation}
but now with
\begin{equation}
        \pQ(\tau) 
=
        \mz \uQ(\tau) 
\, ,\label{eq:MinkEnMomSL}
\end{equation}
and with $\qv{U}$ given in \refeq{eq:UeqSL}.
	Notice that the structure of $\fm$ does not figure in this 
spinless LED, which is therefore identical to the spinless version of
Nodvik's LED, due to Spohn~\cite{spohn}, in which $\fm$ is a point mass 
distribution from the outset. 
	The question then is: What happens if we let $\fe$ tend to a delta 
function concentrated at the same point?
	The first step to the answer is the rigorous control of this spinless 
massive LED which, however, is not very developed yet.  


	\subsection{Semi-relativistic LED}

        In this Appendix we give a brief summary of semi-relativistic  LED.
        This material can also be used as a primer for the relativistic 
theory. 
	We begin with the massive model with spin. 
	Its limiting cases of the more familiar Abraham model 
and the popular `model without spin' are discussed in subsequent 
subsections.

\subsubsection{Massive model with spin}
%

        As in the relativistic theory, the electric field
$\EV (\vect{x},t)\in{\mathbb R}^3$ and the magnetic field
$\BV(\vect{x} ,t)\in{\mathbb R}^3$ at the point 
$\vect{x}\in{\mathbb R}^3$ at time $t\in{\mathbb R}$ 
satisfy the classical Maxwell--Lorentz equations,
\begin{alignat}{1}
        \nab\cdot \BV(\vect{ x}, t)  
&= 
        0\, ,
\label{eq:MLdivB}
\\
        \frac{1}{c}\ddt{\BV(\vect{x}, t)}
        + \nab\times\EV(\vect{x}, t)   
&= 
        \vect{0}\, ,
\label{eq:MLrotE}
\\
        \nab\cdot\EV(\vect{ x}, t)  
&=
        4 \pi \rho (\vect{ x}, t) \, ,
\label{eq:MLdivE}
\\
        - \frac{1}{c}\ddt{\EV(\vect{x},t)}
        + \nab\times\BV(\vect{ x}, t)  
&= 
        4\pi \rho (\vect{ x}, t)\frac{1}{c} \vect{v}(\vect{x},t) \, ,
\label{eq:MLrotB}
\end{alignat}
        Although not manifestly Lorentz co-variant,
the Maxwell--Lorentz equations \refeq{eq:MLdivB}, \refeq{eq:MLrotE},
\refeq{eq:MLdivE}, \refeq{eq:MLrotB} appear to have  a 
relativistic format already, but whether in fact they do satisfy
relativistic Lorentz transformation laws or not depends on the 
model for the charge distribution. 
        In the classical works of Abraham~\cite{abraham, abrahamBOOK} 
before the dawn of 
special relativity~\cite{lorentzACAD, poincare, einsteinA, minkowski},
the electrical charge and current densities formally 
combine as $\rho(\vect{x},t)\bigr(c,\vect{v}(\vect{x},t)\bigr)$ but
do \emph{not} transform like a relativistic four-current density vector.

        In the semi-relativistic formulation of Abraham,
the charge of a particle ($-e$ for a model of the electron) 
is distributed around the instantaneous location  
$\qV(t)\in {\mathbb R}^3$ of 
the particle by a charge density $\fe$ with $SO(3)$ symmetry
and compact support, satisfying 
$\int_{{\mathbb R}^3}\fe(|\xV|){\dd}^3x = -e$. 
        This charge density is rigidly carried along 
by the particle with linear velocity $\dot{\vect{q}}(t)$
and rotating rigidly with angular velocity $\vect{\omega}(t)$.
        Thus, with a single charged particle source  in \refeq{eq:MLdivE},
\refeq{eq:MLrotB}, we have
\begin{equation}
        \rho (\vect{x}, t) 
 = 
        \fe(|\vect{x} -\vect{q}(t)|)\, ,
\label{eq:ALcharge}
\end{equation}
and
\begin{equation}
        \vect{v}(\vect{x},t) 
= 
        \dot{\vect{q}}(t) 
        + \vect{\omega}(t)\times \big(\xV -\qV(t)\big)\, .
\label{eq:ALvfield}
\end{equation}

        The evolution equations for the dynamical variables of the
particle, i.e. position $\vect{q}(t)$, linear velocity $\dot{\vect{q}}(t)$,
and angular velocity $\vect{\omega}(t)$, are likewise not Lorentz
covariant and given by Newton's and Euler's evolution equations 
equipped with the Abraham--Lorentz expressions for the volume-averaged 
Lorentz force and torque \cite{lorentzFORCE}, respectively, that act on 
the charged particle.
        Newton's equation of motion here reads
\begin{equation}
        \frac{\dd \pVb}{\dd t}
= 
        \int_{{\mathbb R}^3}
        \left[\EV(\xV,t) + \frac{1}{c}\vV(\xV,t) \times\BV(\xV,t)\right]
        \fe(|\xV -\qV(t)|) \, \dd^3x\, ,
\label{eq:ALeqTmb}
\end{equation}
where
\begin{equation}
        \pVb
= 
	\mz \dot\qV 
\label{eq:pDEFnewton}
\end{equation}
is the Newtonian linear particle momentum,  with $\mz$ the bare mass 
(`material mass' in \cite{lorentzBOOKb}).
	Euler's equation of motion reads
\begin{equation}
        \frac{\dd {\vect{s}}_{\text{b}}}{\dd t}
= 
        \int_{{\mathbb R}^3} \big(\xV - \qV(t)\big)\times
        \left[\EV(\vect{ x},t) + \frac{1}{c}\vect{v}(\vect{ x},t)
        \times\BV(\vect{ x},t)\right]
        \fe(|\vect{ x} -\vect{q}(t)|) \, \dd^3x\, ,
\label{eq:ALeqRmb}
\end{equation}
where
\begin{equation}
        \vect{s}_{\text{b}} 
= 
        \Iz\vect{\omega} 
\label{eq:sDEFeuler}
\end{equation}
is the classical particle spin associated with the bare moment of 
inertia $\Iz$.

        With the bare inertias $\mz\neq 0$ and $\Iz\neq 0$, 
the semi-relativistic equations listed above pose a Cauchy problem 
for the following initial data at time $t =t_0$: 
for the mechanical variables of the particles, the data are 
$\qV(t_0)$, $\dot\qV(t_0)$ (equivalently: $\pVb(t_0)$), 
and $\vect{\omega}(t_0)$ (equivalently: $\sVb(t_0)$); 
for the fields, the data are $\BV(\vect{ x},t_0)$ satisfying 
\refeq{eq:MLdivB} at $t=t_0$, 
and $\EV(\vect{ x},t_0)$ satisfying \refeq{eq:MLdivE} at $t=t_0$. 
        Notice that \refeq{eq:MLdivB} and \refeq{eq:MLdivE}
are merely initial constraints, which then 
remain satisfied by the fields $\BV(\vect{ x},t)$ 
and  $\EV(\vect{ x},t)$ in the ensuing evolution.
        For \refeq{eq:MLdivB} this is seen by taking the 
divergence of \refeq{eq:MLrotE}. 
	For \refeq{eq:MLdivE} this is seen by taking the divergence of
\refeq{eq:MLrotB} and the time-derivative of \refeq{eq:MLdivE}, then 
using the continuity equation for the charge, 
\begin{equation}
	\partial_t \rho(\vect{x}, t) 
	+ \nab\cdot \big(\rho(\vect{x},t) \vV(\vect{x},t) \big)
=
	0
\, ,\label{eq:continuityLAW}
\end{equation}
which is demanded by the Maxwell--Lorentz equations and 
indeed satisfied for $\rho(\vect{x},t)$ given by \refeq{eq:ALcharge} 
and $\vect{v}(\vect{x},t)$ by \refeq{eq:ALvfield}; cf.~\cite{kiessling}. 

        The semi-relativistic massive LED with spin has the 
following conservation laws~\cite{kiessling}.
        Charge, 
\begin{equation}
-e = \int_{\mathbb R^3} \rho(\vect{x},t)\, \dd^3x  \, ,
\label{eq:totalcharge}
\end{equation}
energy, 
\begin{equation}
W = 
\frac{1}{8\pi}\int_{\mathbb R^3}\bigl(|\EV|^2 +|\BV|^2\bigr)\,\dd^3x 
+  \frac{1}{2}\frac{|\sVb |^2}{\Iz} 
+        
 \frac{1}{2}\frac{|\pVb |^2}{\mz} \, , 
\label{eq:totalenergy}
\end{equation}
linear momentum, 
\begin{equation}
	\PV 
= 
	\frac{1}{4\pi c} \int_{\mathbb R^3} \EV\times\BV \, \dd^3x + \pVb
\, , \label{eq:totalimpuls}
\end{equation}
and angular momentum,
\begin{equation}
	\LV
 = 
	\frac{1}{4\pi c} \int_{\mathbb R^3} \xV\times(\EV\times\BV) \,\dd^3x 
	+ \qV\times\pVb  + \sVb
\, \label{eq:totaldrehimpuls}
\end{equation}
are all conserved during the evolution. 

	More detailed studies of the coupled dynamics of the massive
semi-relativistic LED with spin have only recently begun.

	\subsubsection{Massive model `without spin'}

	A mathematically regular limit which, unfortunately, 
is not quite physical but brings about some considerable simplification 
of the dynamical equations is to send $\Iz\to \infty$ while $\mz$ is
fixed at some finite value, in such a way that
$\sV_{\text{b}}$ converges to some finite vector-valued 
function of $t$, which of course implies that $\vect{\omega}\to \vect{0}$.
        The semi-relativistic Maxwell--Lorentz equations are again
given by \refeq{eq:MLdivB}, \refeq{eq:MLdivE}, \refeq{eq:MLrotE},
\refeq{eq:MLrotB}, and the charge density is still \refeq{eq:ALcharge},
but the current density vector now simplifies to 
\begin{equation}
	\vect{j}(\xV,t) 
= 
\rho(\xV,t) \dot{\qV}(t) 
\, .\label{eq:ALcurrentNOgyro}
\end{equation} 
	Moreover, the Newtonian equation of motion \refeq{eq:ALeqT} reduces to
\begin{equation}
        \frac{\dd {\pV}_{\text{b}}}{\dd t}
= 
        \int_{{\mathbb R}^3}
	        \left[\EV(\xV,t) 
			+ \frac{1}{c}\dot{\qV}(t)\times\BV(\xV,t)
		\right]
        \fe(|\xV -\qV(t)|) \, 
	\dd^3x
\, , \label{eq:ALeqTinfI}
\end{equation}
with $\pV$ still given by \refeq{eq:pDEFnewton}, and the Eulerian equation 
\refeq{eq:ALeqR} reduces to 
\begin{equation}
        \frac{\dd {\vect{s}}_{\text{b}}}{\dd t}
= 
        \int_{{\mathbb R}^3} \big(\xV - \qV(t)\big)\times
        \left[\EV(\xV,t) + \frac{1}{c}\dot{\qV}(t)
        \times\BV(\xV,t)\right]
        \fe(|\xV -\qV(t)|) \, \dd^3x
\, .\label{eq:ALeqRinfI}
\end{equation}
	In this limit, the Maxwell--Lorentz equations 
\refeq{eq:MLdivB}, \refeq{eq:MLdivE}, \refeq{eq:MLrotE},
\refeq{eq:MLrotB}, equipped with the charge density \refeq{eq:ALcharge}
and current density vector \refeq{eq:ALcurrentNOgyro}, and the 
Newtonian equation of motion \refeq{eq:ALeqTinfI}, form a 
non-linearly coupled, \emph{closed} system of equations, with
initial data at $t=t_0$ given by: 
$\qV(t_0)$ and $\dot\qV(t_0)$ (equivalently: $\pVb(t_0)$) for the
particle; for the fields, the data are $\BV(\xV,t_0)$ 
satisfying \refeq{eq:MLdivB} at $t=t_0$, 
and $\EV(\xV,t_0)$ satisfying \refeq{eq:MLdivE} at $t=t_0$. 
	Spin in turn adjusts passively to this 
combined field-particle dynamics according to a linear equation 
\refeq{eq:ALeqRinfI}, which has to be supplemented by the 
initial data $\sVb(t_0)$ (now of course no longer equivalent 
to prescribing $\vect{\omega}(t_0)$). 
	Consequently, in this limit spin drops from the conservation laws 
--- with the exception of the law for the angular momentum. 
	Thus, the conserved energy $W$, linear 
momentum $\PV$, and angular momentum $\LV$  are now given by
\begin{equation}
	W 
= 
	\frac{1}{8\pi}\int_{\mathbb R^3}\bigl(|\EV|^2 +|\BV|^2\bigr)\,\dd^3x 
	+        
	 \frac{1}{2}\frac{|\vect{p}_{\text{b}} |^2}{\mz} 
\, , \label{eq:totalenergyinfI}
\end{equation}
\begin{equation}
	\PV 
= 
	\frac{1}{4\pi c} \int_{\mathbb R^3} \EV\times\BV \, \dd^3x 
	+ \vect{p}_{\text{b}} 
\, , \label{eq:totalimpulsinfI}
\end{equation} 
and
\begin{equation}
	\LV
 = 
	\frac{1}{4\pi c} \int_{\mathbb R^3} \xV
	\times (\EV\times\BV) \, \dd^3x + 
	\qV\times\vect{p}_{\text{b}}  
+ \sVb
\, .\label{eq:totaldrehimpulsinfI}
\end{equation}
        Notice that the laws for energy and linear
momentum do not feature $\sVb$, but the conservation law 
for the angular momentum  does.
	Furthermore, the continuity equation simplifies to
\begin{equation}
	\partial_t \rho(\xV, t) 
	+ \nab\cdot \big(\rho(\xV,t) \dot\qV(t) \big)
=
	0
\, ,\label{eq:continuityLAWinfI}
\end{equation}
but this does not affect the law of charge conservation, which 
is still given by \refeq{eq:totalcharge}.

	Since, in the limit $\Iz\to\infty$, spin is of no dynamical relevance,
apart from the law of angular conservation, \refeq{eq:ALeqRinfI} 
is frequently omitted from the system of equations.
        The resulting subset of equations defines the semi-relativistic 
massive LED `without spin.' 
	This model without spin is still invariant under rotations, and 
by E. Noether's theorems there exists a conserved  quantity associated 
with this invariance; only this quantity does not qualify as the physical 
angular momentum in the conventional sense.
	The observation that discarding the gyroscopic equation 
\refeq{eq:ALeqRinfI} violates the conventional law of angular 
momentum conservation \refeq{eq:totaldrehimpulsinfI} was made in 
\cite{kiessling}.

	The semi-relativistic massive LED without spin is 
the most thoroughly understood version of LED.  
	In particular, we mention the recent rigorous treatments in 
\cite{bauerduerr, komechspohn, kunzespohnA, kunzespohnB, kunzespohnC}, 
where for technical reasons the Einsteinian linear momentum 
\begin{equation}
        \vect{p}_{\text{b}} 
=
 \displaystyle\frac{\mz\dot{\qV}}{ \sqrt{1- |\dot{\qV}|^2/c^2}}
\label{eq:pDEFeinstein}
\end{equation}
with $\mz\neq 0$ is used instead of the Newtonian \refeq{eq:pDEFnewton}
(see also \cite{komechspohnkunze, komechkunzespohn} for work on 
a simpler scalar field-particle theory).  
        In this case, of course also the Newtonian kinetic energy 
in \refeq{eq:totalenergy} is to be replaced by its Einsteinian 
counterpart
\begin{equation}
	\frac{1}{2}\frac{|\vect{p}_{\text{b}} |^2}{\mz}
\longrightarrow
	{\mz c^2 \sqrt{1 + \frac{|\vect{p}_{\text{b}} |^2}{ \mz^2 c^2}}}
\, .\label{eq:einsteinENERGY}
\end{equation}
        In \cite{bauerduerr, komechspohn}, the global existence and 
uniqueness for the Cauchy problem of a particle without spin and 
$\mz\neq 0$ was proven. 
        Moreover, in \cite{bauerduerr} it was shown that the motion
is stable if $\mz >0$ and unstable if $\mz <0$. 
        The papers \cite{komechspohn, kunzespohnA, kunzespohnB, kunzespohnC} 
address the long time asymptotics of the slow motion of a spinless 
particle with $\mz >0$ in slowly varying external fields.
	Using center manifold theory, these authors derive 
effective equations of motion of second order in the time derivative. 
	Originally these effective second-order equations were obtained 
in Landau--Lifshitz~\cite{landaulifshitz} by applying a heuristic 
closure argument to the (in)famous third-order term in the 
Abraham--Lorentz--Dirac (ALD) equation \refeq{eq:ALDeq}, 
which works equally well for the relativistic and for the 
semi-relativistic version.
        As far as we know, the rigorous work of Spohn and collaborators 
gives the first clean derivation of the Landau--Lifshitz closure for the 
ALD equation, so far in its semi-relativistic setting. 

\subsubsection{The Abraham model with spin}
%

	Abraham's~\cite{abraham, abrahamBOOK}  purely electromagnetic model
obtains from the massive semi-relativistic LED with 
spin in the formal double limit $\mz\to 0$ and  $\Iz\to 0$ together with
the assumption that the velocity $\dot{\qV}$ and angular velocity
$\vect{\omega}$ of the particle remain bounded.
	The Maxwell--Lorentz equations are not affected by this 
limit and still given by \refeq{eq:MLdivB}, \refeq{eq:MLrotE},
\refeq{eq:MLdivE}, \refeq{eq:MLrotB} together with 
\refeq{eq:ALcharge} and \refeq{eq:ALvfield}.
	However, the bare Newtonian momentum $\pVb \to \vect{0}$ 
and bare Eulerian spin  $\sVb \to \vect{0}$
identically in this limit. 
	As a consequence, the evolution equations  \refeq{eq:ALeqTmb}
and \refeq{eq:ALeqRmb} degenerate into 
\begin{equation}
        \vect{0}
= 
        \int_{{\mathbb R}^3}
        \left[\EV(\xV,t) + \frac{1}{c}\vV(\xV,t) \times\BV(\xV,t)\right]
        \fe(|\xV -\qV(t)|) \, \dd^3x
\label{eq:ALeqT}
\end{equation}
and
\begin{equation}
        \vect{0}
= 
        \int_{{\mathbb R}^3} \big(\xV - \qV(t)\big)\times
        \left[\EV(\xV,t) + \frac{1}{c}\vV(\xV,t)\times\BV(\xV,t)\right]
        \fe(|\xV -\qV(t)|) \, \dd^3x
\, . \label{eq:ALeqR}
\end{equation}
	The conservation laws for energy, linear and angular
momentum simplify accordingly, because $\pVb$ and 
$\sVb$ vanish quadratically fast with $\mz\to 0$ 
and $\Iz\to 0$ while $\dot\qV$ and $\vect{\omega}$ remain
bounded (by hypothesis!). 
	Thus, the conserved energy $W$, linear momentum
$\PV$, and angular momentum $\LV$ are now given by
\begin{equation}
	W 
= 
	\frac{1}{8\pi}\int_{\mathbb R^3}\bigl(|\EV|^2 +|\BV|^2\bigr)\,\dd^3x 
\, , \label{eq:totalenergyAL}
\end{equation}
\begin{equation}
	\PV 
= 
	\frac{1}{4\pi c} \int_{\mathbb R^3} \EV\times\BV \, \dd^3x 
\, , \label{eq:totalimpulsAL}
\end{equation}
and 
\begin{equation}
	\LV 
 = 
	\frac{1}{4\pi c}\int_{\mathbb R^3}\xV\times(\EV\times\BV)\,\dd^3x 
\, . \label{eq:totaldrehimpulsAL}
\end{equation}
	The continuity equation \refeq{eq:continuityLAW} and its offspring,
the law for the charge \refeq{eq:totalcharge}, still hold as stated.

	This purely electromagnetic Abraham model with spin is not
well-studied at all. 
	In fact, it is hardly mentioned in textbooks and monographs on 
classical electromagnetism, with the notable exception of the book by
Miller~\cite{millerBOOK}. 

	Subjecting equations \refeq{eq:ALeqT} and \refeq{eq:ALeqR} to 
closer scrutiny reveals that Abraham's purely electromagnetic 
semi-relativistic LED with spin is qualitatively a very different 
dynamical theory from the massive semi-relativistic LED with spin.  
	In fact, since the coefficients in front of the time derivative of 
both $\dot\qV$ and $\vect{\omega}$ vanish, these equations constitute a 
\emph{singular limit} of the regular evolution equations \refeq{eq:ALeqTmb} 
and \refeq{eq:ALeqRmb}, a limit in which Cauchy data for the 
dynamical particle variables $\dot\qV$ and $\vect{\omega}$ can 
no longer be prescribed freely; also the field data are severely 
restricted. 
	Only the initial particle position can be prescribed arbitrarily 
in Abraham's model. 
	The situation is similar as in Nodvik's relativistic model,
only worse.
	
	More specifically, choose the initial time to be $t=0$ and
the initial particle position to be $\qV(0) = \vect{0}$, which can 
always be achieved by at most a time and a space translation. 
	As before, denote the initial fields by $\EV_0(\xV)$ 
and $\BV_0(\xV)$. 
	Then \refeq{eq:ALeqT} demands that at $t=0$, we have
\begin{equation}
	c\langle\EV_0\rangle 
+	\dot\qV_0\times \langle\BV_0\rangle 
-	\big(\langle \xV\tens\BV_0\rangle 
		- ( {\rm tr}\, \langle \xV\tens\BV_0\rangle )\tenseur{I}
	\big)\cdot\vect{\omega}_0
= 
	\vect{0}
\, ,\label{eq:ALinitT}
\end{equation}
while \refeq{eq:ALeqR} demands (as for Nodvik's model) that at $t=0$ we have 
\begin{equation}
	c\langle \xV\times\EV_0\rangle 
+	\vect{\omega}_0\times \langle \xV\, \xV\cdot\BV_0\rangle 
= 
	\vect{0}
\, .\label{eq:ALinitR}
\end{equation}
	Here, the averaging is defined by $\langle g\rangle \defeg
(-e)^{-1}\int_{{\mathbb R}^3}g(\xV) \fe(|\xV|)\dd^3 x$, where
we assume with Abraham that $(-e)^{-1}\fe(|\xV|)\geq 0$; furthermore,
$\tens$ now means the three-tensor product.
	One readily deduces from  \refeq{eq:ALinitT} and
\refeq{eq:ALinitR} that only the following subset of the 
originally admissible field data is now admitted. 
	Namely, if $\langle \xV\, \xV\cdot\BV_0\rangle = \vect{0}$, 
then also $\langle \xV\times\EV_0\rangle = \vect{0}$ must hold; 
if $\langle \xV\, \xV\cdot\BV_0\rangle \neq \vect{0}$, 
then 
$\langle \xV\times\EV_0\rangle \cdot \langle \xV\,\xV\cdot\BV_0\rangle = 0$
must hold;
if $\langle\BV_0\rangle = \vect{0}$, then $\langle\EV_0\rangle$
must be in the orthogonal complement of the left kernel space of 
$\langle \xV\tens\BV_0\rangle -
({\rm tr}\, \langle \xV\tens\BV_0\rangle )\tenseur{I}$, unless this 
tensor is the zero tensor, in which case  $\langle\EV_0\rangle$
must  vanish also;
if $\langle\BV_0\rangle \neq \vect{0}$ is in the 
left kernel space of $\langle\xV\tens\BV_0\rangle
-({\rm tr}\,\langle\xV\tens\BV_0\rangle)\tenseur{I}$, 
then also $\langle\EV_0\rangle\cdot \langle\BV_0\rangle = 0$ must
hold. 

\emph{Remark:} The above set of conditions on the initial fields imply
in particular that the Abraham equations admit \emph{no solution} for the
following setup: $\qV_0=\vect{0}$, $\BV_0 = \vect{0}$, 
and $\EV_0 = \EV^{\text{hom}} + \EV^{\text{\textsc{Coulomb}}}$,
i.e. the linear superposition of the charge's radial symmetric Coulomb 
field with a homogeneous external field. 
	In this case no choice of $\dot\qV_0$ and  $\vect{\omega}_0$ will
rescue \refeq{eq:ALinitT}; in short, these perfectly physical
initial conditions are \emph{inconsistent} for the Abraham model. 

	We continue with the discussion of consistent initial data.
	Given now consistent data for the fields in accordance with 
the restrictions just explained, the initial data for
$\dot\qV_0$ and  $\vect{\omega}_0$ are arbitrary only if 
$\langle\BV_0\rangle = \vect{0}$ and 
$\langle\xV\,\xV\cdot\BV_0\rangle = \vect{0}$, respectively;
otherwise they must be of the following form.
	If $\langle\xV\,\xV\cdot\BV_0\rangle \neq\vect{0}$, then
\begin{equation}
	\omega_0
= 
	c\frac{\langle \xV\times\EV_0\rangle \times
\langle\xV\,\xV\cdot\BV_0\rangle}{|\langle\xV\,\xV\cdot\BV_0\rangle|^2} 
 + \gamma \langle\xV\,\xV\cdot\BV_0\rangle
\, ,
\end{equation}
which leaves only one degree of freedom to choose from instead of three, 
viz. the coefficient $\gamma$ is at our disposal.
	If $\langle\BV_0\rangle \neq \vect{0}$, then 
\begin{equation}
	\dot\qV_0
= 
	\frac{1}{|\langle \BV_0\rangle|^2} 
	\Big(c\langle \EV_0\rangle  
+ 		\big(\langle\xV\tens\BV_0\rangle 
		- ({\rm tr}\, \langle \xV\tens\BV_0\rangle )\tenseur{I}
		\big)\cdot\omega_0
	\Big)\times\langle\BV_0\rangle
 + 	\beta \langle\BV_0\rangle
\, ,
\end{equation}
which again leaves only one degree of freedom to choose from instead of 
three, here the coefficient $\beta$ is at our disposal.
	The extent to which such consistently restricted initial 
data now launch a unique regular solution of the system of Abraham 
equations \refeq{eq:ALeqT}, \refeq{eq:ALeqR}, \refeq{eq:MLdivB}, 
\refeq{eq:MLrotE}, \refeq{eq:MLdivE}, and \refeq{eq:MLrotB}, with 
\refeq{eq:ALcharge} and \refeq{eq:ALvfield}, is currently not known. 

	\subsubsection{The Abraham model without spin}

	The purely electromagnetic Abraham model without spin 
\cite{lorentzBOOKa, lorentzENCYCLOP}
obtains from the Abraham model with spin by simply discarding 
\refeq{eq:ALeqR} and setting $\vect{\omega}=\vect{0}$ identically 
in \refeq{eq:ALvfield}, so that \refeq{eq:ALeqT} goes over into
\begin{equation}
        \vect{0}
= 
        \int_{{\mathbb R}^3}
        \left[\EV(\xV,t) + \frac{1}{c}\dot\qV(t) \times\BV(\xV,t)\right]
        \fe(|\xV -\qV(t)|) \, \dd^3x\, ,
\label{eq:ALeqTnoGYRO}
\end{equation}
while the Maxwell--Lorentz equations for this model are the same as 
for the massive model without spin, i.e. again given by \refeq{eq:MLdivB}, 
\refeq{eq:MLdivE}, \refeq{eq:MLrotE}, \refeq{eq:MLrotB}, with the charge 
density \refeq{eq:ALcharge} and with the current density vector 
\refeq{eq:ALcurrentNOgyro}. 
	The continuity equation for the charge is the simpler 
\refeq{eq:continuityLAWinfI}, while the law of charge conservation  
is of course just \refeq{eq:ALcharge}.
	The conserved  energy, linear and angular momentum
are  given again by the purely electromagnetic expressions 
 \refeq{eq:totalenergyAL}, \refeq{eq:totalimpulsAL}, and 
\refeq{eq:totaldrehimpulsAL} that apply to the Abraham model with spin.
	The Abraham model without spin can also be obtained as 
the singular limit $\mz\to 0$  of the massive semi-relativistic 
model without spin under the hypothesis that $\dot\qV$ remains
bounded.

	For the Abraham model without spin a similar assessment holds
as for the Abraham model with spin.
	Again choosing the initial time to be $t=0$ and
the initial particle position to be $\qV_0 = \vect{0}$ 
(after at most a time and a space translation), the restrictions
for the field data are now determined by \refeq{eq:ALinitT} with 
$\vect{\omega}_0 =\vect{0}$ (and ignoring the discarded \refeq{eq:ALinitR}).
	We conclude that, if  $\langle\BV_0\rangle=\vect{0}$ then 
also  $\langle\EV_0\rangle=\vect{0}$ must hold, and if 
$\langle\BV_0\rangle \neq \vect{0}$, then 
$\langle\EV_0\rangle\cdot\langle\BV_0\rangle ={0}$ must hold. 
	Although these are milder restrictions than those for the Abraham 
model with spin, these restrictions on the field data in the Abraham 
model without spin still rule out various perfectly physical initial data, 
for instance those remarked on in the previous subsection, i.e.
$\qV_0=\vect{0}$, $\BV_0 = \vect{0}$, and 
$\EV_0 = \EV^{\text{hom}} + \EV^{\text{\textsc{Coulomb}}}$.

	As for the initial velocity of the particle $\dot\qV_0$ in 
given  consistently restricted initial fields, the following rules hold. 
	The initial velocity $\dot\qV_0$ is at our disposal only if  
$\langle\BV_0\rangle=\vect{0}$, and if 
$\langle\BV_0\rangle\neq\vect{0}$ it is of the form
\begin{equation}
	\dot\qV_0
= 
	c
\frac{\langle\EV_0\rangle \times\langle\BV_0\rangle}{|\langle \BV_0\rangle|^2} 
 + 	\beta \langle\BV_0\rangle
\, ,
\end{equation}
which, once again, leaves us only the choice of $\beta$.  

	Whether such a restricted, consistent set of initial 
data can now be continued into a unique regular solution of 
\refeq{eq:ALeqTnoGYRO}, \refeq{eq:MLdivB}, \refeq{eq:MLdivE}, 
\refeq{eq:MLrotE}, and \refeq{eq:MLrotB}, with \refeq{eq:ALcharge} 
and \refeq{eq:ALcurrentNOgyro}, is currently not rigorously known.
	As a matter of fact, there are good reasons to doubt that 
the evolution problem is well-defined even given consistent initial data.
	For notice that \refeq{eq:ALeqTnoGYRO} states that at 
\begin{equation}
	\langle\EV\rangle (t)
+ 	
	\frac{1}{c}\dot\qV(t)\times \langle\BV\rangle (t)
= 
	\vect{0}
\, ,\label{eq:ALeqTnoGYROt}
\end{equation}
for \emph{all} $t$, where the averages here are now defined by 
\begin{equation}
	\langle g\rangle(t)
 \defeg
	\frac{1}{-e}\int_{{\mathbb R}^3}g(\xV,t) \fe(|\xV-\qV(t)|)\dd^3 x
\, .
\end{equation}
	Hence, we can repeat our discussion of the conditions on the 
initial data essentially verbatim to conclude that:
\begin{itemize}
\item
	If at any time $t$ we have $\langle\BV\rangle(t) =\vect{0}$,
	then \refeq{eq:ALeqTnoGYROt} has no solution $\dot\qV(t)$ 
	at that instant $t$ unless also  $\langle\EV\rangle(t)=\vect{0}$;
\item
	If $\langle\BV\rangle(t) \neq \vect{0}$, then \refeq{eq:ALeqTnoGYROt}
	has no solution $\dot\qV(t)$ at $t$ unless also
	$\langle\EV\rangle(t)\cdot\langle\BV\rangle(t) ={0}$;
\item
	If $\langle\BV\rangle(t) \neq \vect{0}$ and 
	$\langle\EV\rangle(t)\cdot\langle\BV\rangle(t) ={0}$, then 
	\refeq{eq:ALeqTnoGYROt} states that $\dot\qV(t)$ is of the form
\begin{equation}
	\dot\qV(t) 
= 
	c
	\frac{\langle\EV\rangle(t)\times\langle\BV\rangle(t)}
	{|\langle\BV\rangle(t)|^2} 	
 + 	\beta(t) \langle\BV\rangle(t)
\, 
\end{equation}
with an \apriori\ undetermined coefficient $\beta(t)$.
\end{itemize}
	It does not seem very likely that this system of equations, coupled to 
the Maxwell--Lorentz equations, evolves a unique regular solution from a
consistent set of initial data. 

	While there is hardly any literature on the Abraham model with spin, 
the Abraham model without spin is discussed in most major texts since
Abraham~\cite{abrahamBOOK}, 
e.g.~\cite{lorentzBOOKb, panofskyphillipsBOOK, jacksonBOOK, 
rohrlichBOOK, yaghjianBOOK, sommerfeldBOOK}, 
and also in the more leisurely text of Peierls~\cite{peierlsBOOK}.
	However, none of these texts makes the elementary observation 
that the initial value problem for the purely electromagnetic Abraham 
model(s) is singular in the sense explained above, nor that Abraham's
`evolution' equation \refeq{eq:ALeqTnoGYROt} preserves this singular 
structure in time; some of the main conclusions drawn in
\cite{jacksonBOOK, lorentzBOOKb, panofskyphillipsBOOK, rohrlichBOOK, 
sommerfeldBOOK, yaghjianBOOK}  are even 
in open conflict with the singular nature of Abraham's evolution equations. 
	A re-assessment of these traditional treatments of the Abraham model 
seems to be called for.



\end{document}